\documentclass[letterpaper,12pt]{article}
\pdfoutput=1

\usepackage{jheppub}
\usepackage{amsmath,mathtools}
\usepackage{amssymb}
\usepackage{graphicx}
\usepackage{cancel}
\usepackage[dvipsnames]{xcolor}
\usepackage{color}
\usepackage{mathrsfs}
\usepackage{amsfonts}
\usepackage{dsfont}

\usepackage{subcaption}
\usepackage{float}
\usepackage{afterpage}
\allowdisplaybreaks

\def\({\left(}
\def\){\right)}
\def\[{\left[}
\def\]{\right]}

\newcommand{\be}{\begin{equation}}
\newcommand{\ee}{\end{equation}}

\newcommand{\ban}[1]{\begin{align}#1\end{align}}

\newcommand{\corr}[1]{\left< #1\right>}
\newcommand{\Tr}{\text{Tr}}
\newcommand{\ket}[1]{\left| #1\right>}
\newcommand{\bra}[1]{\left< #1\right|}

\title{The Torus Operator in Holography}

\author{Donald Marolf and Jason Wien}

\affiliation{Department of Physics, University of California, Santa Barbara, CA 93106, USA}

\emailAdd{marolf@physics.ucsb.edu}
\emailAdd{jswien@physics.ucsb.edu}

\abstract{
We consider the non-local operator  ${\mathcal T}$ defined in 2-dimensional CFTs by the path integral over a torus with two punctures. Using the AdS/CFT correspondence, we study the spectrum and ground state of this operator in holographic such CFTs in the limit of large central charge $c$. In one region of moduli space, we argue that the operator retains a finite gap  and has a ground state that differs from the CFT vacuum only by order one corrections.  In this region the torus operator is much like the cylinder operator. But in another region of moduli space we find a puzzle.  Although our ${\mathcal T}$ is of the manifestly positive form $A^\dagger A$, studying the most tractable phases of $\Tr( {\mathcal T}^n)$ suggests that ${\mathcal T}$ has negative eigenvalues.  It seems clear that additional phases must become relevant at large $n$, perhaps leading to novel behavior associated with a radically different ground state or a much higher density of states.   By studying the action of two such torus operators on the CFT ground state, we also provide evidence that, even at large $n$, the relevant bulk saddles have $t=0$ surfaces with small genus.
}

\makeatletter
\def\@fpheader{\relax}
\makeatother

\begin{document}
\maketitle

\section{Introduction}
\label{section:intro}

The Euclidean time-evolution operator $e^{-\beta H}$ is a central object in quantum field theory (QFT). We focus on conformal field theories in $d=2$ spacetime dimensions so that the matrix elements of $e^{-\beta H}$ are given by Euclidean path integrals over a cylinder of height $\beta$ and radius $r_0$, with the latter chosen to agree with the radius of the circle on which the QFT is defined. We thus refer to $e^{-\beta H}$ as the cylinder operator $C(\beta) = e^{-\beta H}$ below.   Interesting properties of this operator include the ground state of $\beta H = - \ln C(\beta)$, the partition function ${\cal Z} = {\rm Tr} \left( C(\beta) \right)$, the gap between the ground state and the first excited state, and other properties of the spectrum.

We restrict our attention to holographic CFTs; i.e., to families of CFTs that have a bulk AdS$_3$ dual in the limit of large central charge $c$.  For such theories the AdS$_3$ analogue of the Hawking-Page phase-transition \cite{Hawking:1982dh} is associated with the fact that the above-mentioned gap is of order $1$ at large $c$, while above some energy threshold $H$ has a black-hole-like density of states with entropy $S={2 \pi} \sqrt{E\, \ell\, c/3}$, assuming that the CFT spectrum is sufficiently sparse in the manner of \cite{Hartman:2014oaa}.

\begin{figure}[ht!]
\centering
\includegraphics[width=0.5\textwidth]{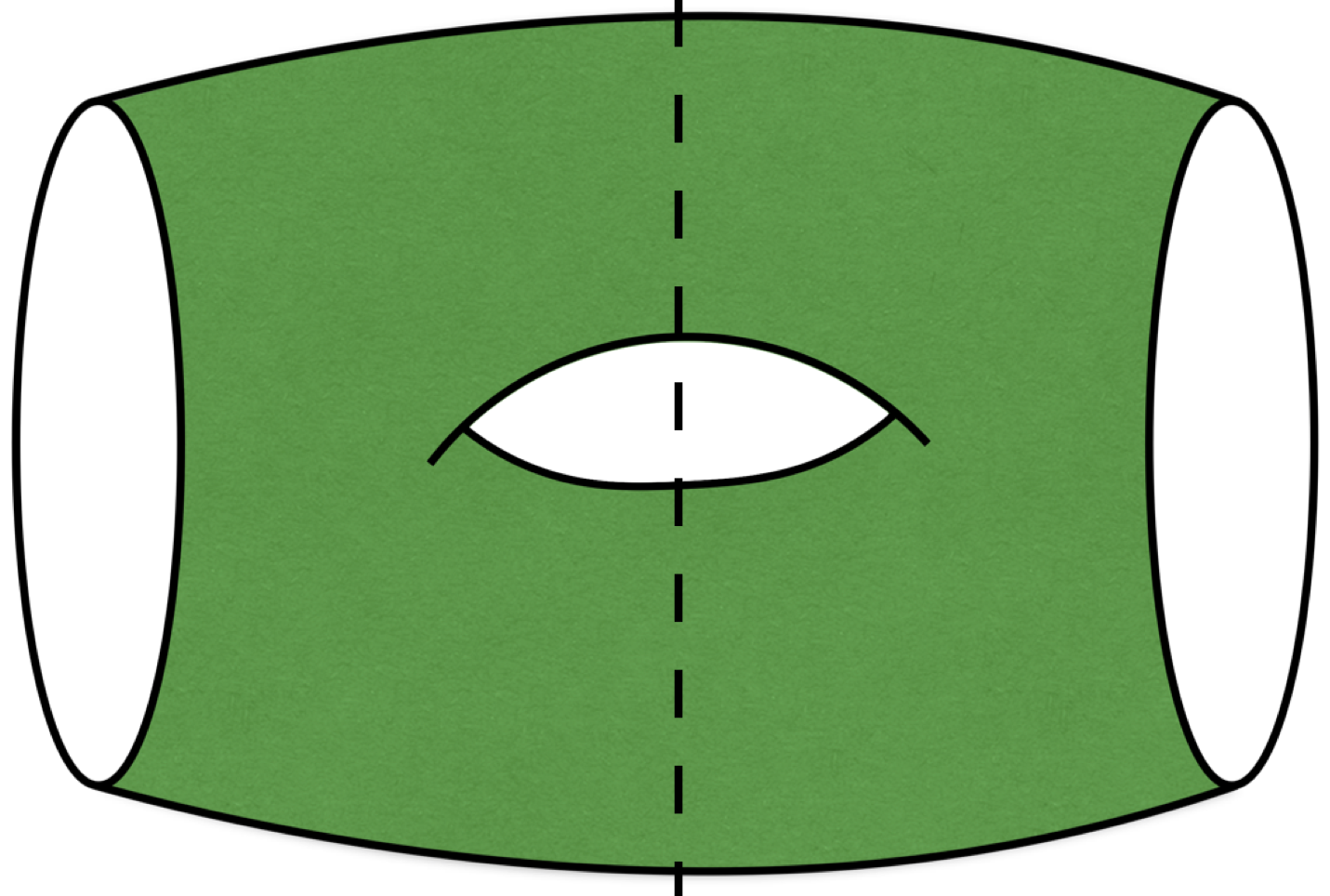}
\caption{
The surface $\mathscr T$ used to define the torus operator $\mathcal T$.  The surface has three ${\mathds Z}_2$  symmetries, acting on the page by reflecting front-to-back, right-to-left across the vertical dashed line, and top-to-bottom.\label{fig:operator} Cutting the path integral along the plane defined by the normal to the page and the vertical axis (dashed line) decomposes $\mathcal T$ into the product $A^\dagger A$ where $A$ maps our CFT Hilbert space ${\cal H}$ to   ${\cal H} \otimes  {\cal H}$.
}
\end{figure}

The goal of the present work is to study the effect of adding topology to the above path integrals.  In particular, we investigate CFT operators ${\cal T}$ defined by Euclidean path integrals over some twice-punctured torus $\mathscr T$ as drawn in figure \ref{fig:operator}.   We wish to understand how such properties depend on the moduli of $\mathscr T$, though we suppress these moduli from our notation. For simplicity we impose three ${\mathds Z}_2$ symmetries on $\mathscr T$ which act on figure \ref{fig:operator} by reflecting the page front-to-back, top-to-bottom, and right-to-left across the vertical dashed line.   This reduces the dimension of the moduli space to two.  Furthermore, by cutting the path integral along the dashed vertical line in figure \ref{fig:operator}, one sees that the reflection exchanging the two resulting pieces implies that $\mathcal T = A^\dagger A$ for the operator $A$ that maps our CFT Hilbert space ${\cal H}$ to   ${\cal H} \otimes  {\cal H}$ as
defined by the path integral over the right piece. It follows that $\mathcal T$ is manifestly non-negative and we may define the corresponding ``Hamiltonian'' $K := - \log \mathcal T$.  The largest eigenvalue of $\mathcal T$ corresponds to the ground state of $K$. We call this state $\ket 0_K$ and often refer to it simply as the torus operator ground state.  We also refer to eigenvalues of $K$ as ``$K$-energies."

A closely related problem was recently considered in \cite{MRW}, which studied path integrals over tori with a single puncture in holographic CFTs.  To understand the precise connection, recall that the CFT vacuum $|0\rangle$ is given by the Euclidean path integral over a disk.  The state ${\cal T}|0\rangle$ is thus computed by sewing a disk into the right-hand puncture of the twice-punctured torus $\mathscr T$.  The result is a once-punctured torus $\overline {\mathscr T}$, which is conformally equivalent to the once-punctured tori studied in \cite{MRW}.  In particular, we show in appendix \ref{appendix:moduli} below that our ${\mathds Z}_2$ symmetries do not significantly restrict the resulting moduli space.

\begin{figure}[h!]
\centering
\includegraphics[width=0.5\textwidth]{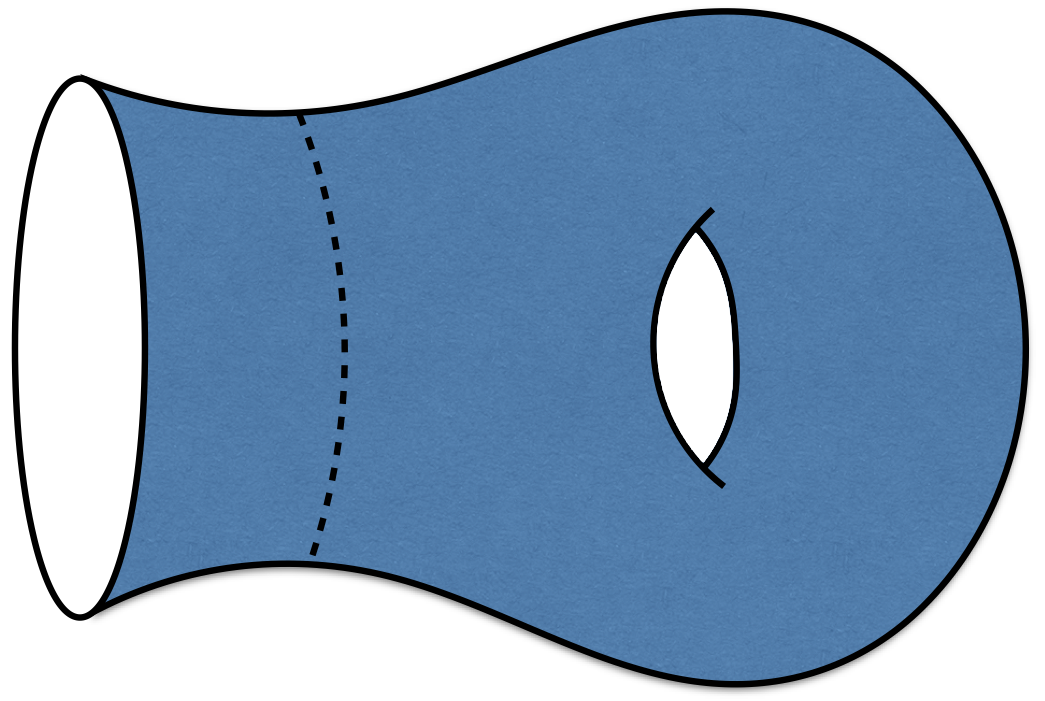}
\caption{
A cartoon of the $t=0$ surface for the toroidal geon state. The geometry consists of a single asymptotically AdS boundary separated from a genus one surface by a black hole horizon (dashed line). \label{figure:geon}
}
\label{fig:Tgeon}
\end{figure}

Now, as we review in greater detail in appendix \ref{appendix:moduli}, ref. \cite{MRW} identified two phases in the moduli space of $\overline {\mathscr T}$. In the first phase, the bulk dual of ${\cal T} |0\rangle$ is empty AdS$_3$  with $O(1)$ excitations for the bulk quantum fields.  But in the second phase the bulk dual of ${\cal T} |0\rangle$ contains a black hole with one asymptotic region and a (punctured) torus inside the horizon; see figure \ref{fig:Tgeon}. Such spacetimes are known as toroidal geons.  This raises the question of whether ${\cal T}$ might have other black-hole-like properties when the bulk dual of $\mathcal T \ket 0$ is a toroidal geon.  Indeed, we will show in section \ref{sec:GRSBP} that this transition also implies a phase transition for the ground state $|0_K \rangle$ of $K = - \log \mathcal T$.  While the actual properties of $|0_K \rangle$ remain unclear in the toroidal geon case, it is natural to ask if $|0_K \rangle$ might also be dual to a bulk black hole, what the internal topology of that black hole might be, and whether the density of states for ${\cal T}$ might become black-hole-like near $|0_K \rangle$ with $S = O(c)$ for large $c$.  Such a density of states would be high enough that we would then refer to it as a ``gapless'' or ``continuous" spectrum.

We begin in section \ref{section:gaps} by discussing how the above issues are related to properties of the partition functions $\Tr(\mathcal T^n)$.  Much of our work will analyze such partition functions using the dual AdS$_3$ system by further developing the techniques described in \cite{Krasnov1, Krasnov2, MRW}.  We therefore review the construction of the relevant (handlebody) bulk solutions in \S\ref{section:handlebody}, along with the computation of the associated bulk actions. We also discuss a heuristic for understanding which phase dominates a given path integral.  Using such techniques to study $\Tr(\mathcal T^n)$ in \S \ref{section:replica} then leads to a puzzle:  although $\mathcal T = A^\dagger A$ is manifestly non-negative, the most tractable phases -- and in particular those directly associated with the phases studied in \cite{MRW} --  suggest $\mathcal T$ to have negative eigenvalues.   In \S \ref{section:states} we then consider the action of $\mathcal T$ on states containing black holes with toroidal interiors.  We close with some discussion of possible resolutions in \S\ref{section:discuss}. Further technicalities related to our use of results from \cite{MRW} appear in appendix \ref{appendix:moduli}, and comments on numerical errors are relegated to appendix \ref{appendix:error}.

\section{Gaps, ground states, and traces}
\label{section:gaps}

\begin{figure}[ht!]
	\centering
\includegraphics[width=0.55\textwidth]{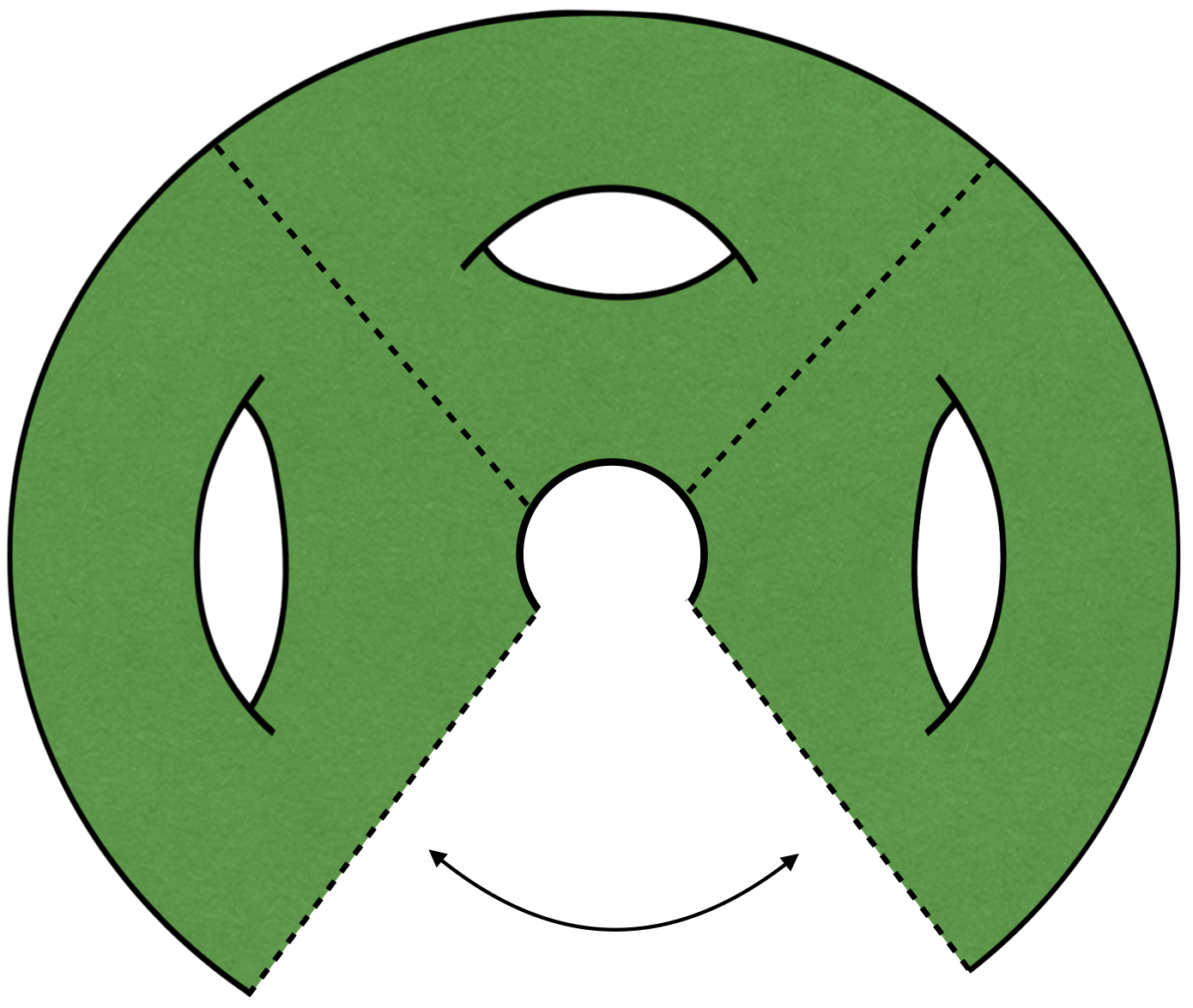}
\put(-115,7){\makebox(0,0){$n$}}
\caption{Partition function of genus $n+1$ that computes $\Tr(\mathcal T^n)$ in the CFT. \label{fig:genusN}}
\end{figure}

As mentioned above, a key question of interest is the density of $K$-eigenstates near the torus ground state $|0_K\rangle.$  In particular, we wish to distinguish between the case where $K$ has a gap $\Delta$ that does not vanish at large $c$ and the case where $\Delta \rightarrow 0$ as $c \rightarrow \infty$. We will probe this issue in section \ref{section:replica} by studying $\Tr(\mathcal T^n)$ for large $n$.  This trace is computed by evaluating the path integral over a Riemann surface consisting of $n$ copies of $\mathcal T$ glued together along the seams. As shown in figure \ref{fig:genusN}, the result is a Riemann surface of genus $n+1$ with an $n-$fold ``replica symmetry'' acting by a $2\pi/n$ rotation.

In particular, as explained in more detail in section \ref{section:handlebody}, we will further translate $\Tr(\mathcal T^n)$ into a bulk Euclidean path integral. For the 2d CFTs on which we focus, taking the limit of large central charge $c$ corresponds to taking $G_N  = \frac{3 \ell}{2c} \to 0$ with fixed AdS scale $\ell$. In this limit we may evaluate the gravitational partition function via the saddle point approximation using only the dominant bulk solution ($\text{dom.}$):
\ban{
\label{eq:Trn}
-\log \Tr(\mathcal T^n)=S_\text{EH}[g_\text{dom.}]\, .
}
In the rest of this section we describe how studying \eqref{eq:Trn} can both determine the value of $\Delta$ and characterize the ground state $|0_K \rangle.$

\subsection{Gapped or Gapless?}

To understand how properties of $\Delta$ relate to properties of the dominant bulk saddle, let us first suppose that $\Delta$ does not vanish at large $c$.  Then for large $n$ we have
\ban{
\label{eq:corr}
\Tr(\mathcal T^n) = e^{-\lambda_0 n}\left(1+O(e^{-\Delta n})\right)\, 	
}
where $\lambda_0$ is the smallest eigenvalue of $K$. Since exponential corrections as in \eqref{eq:corr} correspond to $O(c^0)$ effects in the bulk (for $\Delta = O(c^0)$) or sub-dominant bulk solutions (for $\Delta = O(c)$), for such cases there should be an $n_0$ of order $1$ such that the leading-order gravitational action of the dominant bulk solution becomes precisely linear for $n> n_0$.  We refer to this behavior as ``gapped,'' and in this case we can write for $n>n_0$
\ban{
- \ln {\rm Tr} \left( \mathcal T^{n} \right) \sim    n \lambda_0  + O(e^{-\Delta n}) \cdots \, .
}

On the other hand, if $\Delta$ vanishes as $c\rightarrow \infty$ then the corrections in \eqref{eq:corr} can be neglected only for $n \gtrsim 1/\Delta$. For $\Delta$ vanishing as a power of $c$ this should result from perturbative corrections in the bulk. It is difficult to imagine how this could be the case, so we instead focus on the case where the density-of-states becomes black-hole like and the gap is exponentially small (with $- \ln \Delta = O(c)$).  The resulting near-continuum of low energy states means that evaluating the trace requires us to sum over all the states in the associated band:
\ban{
\label{eq:nmIP}
{\rm Tr} \left( \mathcal T^{n} \right) &= \sum_{k=0}^\infty e^{-n( \lambda_0 + g(k)e^{-c\, \hat s} )}\, ,
}
where we have taken the spectrum of excitations above the ground state $k=0$ to be $g(k) e^{-c\, \hat s}$; i.e. to leading order the density of states is $s= c \, \hat s$ for some constant $\hat s$ with further $O(1)$ details determined by $g(k)$.

To understand the implications for a bulk dual, recall that in a theory with $c$ degrees of freedom it is natural to expect the number of states at each $k$ to scale like the volume $k^{c-1}$ of a $(k-1)-$sphere so that $g(k)\sim k^{1/c}$ at large $c$, or more generally that $g(k)\sim k^{1/(\alpha c)}$ for some $\alpha$ of order $1$. For small $ne^{-c \,\hat s}$ we can then approximate the sum in \eqref{eq:nmIP} by an integral to find
\ban{
\label{eq:asymptlin}
- \ln {\rm Tr} \left( \mathcal T^{n} \right) \sim    n \lambda_0 + \alpha \, c \log(n) + O(1) \cdots  \, .
}

Note that since the bulk saddle-point approximation is valid only for asymptotically large $c$, in this approximation $c$ is always taken large relative to $n$ and we may use \eqref{eq:asymptlin}  no matter how large $n$ may be, as $\lambda_0$ also scales with $c$.  The gravitational action of the dominant bulk saddle should thus also be given by \eqref{eq:asymptlin} at large $n$; i.e., it is never precisely linear in $n$, but only becomes approximately linear as $n\rightarrow \infty$. We refer to this behavior as ``gapless.''

\subsection{Characterizing the ground state}

Whether gapped or gapless, for large $n$ the operator ${\mathcal T}^n$ approximates $e^{-n\lambda_0}$ times the projector $|0_K \rangle \langle 0_K |$.  At least for even $n$, the analogous statement also holds for ${\mathcal T}^{n/2}$.  Note that this operator is defined by path integrals over an $n/2$-torus with two punctures.  Indeed, sewing together two copies of this Riemann surface along the two pairs of punctures gives the partition function ${\cal Z}(n) = {\rm Tr} ({\mathcal T}^{n} )$.  Reversing this logic, we see that ${\mathcal T}^{n/2}$ is obtained by cutting open the path integral for ${\cal Z}(n)$ along a pair of circles $C_1,C_2$.

Now, it is well known that cutting open a partition function path integral yields a state.  For $n$ even, let $|\psi_{{\mathcal T},n/2} \rangle$ be the state defined in this way by cutting ${\cal Z}(n)$ along the above two circles ($C_1,C_2$).  The relation between ${\mathcal T}^{n/2}$ and $|\psi_{{\mathcal T},n/2} \rangle$ is described by the time-reversal operator ${\mathbb T}$.
While time-reversal is often described as an anti-linear operator on the Hilbert space, it may be equivalently characterized as a map from ket-vectors $\langle \alpha|$ to bra-vectors $|{\mathds T} \alpha \rangle$.  We shall use the latter description.  As a result we may use ${\mathds T}$ to recode the information in any operator ${\cal O}$ as a state on two copies of the system. Choosing a basis $\{ | i \rangle\}$,
 we may write this recoding in the form
\begin{equation}
\label{eq:TactO}
|\psi_{\cal O} \rangle = {\mathds T} {\cal O} = {\mathds T} \sum_{i,j} |i \rangle {\cal O}_{ij} \langle j| := \sum_{i,j} {\cal O}_{ij} (|i\rangle \otimes |{\mathds T} j \rangle),
\end{equation}
where we have defined ${\mathds T}$ on $|i\rangle \langle j|$ to act trivially on the ket-vector and to map $\langle j|$ into $| {\mathds T}j \rangle$.
When ${\mathds T}$ is a symmetry of the system, it commutes with the Hamiltonian and we may choose a basis of energy eigenstates $\{ |E \rangle \}$ such that $|{\mathds T} E \rangle = |E \rangle$.  Equation \eqref{eq:TactO} then gives the well-known relation between the cylinder operator $C(\beta/2) = e^{-\beta H/2}$ and the thermofield double state $|TFD(\beta)\rangle$ at temperature $1/\beta$:
\begin{equation}
\label{eq:pTFD}
|\psi_{C(\beta/2)} \rangle = {\mathds T} e^{-\beta H/2} = \sum_E e^{-\beta E/2} (|E\rangle \otimes |E\rangle) = |TFD(\beta)\rangle.
\end{equation}
In the same way, we have
\begin{equation}
\label{eq:pTn2}
|\psi_{{\mathcal T},n/2} \rangle = {\mathds T}  {\mathcal T}^{n/2}.
\end{equation}
Due to the analogy between \eqref{eq:pTFD} and \eqref{eq:pTn2}, we refer to the latter as a thermofield-double-like state.

In \cite{maldTFD}, Maldacena studied the bulk dual of $|\psi_{C(\beta/2)} \rangle =|TFD(\beta) \rangle$ by cutting open the corresponding bulk path integral.
We will do the same below for the TFD-like states $|\psi_{{\mathcal T},n/2} \rangle$. Now, at large enough $n$, the operator ${\mathcal T}^{n/2}$ approaches $e^{-n\lambda_0/2} |0_K \rangle \langle 0_K|$, so (choosing an appropriate phase for $|0_K\rangle$) we find
\begin{equation}
\label{eq:TFDL1}
|\psi_{{\mathcal T},n/2} \rangle = {\mathds T} {\mathcal T}^{n/2} = e^{-n\lambda_0/2} \left( |0_K \rangle |0_K\rangle + O(e^{-n\Delta/2} \right).
\end{equation}
And as discussed above, for $\Delta$ of order $1$ or larger at large $c$, the bulk semi-classical approximation will not capture the exponentially small corrections.  So above some $n_0$, the state defined by cutting open the bulk path integral for ${\cal Z}(n)$ should be a product state proportional to two copies of $|0_K \rangle$.   One thus expects to find a disconnected pair of bulk spacetimes analogous to the pair of empty AdS$_3$ spacetimes dual to the thermofield double $|\psi_{C(\beta/2)} \rangle = TFD(\beta) \rangle$ at temperatures lower than the AdS$_3$ version \cite{maldTFD} of the Hawking-Page transition\footnote{As in \cite{maldTFD}, our Euclidean bulk saddles will all be invariant under a (Euclidean) time-reversal symmetry that fixes a bulk surface we call $t=0$. In the bulk semi-classical approximation, cutting open the path integral at $t=0$ amounts to using the induced metric at $t=0$ as initial data to construct a Lorentz-signature spacetime that is similarly invariant under time-reversal. In particular, the resulting spacetime has vanishing extrinsic curvature at $t=0$.}.

On the other hand, much as in our discussion above, for a black-hole-like density of states the semiclassical limit cannot access large enough $n$ to see the corrections in \eqref{eq:TFDL1} become small.  At any $n$ it will thus characterize
$|\psi_{{\mathcal T},n/2} \rangle $ as a highly entangled state.  The amount $S(n)$ of such entanglement then quantifies the density of states and should take on values typical of a bulk black hole for each $n$, with $S(n) \rightarrow 0$ as $n \rightarrow \infty$.  In such cases, the bulk dual of $|\psi_{{\mathcal T},n/2} \rangle $ should be a connected wormhole-like spacetime having two asymptotic regions, and where the corresponding entanglement wedges meet at a Hubeny-Rangamani-Takayanagi (HRT) surface \cite{HRT} of area $4GS(n)$.  Topological censorship theorems \cite{Galloway:1999bp} then imply that this wormhole lies inside a black hole.

As $n \rightarrow \infty$, the area of this HRT surface shrinks to zero and the spacetime approximately splits into two parts, each of which should again be closely related to the bulk dual of $|0_K \rangle.$  In particular, the topology of the bulk dual to $|0_K \rangle$ can be determined in this way.


\section{Handlebody Phases}
\label{section:handlebody}

We now review the use of bulk handlebodies to compute partition functions $Z[X]$ for $d=2$ holographic CFTs on Riemann surfaces $X$.  That the CFT is holographic means \cite{GKP,wittenholo} that we have,
\ban{
\mathcal Z_\text{CFT} = \mathcal Z_\text{grav.}\, , \label{eq:GKP}
}
where $\mathcal Z_\text{grav.} = \int \mathcal D g\, e^{-S_{EH}[g]}$ and $S_{EH}$ is the Einstein-Hilbert action with Newton constant $G_N$ and negative cosmological constant.\footnote{For exact equality, $\mathcal Z_\text{grav.}$ should be the full quantum gravity partition function including other fields, strings, higher derivative corrections, etc. But we will largely restrict attention below to the so-called universal sector in which the bulk is described by $S_{EH}[g]$ alone at leading order in large central charge $c$. As discussed in \cite{Seiberg:1999xz}, for known holographic dualities involving AdS$_3$, instabilities involving ``long strings" often imply that the dominant bulk solution does not in fact lie in the universal sector.  But at least when large black holes are present, we may expect the associated temperature to remove such instabilities in much the same way as the mass deformations described in \cite{MaldMaoz}.  We therefore ignore such instabilities until \S \ref{section:discuss}.}

The gravitational path integral is over Asymptotically Locally Anti-de Sitter (AlAdS) Euclidean bulk manifolds $M$ that satisfy the boundary condition $\partial M = X$, where $\partial M$ is the conformal boundary of $M$.
All solutions of this theory are quotients of global AdS$_3$, even if the solution features non-trivial topology and/or multiple boundary regions. An interesting class of bulk solutions are the so-called handlebodies, described by choosing an appropriate set of cycles on the Euclidean boundary to be contractible in the bulk. As above, we consider boundary conditions in Euclidean signature given by a compact Riemann surface $X$, where $X$ has at least one reflection symmetry, which we will call time-reflection. The bulk surface $\Sigma$ invariant under this reflection will be called the $t=0$ surface. We leave implicit any analytic continuation to Lorentzian signature.

For a given boundary Riemann surface $X$ of genus $g$, we can choose a basis $\{\alpha_i, \beta_j\}$ of cycles\footnote{Throughout this paper we use the term ``cycle'' to mean not only an element of $\pi_1(X)$ but in fact a particular curve in $X$ belonging to the associated equivalence class.} for the homotopy group $\pi_1(X)$, such that $\alpha_i \cap \beta_j = \delta_{ij}$ and
\ban{
\prod_{i=1}^g \alpha_i^{-1} \beta_i^{-1} \alpha_i \beta_i = 1 \, .
}
Given a such basis, we can define a bulk manifold with boundary $X$ by declaring the cycles $\{\alpha_i\}$ to be contractible in the bulk while the cycles $\{\beta_i\}$ remain non-contractible. To be explicit, we take the smallest normal subgroup $\Gamma$ of $\pi_1(X)$ generated by the $\alpha_i$  and define the bulk to have homotopy group $\pi_1(M) \coloneqq \pi_1(X)/\Gamma$; see \cite{Ford,TZ} for further discussion of these Schottky groups.
It is known that all such $\pi_1(M)$ can be embedded in the group of AdS${}_3$ isometries so that we may construct $M$ as the quotient $\text{AdS}_3/\pi_1(M)$.

In this way, each possible basis for the boundary Riemann surface $X$ defines a handlebody geometry with boundary $X$, and we refer to these bulk solutions as the set of handlebody phases. While they are not the only solutions for a particular set of boundary conditions, it has been conjectured that the non-handlebody solutions are always sub-dominant \cite{Yin}, and indeed certain non-handlebody solutions are known to be forbidden by AdS/CFT \cite{MaldMaoz}.

We wish to compute the Euclidean gravitational path integral using a saddle point approximation, so for each boundary $X$ we need to find the bulk phase with least action. It is often assumed that phases which break the symmetry of the boundary are sub-dominant to the ones which preserve the symmetry, and for simplicity we will often focus on bulk manifolds with fundamental groups $\pi_1(M)$ which preserve the symmetries of $\pi_1(X)$.  We will use a normalization in which the action for certain non-handlebody solutions vanishes, and we will see explicitly that the dominant action is always negative for the subspace of moduli space considered.

\begin{figure}[ht!]
\centering
\includegraphics[width=0.5\textwidth]{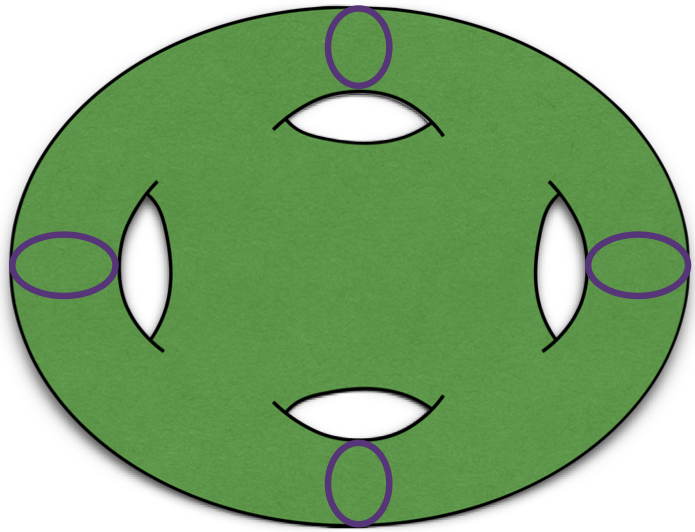}
\caption{
The naive handlebody phase for a genus 4 boundary with the $\alpha$ cycles drawn in purple. The four $\beta$ cycles are not drawn but each circle one of the four holes. \label{fig:naive}
}
\end{figure}

We can generate a large class of phases by the following algorithm. For simplicity we work with the homology group, and therefore we might miss phases with the same bulk homology but different bulk homotopy. However, we do not expect this restriction to affect our main results. First, we embed $X$ into $\mathbb R^3$ and choose a standard basis in which the $\alpha$ cycles go around the ``handles'' and the $\beta$ cycles go around the ``holes'' as shown in Figure \ref{fig:naive}.
This phase is referred to as the ``naive handlebody,'' as it corresponds to filling in the bulk manifold with a solid handlebody as suggested by the picture. We then construct new phases by considering the image of these cycles under an arbitrary element $M$ of the mapping class group and choosing the cycles $M(\{\alpha\})$ to be contractible in the bulk. Note that this procedure can generate phases that break symmetries of $X$. One might expect such phases to be subdominant, though this is not always the case.

Some of the resulting phases were originally described in \cite{Brill1,Brill2}, with their application to holography described in \cite{Skenderis}. A more systematic treatment in terms of Schottky uniformization and a recipe for computing their actions were then derived in \cite{Krasnov1, Krasnov2}. Numerical techniques for evaluating these actions and matching moduli between phases were later  introduced in \cite{MRW}. The remainder of this section reviews these techniques for use in \S\ref{section:replica} and \S\ref{section:states}.

\subsection{Schottky Uniformization}
\label{sec:SU}

All solutions of Einstein's equation in 2+1 dimensions with negative cosmological constant can be constructed by taking a quotient of Euclidean AdS$_3$ by some subset of the symmetry group PSL$(2,\mathbb C)$. In the half-space representation of Euclidean AdS$_3$, the quotient operations act to identify pairs of hemispheres anchored to the boundary. If we think of the boundary as the complex plane $\mathbb C$, then on the boundary the identifications are given by M\"obius transformations which identify pairs of circles. We can thus specify a particular bulk quotient by describing these pairs of boundary circles, with the quotient extending into the bulk by acting on hemispheres anchored to each such boundary circle. Any connected part of the bulk region that remains after these hemispheres are removed can serve as a fundamental domain for the quotient handlebody.

Representing a Riemann surface in terms of pairs of circles identified on $\mathbb C$ is known as a Schottky uniformization. In general, this uniformization is not unique. More explicitly, we think of $\mathbb C \cup \{\infty \}$ as the Riemann sphere and
consider the domain defined by removing $2g$ non-intersecting closed disks.  The boundaries of these disks are $2g$ circles, which we group into $g$ pairs $(C_i,C_i')$. Further let the M\"obius transformation $L_i$ map the interior of $C_i$ (defined by the embedding in $\mathbb C$) to the exterior of $C_i'$.

As defined, each $L_i$ is a loxodromic transformation, meaning that it is conjugate to the transformation $w\mapsto q_i^2 w$ for some $q_i \in \mathbb C$ with $0<|q_i|<1$. Taking the quotient by the group generated by $\{L_i\}$ defines a Riemann surface of genus $g$ with a fundamental domain $D$ given by the region exterior to all of the circles. Each transformation $L_i$ is associated with three complex free parameters, while we are free to choose overall normalizations which fix three of these $3g$ parameters. The space of surfaces described in this manner is thus $\mathbb C^{3g-3}$, which matches the moduli space of a genus $g$ Riemann surface. In fact, the Koebe retrosection theorem \cite{koebe} tells us that all compact Riemann surfaces can be described in this way.

To connect the choice of Schottky uniformization for the boundary Riemann surface to the choice of bulk handlebody phase, note that the circles $C_i$ on the boundary are contractible in the bulk since each $C_i$ can be shrunk to a point by contracting the cycle along the corresponding bulk hemisphere.  As a result, to construct a bulk in which $\{\alpha_i\}$ are contractible, we need only find a Schottky uniformization with $C_i = \alpha_i$.

The cycles $\beta_i$ that are dual to $\alpha_i$ remain non-contractible, as these are the cycles that run from $C_i$ to $C_i'$ on the boundary. Each such cycle is associated with the transformation $L_i$, and any non-contractible bulk cycle given by a product of $\beta_i$ can be associated with some $L$ which is correspondingly a product of the $L_i$. Additionally, for any non-contractible cycle there is a unique geodesic representative of the associated homotopy class. Since all loxodromic $L$ have $\Tr\, L >2$,  as shown in \cite{Maxfield3D} the length of this geodesic is
\ban{
\ell(L) = 2 \cosh^{-1} \frac{\Tr \,L}{2} \, \label{eq:geodL}.
}
This technology provides a useful way to understand the bulk topology and compute the lengths of various horizons and geodesics.

Given a Schottky representation of a boundary Riemann surface, the topology of the ``$t=0$'' bulk slice can be determined as follows. Suppose $n$ of the $g$ pairs of circles $\{C_i,C_i'\}$ lie on the boundary $t=0$ slice.\footnote{Since the bulk $t=0$ surface is invariant under a reflection symmetry of the boundary, if $C_i$ lies on the boundary slice then so must $C_i'$.} Additionally, suppose these identifications create $b$ disjoint boundary circles at $t=0$. The number of holes in the bulk $t=0$ slice is then given by
\ban{
g_{t=0} =\frac{ n-b+1}2\, \label{eq:bulkgenus}.
}
This formula follows from an analog of the ``doubling'' construction $X = 2Y$ as pointed out in \cite{Krasnov2}, using only the subset of identifications that act on the $t=0$ slice. As the boundary identifications are extended into the bulk along hemispheres, the quotient of the bulk $t=0$ slice is determined precisely by the identifications which act on the intersection of this slice with the boundary.

To construct a Schottky domain for a particular bulk handlebody, however, it is more convenient to reverse this procedure. Given a handlebody phase described by cycles $\{\alpha_i\}$ on $X$ to be made contractible in the bulk, we can construct a Schottky uniformization as follows. First, for each $i$ cut along the geodesics homologous to $\alpha_i$, calling each side of the cut $C_i$ and $C_i'$. Cutting the Riemann surface in this manner defines a sphere with $2g$ punctures. Next, project this punctured sphere into the plane $\mathbb C$. The resulting circles and maps $L_i$ identifying $C_i$ and $C_i'$ are precisely the ingredients needed for Schottky uniformization. In \S \ref{section:replica} and \S\ref{section:states} we provide some explicit examples; further examples can be found in \cite{MRW, cones}.

Finally, in comparing different handlebody phases with the same boundary conditions we must be sure that the moduli of the two boundary Riemann surfaces agree. Determining the moduli of a Riemann surface from its Schottky representation requires defining a boundary conformal frame in which to compute cycle lengths. While some of the moduli are fixed by symmetry, others must be fixed by computing the lengths of certain cycles. In practice this moduli matching problem is the most difficult part of constructing a phase diagram for higher genus partition functions. However, there is a useful heuristic which can sometimes be used as a shortcut.

From the numerical results of \cite{MRW, cones} and results presented in \S \ref{section:replica} and \S\ref{section:states}, one observes that the action tends to be a monotonic function of the sum of the lengths of boundary cycles chosen to be contractible. When this sum is large the action tends to be more positive, and when this sum is small the action tends to be more negative. We therefore posit the heuristic that for a particular Riemann surface $X$, the phase which dominates the partition function is the one in which the sum of contractible cycle lengths is minimized, or simply ``small boundary cycles like to pinch off.'' We will use this heuristic to try to gain some intuition for the results in \S \ref{section:replica} and \S\ref{section:states}, and we will also test the heuristic against numerical computations of the actions.

One caveat in applying this heuristic is that for a given bulk phase there may be multiple choices of contractible boundary cycles one can use to define it, and so in applying the heuristic one needs to consider the choice with minimal total length. While this heuristic is known not to be exact (for example it fails near the AdS/toroidal geon phase boundary in \cite{MRW}), it is still useful for building intuition about which phase dominates a given partition function. Moreover, in many cases there is a symmetry relating two bulk phases at a particular point in moduli space; see e.g. the example in appendix A.3 of \cite{MBW1}. At the symmetry point there is a choice of basis in which the action and the total length of contractible cycles are equal in each phase. Moving away from this point as the total length of contractible cycles decreases, the action typically becomes more negative, and if for the dual phase the total length of contractible cycles increases then the heuristic is exact.

\subsection{The Boundary Metric and Bulk Action}

Ultimately we will be interested in comparing the actions of different handlebody phases in order to determine the dominant semi-classical bulk geometry. We must therefore regulate the action by choosing a particular conformal frame. We do so by choosing the boundary to have constant Ricci scalar $R_\text{bndy} = -2$ in AdS units.

Using coordinates $w= x + i y$, we can write the boundary metric as
\ban{
ds^2_\text{bndy} = e^{2\phi} |dw|^2\, ,
}
where regularity of the metric under the quotient by $L_i$ requires
\ban{
\phi(L_i(w)) = \phi(w) -\frac 12 \log \left| L'_i(w)\right|^2 \label{eq:bcs} \, .
}
The requirement $R_\text{bndy}=-2$ is equivalent to choosing $\phi$ to satisfy the Liouville equation
\ban{
\nabla^2 \phi = 4 \partial_w \partial_{\bar w}\phi = (\partial_x^2 + \partial_y^2) \phi = e^{2\phi} \label{eq:fieldeqn}\, ,
}
subject to the boundary conditions \eqref{eq:bcs}. In this way, the problem of finding the boundary conformal frame is reduced to solving the scalar field equation \eqref{eq:fieldeqn} on the Schottky domain $D$ with boundary conditions \eqref{eq:bcs}. We will do so in \S\ref{section:replica} and \S\ref{section:states} using the numerical methods described in \cite{MRW} and reviewed in the next subsection.

As shown in \cite{Krasnov1}, with this choice of conformal frame the evaluation of the Einstein-Hilbert action for a particular solution can be written in terms of the Takhtajan-Zograf action \cite{TZ} for the scalar field $\phi$:
\ban{
I = - \frac c{24\pi} \left[ I_\text{TZ}[\phi] - A - 4 \pi (g-1)(1-\log 4 R_0^2)\right]\, ,
}
where $A$ is the area of the boundary and $R_0$ is the radius of the sphere for which the partition function is one. We will set $R_0=1$ in the results section. As explained in \cite{MRW}, if we define $R_k$ to be the radius of $C_k$ and $\Delta_k$ as the distance between the center of $C_k$ and the point $w_\infty^{(k)}$ mapped to $\infty$ by $L_k$, this action reduces to
\ban{
I_{TZ}[\phi] = \int_D d^2 w\left( \left(\nabla \phi\right)^2 + e^{2\phi} \right) + \sum_k \left(\int_{C_k} 4 \phi\, d\theta_\infty^{(k)}  - 4 \pi \log \left |R_k^2 - \Delta_k^2 \right|\right)\,,
}
where $\theta_\infty^{(k)}$ is the angle measured from the point $w_{\infty}^{(k)}$.  If we can further reduce $D$ by some set of symmetries, this action can take an even simpler form as shown in \cite{MRW,cones}. As in \cite{cones}, we introduce Jacobian factors to turn all of the integrals over $\theta^{(k)}_\infty$ into numeric integrals over $\theta^{(k)}_0$, i.e. about the center of each circle.

\subsection{Numerical Methods}
\label{sec:methods}
Equation \eqref{eq:fieldeqn} on the Schottky domain $D$ with boundary conditions \eqref{eq:bcs} is generally difficult to solve analytically. Following \cite{MRW}, we thus proceed numerically using finite element methods and the Newton-Raphson algorithm. See \cite{FEMgentle,FEMlecture} for introductions to finite element methods.

In all cases of interest, we may write the boundary of our domain as $\partial D = \bigcup_i \partial D_i$ where $\partial D_i$ is an arc\footnote{Straight line segments are ``arcs'' of infinite-radius circles.} of a circle with radius $R_i$, where each $\partial D_i$ is the fixed point set of some involution or reflection symmetry of $D$. As shown in appendix A of \cite{MRW}, we can then use the boundary conditions \eqref{eq:bcs} to find
\ban{
\left.\nabla_n \phi \right|_{\partial D_i } = - \frac 1 {R_i} \, \label{eq:bcred}.
}

To solve \eqref{eq:fieldeqn} using the Newton-Raphson algorithm, we first write $\phi = \phi_{(n)} + \delta \phi_{(n)}$ and solve a linearized equation for $\delta\phi_{(n)}$. We then set $\phi_{(n+1)} = \phi_{(n)}+ \delta\phi_{(n)}$ and similarly solve a linearized equation for $\delta\phi_{(n+1)}$. We repeat this process until $||\delta\phi_{(n+1)}||_\infty < 10^{-10}$.

At stage $n$ in the Newton-Raphson algorithm, we can integrate the linearized equation against a test function $\psi$, and integrating by parts to incorporate the reduced boundary conditions \eqref{eq:bcred} we have the following equation for $\delta \phi_{(n)}$:
\ban{
- \int_D \nabla \psi\cdot \nabla \delta \phi_{(n)}- 2 \int_D \psi \, e^{2\phi_{(n)}} \, \delta\phi_{(n)} = \int_D \nabla \psi \cdot \nabla \phi_{(n)} + \int_D \psi \, e^{2\phi_{(n)}} + \sum_{i}\frac {\sigma_i}{R_i} \int_{\partial D_i} {\psi}\, d \theta_i \, ,
}
where $\sigma_i = \pm 1$ when the orientation of $\partial D_i$ as part of $\partial D$ is counter-clockwise or clockwise respectively. With an initial seed of $\phi_{(0)}=0$, we can now use this equation and standard finite element techniques to solve for $\phi$.

To match moduli between different phases, we need to compute the lengths of various geodesics on the boundary. In the case where a geodesic is fixed by a symmetry of the domain $D$, we can explicitly compute its length by numerically integrating the boundary metric over the appropriate curve. However, in some cases the domain $D$ breaks some of the symmetries of $X$ (even though the handlebody solution does not), and in practice it is difficult to numerically solve for the associated geodesic.

However, we can instead compute the boundary geodesic lengths by mapping the domain $D$ to a subset of the Poincar\'e disk. We do so using the numerical solution for $\phi$ to compute the length of each boundary segment $\partial D_i$. By symmetry each $\partial D_i$ is a geodesic that orthogonally intersects the adjacent segment $\partial D_j$. Knowing that the metric has been chosen so that $R_\text{bndy} =-2$, we can then construct a region in the Poincar\'e disk bounded by orthogonally-intersecting geodesic segments of the same lengths. The geometry of this region must then exactly match the geometry of $D$. Since the length of any geodesic segment in the Poincar\'e disk can be computed using a simple analytic formula, we can use this correspondence to easily compute geodesic length in our domain $D$.


\section{Computing $\Tr(\mathcal T^n)$}
\label{section:replica}

We now use the technology of section \ref{section:handlebody} to compute $\Tr(\mathcal T^n)$ and to study the associated TFD-like states ${\mathds T} {\cal T}^{n/2}$ obtained by cutting open the corresponding path integral.  As explained in section \ref{section:gaps}, characterizing the large $n$ bulk duals of these states would tell us if ${\cal T}$ is gapped or gapless, and would also give the bulk dual of the ground state $|0_K\rangle$.

Since any Riemann surface $X$ is associated with an infinite number of bulk handlebody saddles, it will not be possible to study them all in detail.  Indeed, even classifying the full set of possible phases is an onerous task.  We will therefore proceed pragmatically, beginning in section \ref{sec:RSP} with bulk handlebodies that preserve the boundary replica symmetry. We then note in section \ref{sec:GRSBP} that the results of \cite{MRW} imply that at least one phase breaking this replica symmetry has lower action, though the positivity of ${\cal T}$ forbids that particular phase from being the most dominant.  This raises a puzzle, as other natural alternatives for the most dominant phase also suffer from the same issue.

\subsection{Replica Symmetric Phases}
\label{sec:RSP}

\begin{figure}[ht!]
	\centering
	\includegraphics[width=0.55\textwidth]{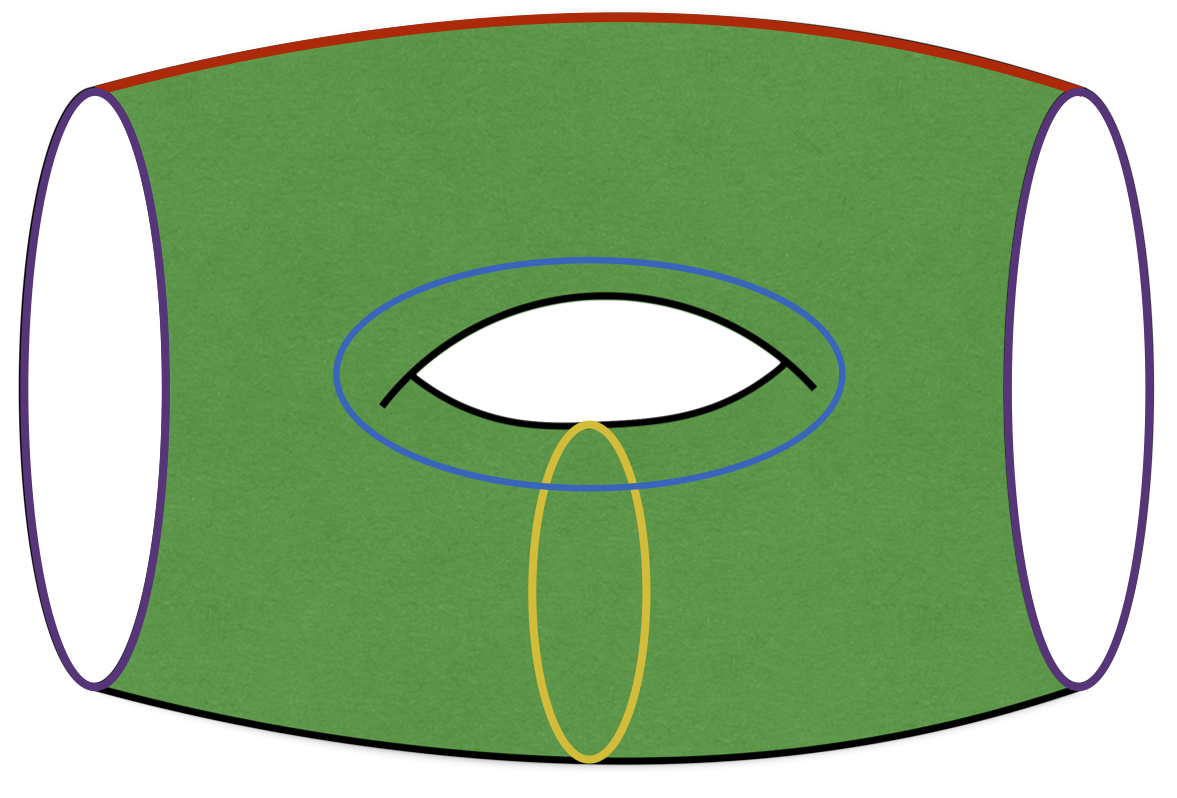}
	\put(-115,-5){\makebox(0,0){$\alpha_i$}}
	\put(-120,90){\makebox(0,0){$\beta_i$}}
	\put(-120,173){\makebox(0,0){$\beta_0$}}
	\put(-220,10){\makebox(0,0){$\alpha_0$}}
	\caption{The fundamental unit $\mathcal T$ with cycles labeled. Only $1/n^{\text{th}}$ of $\beta_0$ is drawn. \label{fig:unit}}
\end{figure}
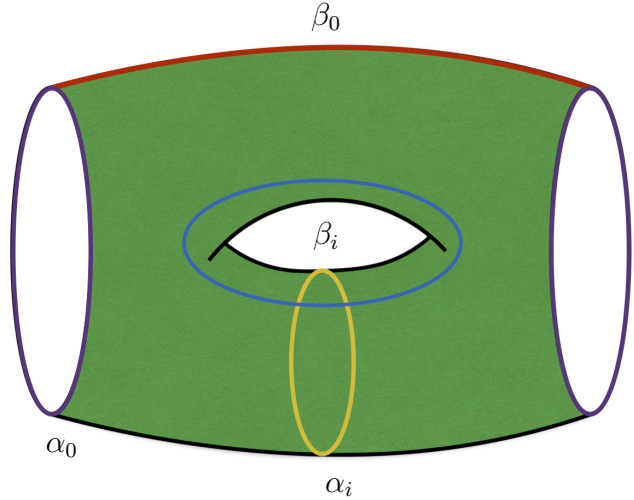

We first consider phases which explicitly preserve the replica symmetry.\footnote{That is, we study phases whose Schottky domains respect the replica symmetry. There are more complicated phases that preserve replica symmetry, even though the symmetry is broken by the Schottky representation.} To catalog such phases we can restrict our attention to the fundamental unit $\mathcal T$ drawn in figure \ref{fig:unit}.
We consider four distinct phases of this partition function, divided into two classes. The first class are called AdS phases, given by the choice of $\{\alpha_0, \alpha_1, \cdots, \alpha_n\}$ or $\{\alpha_0, \beta_1, \cdots, \beta_n\}$ contractible. In both of these phases, if any of the $n$ choices of cycles $\alpha_0$ is contractible, then the image of $\alpha_0$ under $\frac{2\pi}n k$ rotation is also contractible. As a result, each $\alpha_0$ cycle (purple in figure \ref{fig:unit}) bounds a slice of the bulk with vanishing extrinsic curvature and the geometry of the Poincar\'e disk. This is the Lorentzian-signature initial data for global AdS$_3$, so the bulk geometry for $\mathcal T^{n/2}$ (with even $n$) is just a pair of global AdS$_3$ geometries. We refer to such saddles as AdS phases for the TFD-like state.

When $\alpha_0$ is contractible we can use a trick described in \cite{MRW} to build phases for higher genus Riemann surfaces from lower genus phases. The boundary conditions on $\phi$ allow us to glue together Schottky domains along contractible geodesics, and we can glue together $n$ copies of the unit $\mathcal T$ along the contractible geodesics $\alpha_0$. In the bulk, this gluing occurs along the associated hemispheres (which have vanishing extrinsic curvature).
In this way, when $\alpha_0$ is contractible the action is given by $n$ times the action for one of the units. The action for such AdS phases is exactly linear in $n$, and so these are gapped phases. If the operator $\mathcal T$ is gapped we would expect a phase of this general sort to dominate for large $n$, though there remains the possibility that the fundamental unit of the dominant phase could be larger than ${\mathcal T}$. Note that in defining the action of a fundamental unit we are free to include a Gibbons-Hawking boundary term $\frac{1}{8\pi G_N}\int K$ at any finite boundary (e.g., on the plane passing through an $\alpha_0$ cycle at either end of figure \ref{fig:unit}).  Since the extrinsic curvature vanishes on such boundaries, this boundary term does not affect the numerical value of the action.

\begin{figure}[ht!]
	\centering
	\begin{subfigure}{0.47\textwidth}
		\includegraphics[width=\textwidth]{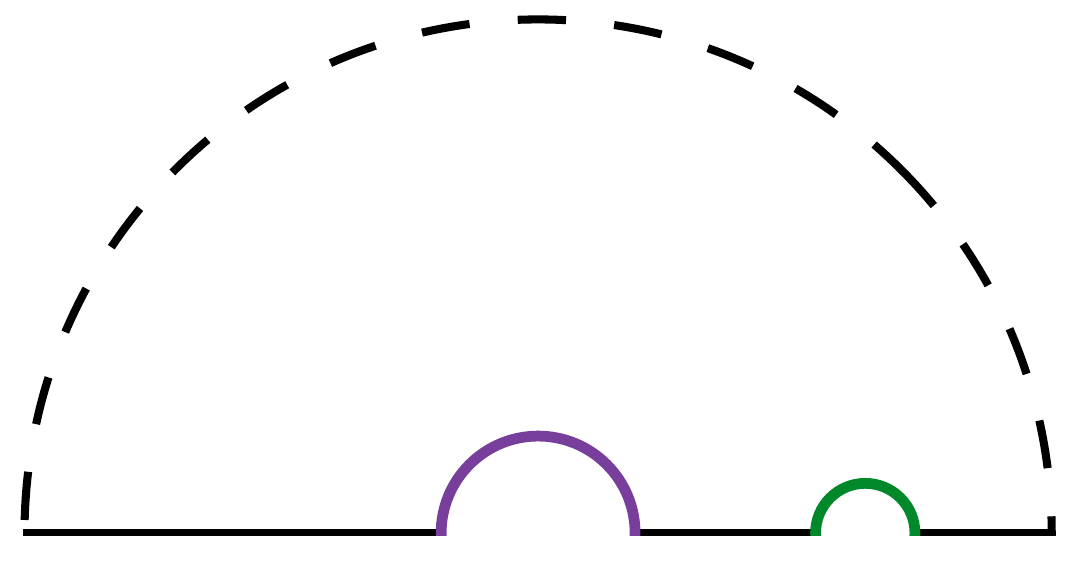}
		\put(-38,5){\makebox(0,0){$\alpha_i$}}
		\put(-100,114){\makebox(0,0){$\tilde \alpha_i$}}
		\put(-102,15){\makebox(0,0){$\alpha_0$}}
		\put(-160,-3){\makebox(0,0){$\beta_0$}}
		\put(-15,-3){\makebox(0,0){$\beta_i$}}
		\put(-65,-3){\makebox(0,0){$\tilde\beta_0$}}
		\vspace{0.1cm}
		\subcaption{AdS phase with $\{\alpha_0, \alpha_{i}\}$ contractible}
		
	\end{subfigure}
	\hfill
	\begin{subfigure}{0.47\textwidth}
		\includegraphics[width=\textwidth]{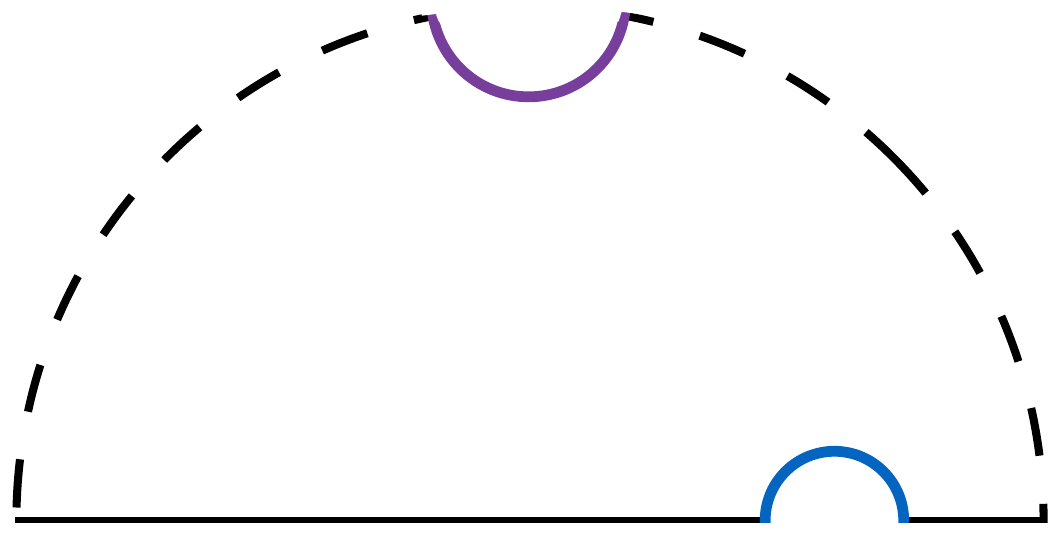}
		\put(-45,7){\makebox(0,0){$\beta_i$}}
		\put(-103,95){\makebox(0,0){$\alpha_0$}}
		\put(-23,80){\makebox(0,0){$\tilde \beta_0$}}
		\put(-177,85){\makebox(0,0){$\beta_0$}}
		\put(-125,-5){\makebox(0,0){$\tilde \alpha_i$}}
		\put(-15,-5){\makebox(0,0){$\alpha_i$}}
		\vspace{0.1cm}
		\subcaption{AdS phase with $\{\alpha_0, \beta_i\}$ contractible}
		
	\end{subfigure}
	\caption{One quarter of the Schottky domains used to construct the two AdS phases, reduced by the reflection symmetries in the x-axis and the inversion symmetry through the unit circle (dashed). Various boundary cycles are labeled in each phase.  }
	\label{fig:AdSphases}
\end{figure}
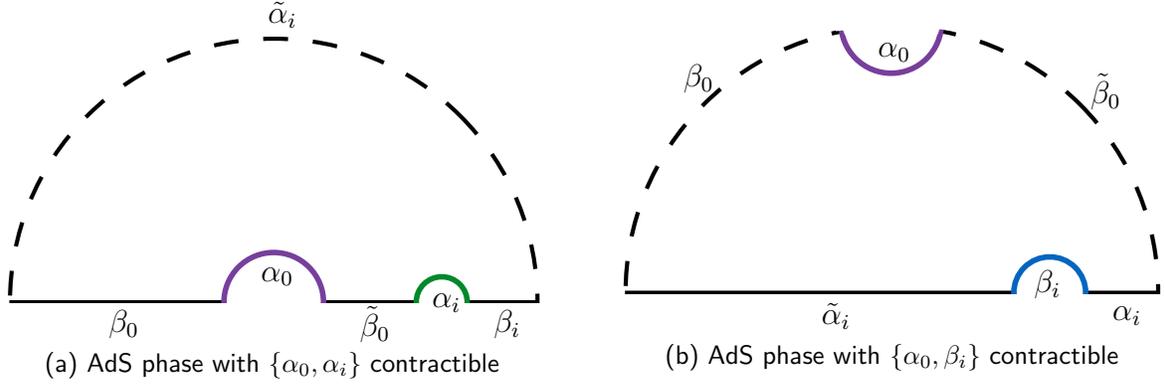

The Schottky domains used to represent the above AdS phases are shown in figure \ref{fig:AdSphases}.
In each of these domains we include the cycles $\tilde \beta_0$ and $\tilde \alpha_i$, which are the images of $\beta_0$ and $\alpha_i$ under reflection across the horizontal line of symmetry in figure \ref{fig:unit}. Each domain has four free real parameters, which are reduced to two by imposing $|\tilde \beta_0| = |\beta_0|$ and $|\tilde \alpha_i| = |\alpha_i|$.

The second class of phases we call BTZ phases, given by the choice of $\{\beta_0, \beta_1, \cdots, \beta_n\}$ or $\{\beta_0, \alpha_1, \cdots, \alpha_n\}$ contractible.\footnote{Note that there is another phase given by $\{\tilde \beta_0,\tilde \alpha_i\}$ contractible, but this phase will have exactly equal action by the symmetry.} In these phases the moment of time symmetry in the bulk looks like a BTZ wormhole, and the action does not have a simple dependence on $n$. The trick we used to paste together AdS phases does not work, as these phases cannot be constructed by pasting together lower genus units along contractible geodesics. We thus expect the BTZ phases to be gapless.

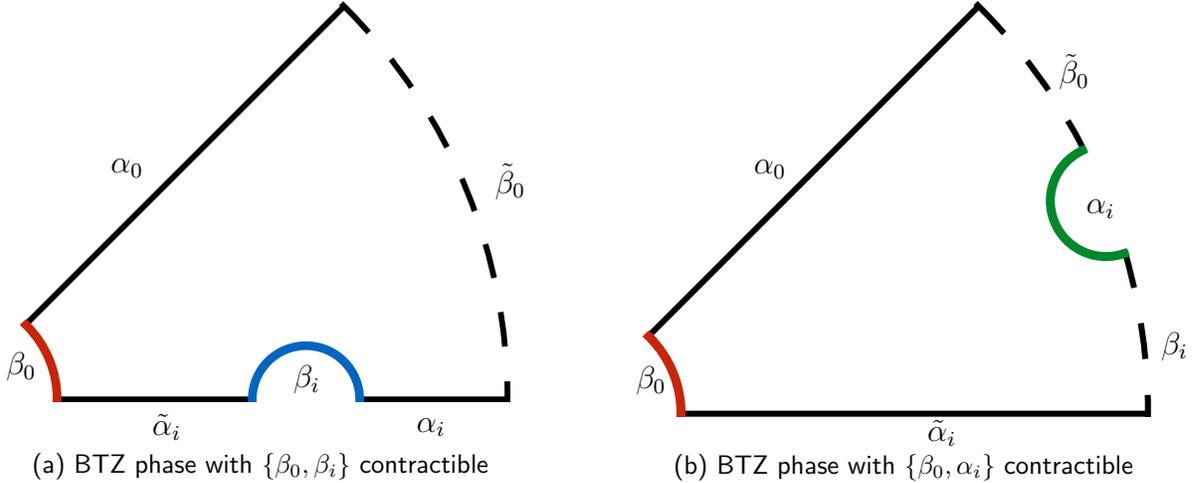
\begin{figure}[ht!]
	\centering
	\begin{subfigure}{0.45\textwidth}
		\includegraphics[width=\textwidth]{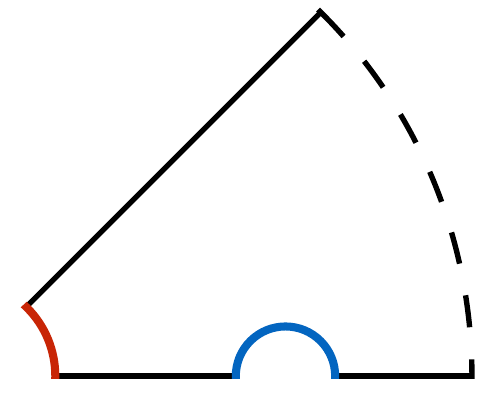}
		\put(-82,20){\makebox(0,0){$\beta_i$}}
		\put(-190,25){\makebox(0,0){$\beta_0$}}
		\put(-5,95){\makebox(0,0){$\tilde\beta_0$}}
		\put(-35,2){\makebox(0,0){$\alpha_i$}}
		\put(-150,100){\makebox(0,0){$\alpha_0$}}
		\put(-135,2){\makebox(0,0){$\tilde\alpha_i$}}
		\subcaption{BTZ phase with $\{\beta_0, \beta_{i}\}$ contractible}
		
	\end{subfigure}
	\hfill
	\begin{subfigure}{0.45\textwidth}
		\includegraphics[width=\textwidth]{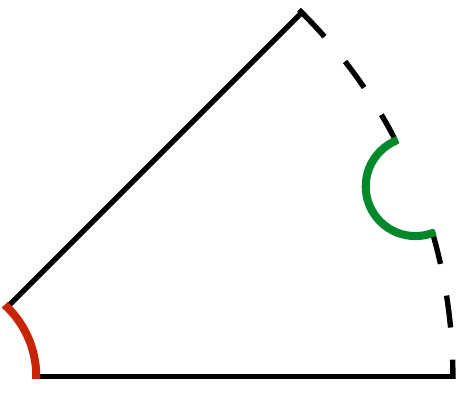}
		\put(3,33){\makebox(0,0){$\beta_i$}}
		\put(-195,20){\makebox(0,0){$\beta_0$}}
		\put(-25,85){\makebox(0,0){$\alpha_i$}}
		\put(-150,100){\makebox(0,0){$\alpha_0$}}
		\put(-35,137){\makebox(0,0){$\tilde\beta_0$}}
		\put(-85,0){\makebox(0,0){$\tilde\alpha_i$}}
		\subcaption{BTZ phase with $\{\beta_0, \alpha_i\}$ contractible}
		
	\end{subfigure}
	\caption{A slice of the Schottky domains used to construct the two BTZ phases, reduced by the reflection symmetry in the x-axis, $2\pi/n$ rotational symmetry about the origin, and the inversion symmetry through the unit circle (dashed). Various boundary cycles are labeled in each phase.  	\label{fig:BTZphases}}
\end{figure}

Instead, we take advantage of the replica symmetry to further reduce the Schottky domain. The reduced domains for these phases are shown in figure \ref{fig:BTZphases}, with the replica symmetry acting by a $2\pi/n$ rotation about the origin.

In practice, we find that it becomes difficult to numerically generate a mesh and solve the requisite differential equations for $\phi$ for $n>4$. Instead we notice that given a solution $\phi_n$ for the metric for a particular BTZ phase at replica number $n$, the solution $\phi_{n+1}$ can be approximated as $\phi_{n+1}(w) = \phi_{n}(w^{n/(n+1)})$. The function $\phi_{n+1}(w)$ is a solution with the correct boundary conditions up to corrections $O(1/n)$.  Under this transformation the line $\theta= \pi/n$ is mapped to $\theta = \pi/(n+1)$, effectively turning the $n-$fold replica symmetry into $(n+1)-$fold replica symmetry. Extending this idea, we can exactly solve for $\phi_2(w)$ then approximate all higher solutions as $\phi_n(w) = \phi_2(w^{2/n})$. This approximation introduces some error into the computation of the action and moduli, and we can estimate this error by explicitly comparing it to the exact solution. {For $n=4$ we find that in the region of interest the error is between $0.1\%$ and $5\%$ with most errors around $1\%$, though detailed comparisons beyond $n=4$ are beyond the scope of our numerics.}

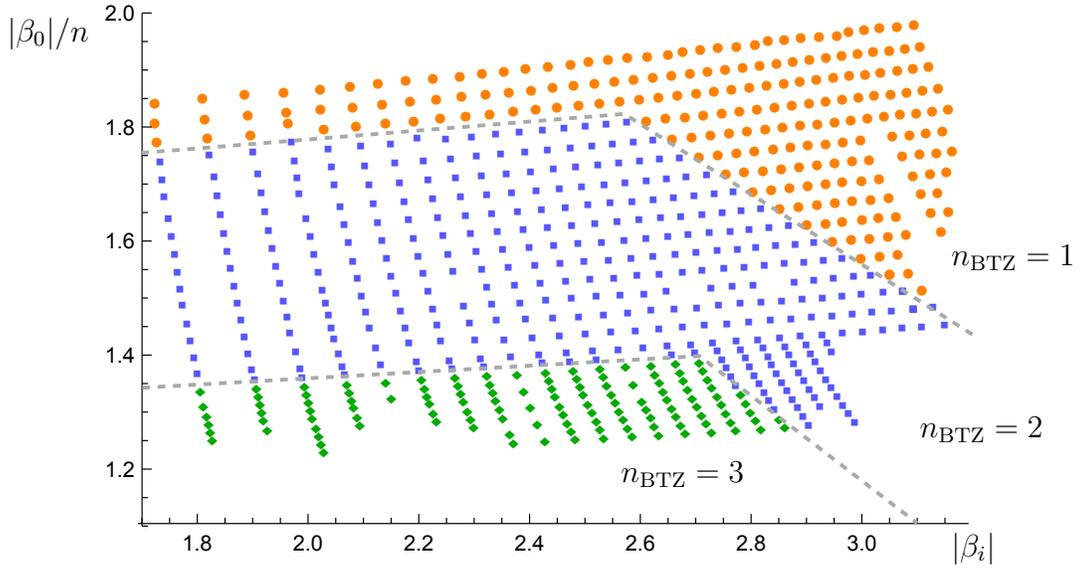
\begin{figure}[ht!]
	\centering
	\includegraphics[width=0.75\textwidth]{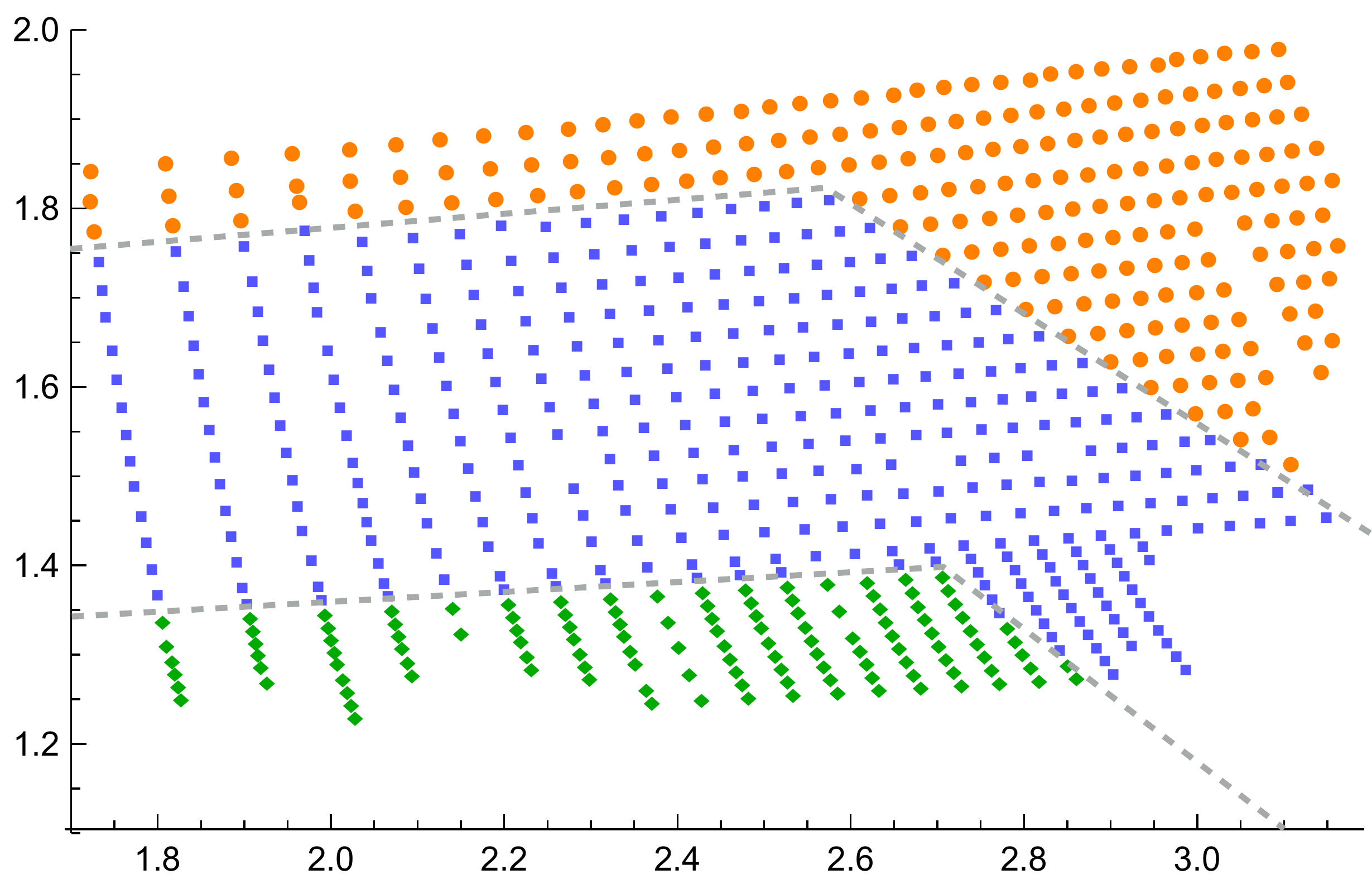}
	\put(0,5){\makebox(0,0){$|\beta_i|$}}
	\put(-350,200){\makebox(0,0){$|\beta_0|/n$}}
	\put(15,115){\makebox(0,0){$n_\text{BTZ}=1$}}
	\put(3,50){\makebox(0,0){$n_\text{BTZ}=2$}}
	\put(-110,33){\makebox(0,0){$n_\text{BTZ}=3$}}
	\caption{Numerical computation of $n_\text{BTZ}$ for different values of the moduli. Suggested phase boundaries are drawn by hand in dashed grey, each consisting of two linear segments. The kink corresponds to a transition in dominance between two distinct AdS phases.
		 \label{fig:replicaresults}}
\end{figure}

In comparing the phases, we first note that our heuristic predicts an AdS phase to dominate at finite $n$. As we increase $n$, the length of the cycle $\beta_0$ is proportional to $n$ by replica symmetry, while the lengths of $\alpha_0$, $\beta_i$, and $\alpha_i$ stay fixed. Therefore as we increase $n$, the sum of lengths $\{\alpha_0, \beta_1, \cdots, \beta_n\}$ will become smaller than $\{\beta_0, \beta_1, \cdots, \beta_n\}$, and the sum of lengths $\{\alpha_0, \alpha_1, \cdots, \alpha_n\}$ will become smaller than $\{\beta_0, \alpha_1, \cdots, \alpha_n\}$.  So at some finite $n$, our heuristic predicts an AdS phase to dominate; thus for $n>n_\text{BTZ}$ we have the TFD-like state
\ban{
\mathbb T \mathcal T^{n/2} \sim \ket{\text{AdS}}\ket{\text{AdS}} \, , \label{eq:tAdS}
}
so that $\ket 0_K = \ket{\text{AdS}}$. The maximum replica number $n_\text{BTZ}$ at which a BTZ phase dominates is a function of the moduli.
Figure \ref{fig:replicaresults} shows $n_\text{BTZ}$ at various points in the two dimensional moduli space, computed numerically using the technology described in the previous subsection.
Consistent with this plot, we expect there to be a corresponding region of moduli space for any $n_\text{BTZ}$, with the area of the region decreasing for large $n_\text{BTZ}$. In this way, taken by themselves the above results suggest that the torus operator is gapped for any choice of moduli.

We can understand this behavior by comparing $\mathcal T$ to the cylinder operator. Taking the cylinder to have circumference $2\pi \ell$ and length $\beta$, the BTZ phase dominates when $\beta \, n < 2 \pi \ell$. In this case the region of moduli with a particular $n_\text{BTZ}$ is given by $2\pi \ell / (n_\text{BTZ}+1) < \beta < 2 \pi \ell / n_\text{BTZ}$. We can make $n_\text{BTZ}$ as large as we like by choosing the moduli appropriately, but the volume occupied decreases with increasing $n_\text{BTZ}$. The torus operator shows a similar behavior, consistent with the conclusion that it too is gapped.

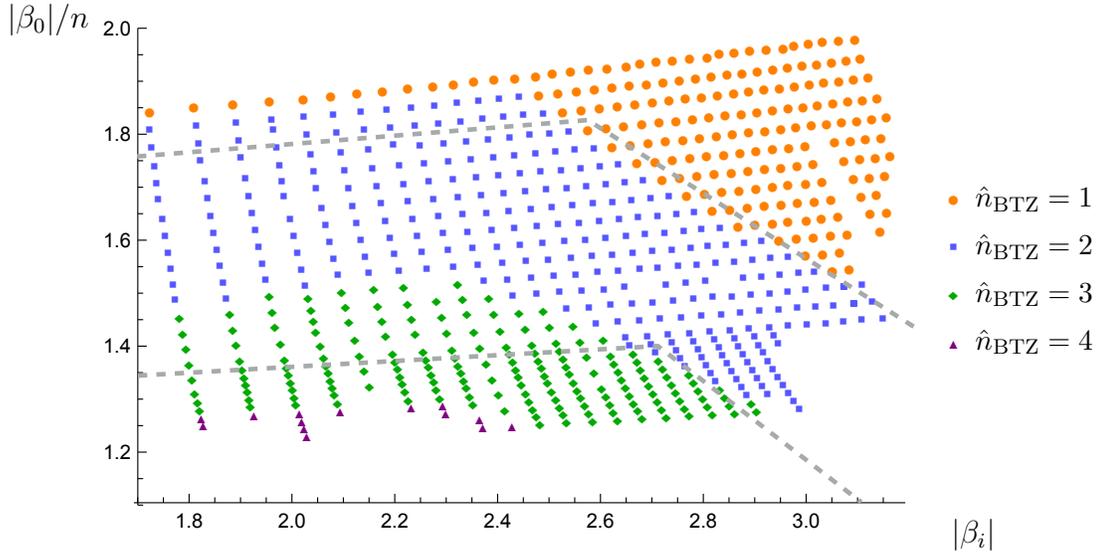
\begin{figure}[ht!]
	\centering
	\includegraphics[width=0.75\textwidth]{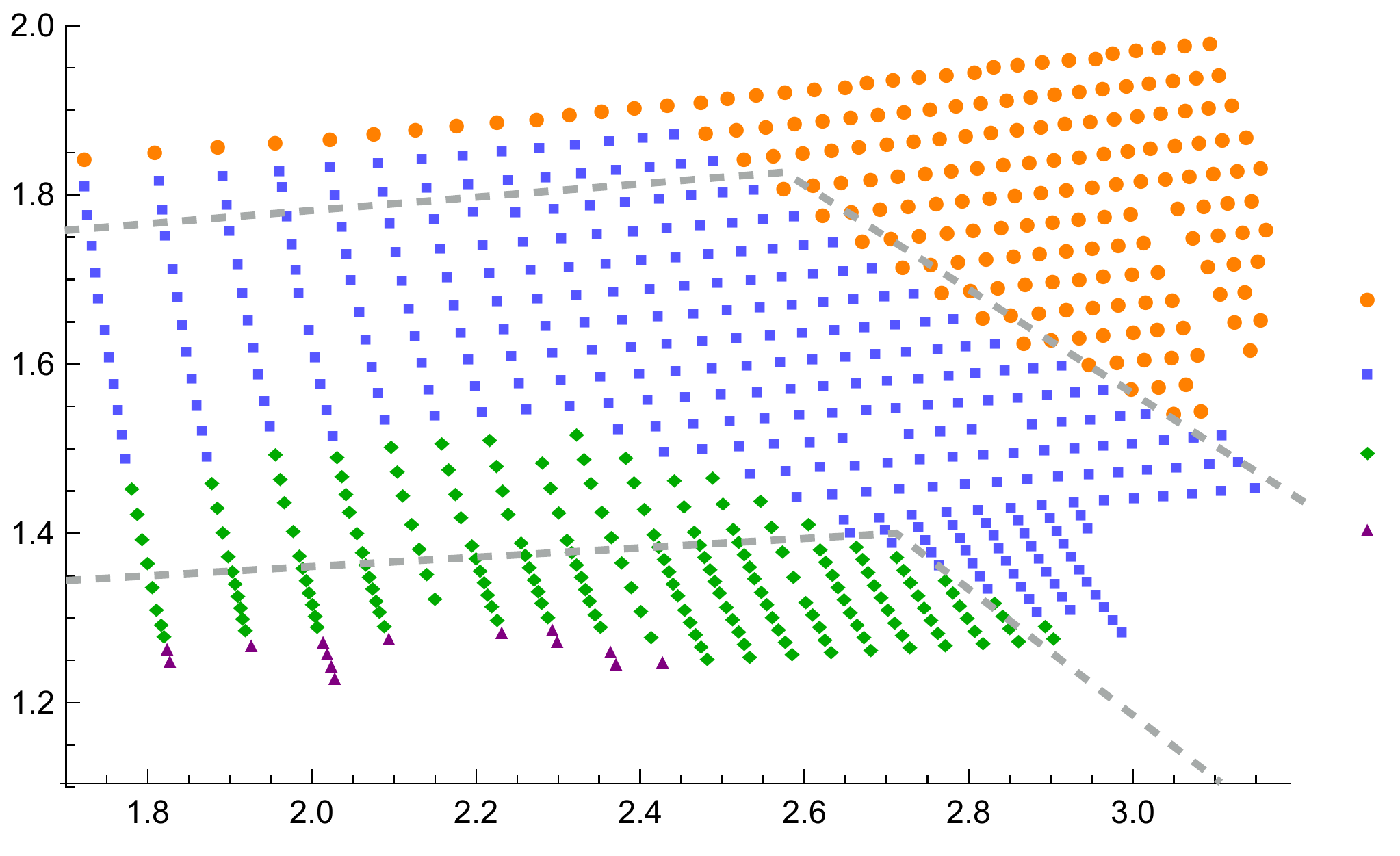}
	\put(0,5){\makebox(0,0){$|\beta_i|$}}
	\put(-350,200){\makebox(0,0){$|\beta_0|/n$}}
	\put(23,132){\makebox(0,0){\small $\hat n_\text{BTZ}= 1$}}
	\put(23,114){\makebox(0,0){\small$\hat  n_\text{BTZ}= 2$}}
	\put(23,96){\makebox(0,0){\small$\hat n_\text{BTZ}= 3$}}
	\put(23,78){\makebox(0,0){\small$\hat  n_\text{BTZ}= 4$}}
	\caption{Heuristic estimation of $\hat n_\text{BTZ}$ for different values of the moduli. We compare these values to the previously determined regions for $n_\text{BTZ}$ by drawing the dashed boundaries from figure \ref{fig:replicaresults}.
		 \label{fig:replica_heuristic}}
\end{figure}

It is illustrative to compare these results with the values of  $n_\text{BTZ}$ predicted by our heuristic, which we denote $\hat n_\text{BTZ}$. In figure \ref{fig:replica_heuristic}, we plot the value of $\hat n_\text{BTZ}$ predicted by the heuristic along with the boundaries previously drawn for $n_\text{BTZ}$ in figure \ref{fig:replicaresults} as determined by the computation of the action.
We see that the heuristic is accurate up to some order one offset, in that the corresponding boundaries have the right qualitative structure but differ from the true boundaries by an order one distance in this space.  However, around $|\beta_i | \sim 2.2$, for the boundary between $\hat n_{BTZ}=2$ and $\hat n_{BTZ}=3$  this offset becomes comparable to the (vertical) width of the $\hat  n_{BTZ}=2$ region.  Note that our heuristic predicts an $\hat n_\text{BTZ} = 4$ region and so we expect that if we were able to push the numerics further we would expect to see this region in figure \ref{fig:replicaresults} as well.

\subsection{Gapped Replica Symmetry Breaking Phases}
\label{sec:GRSBP}

We have thus far restricted our analysis to a particular class of phases which explicitly preserve replica symmetry. We now consider the possibility that replica symmetry is broken at large $n$, perhaps in some mild way. But since the above phases have a total cycle length that scales with $n$, our heuristic suggests that we focus on phases where the total cycle length grows at a similar rate or more slowly.  And based on the results above, it is natural to begin with a study of gapped such phases.  We postpone discussion of gapless symmetry breaking phases until \S \ref{section:discuss}.

One possibility is that the $\mathbb Z_n$ replica symmetry is broken to $\mathbb Z_{n/k}$, i.e. the phase consists of repeating blocks formed from $\mathcal T^k$ units. As in the previous section, our heuristic suggests that gapped phases will dominate at large $n$. Given some set of $k$ cycles $\{\gamma_i\}$ contained in $\mathcal T^k$ which are made contractible in a replica symmetry breaking phase, at large $n$ the total length of $\{n/k \times \gamma_i,\beta_0\}$ will always be larger than that of $\{n/k \times \gamma_i, \alpha_0\}$, as the length of $\beta_i$ grows with $n$ while the length of $\alpha_0$ is constant. So we still expect gapped phases to dominate above some $n$.\footnote{It is possible to break the replica symmetry to $\mathbb Z_{n/k}$ in such a way that choosing $\{\gamma_i\}$ and $\alpha_0$ to be contractible is inconsistent; the gapless phase described in \S\ref{section:discuss} below is an example. In such cases, our heuristic could be consistent with a gapless phase dominating at large $n$. }

\begin{figure}[ht!]
	\centering
	\includegraphics[width=0.55\textwidth]{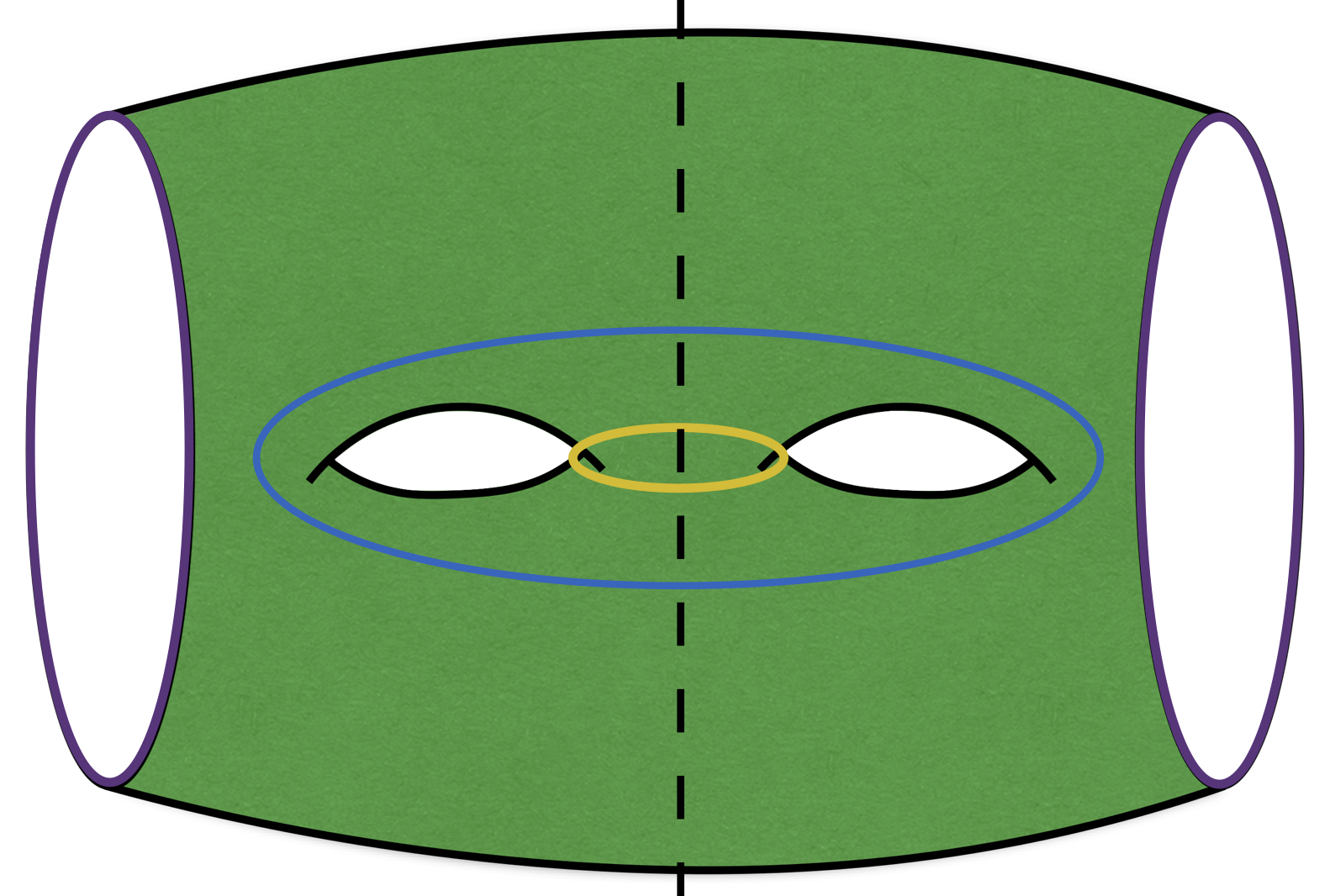}
	\put(-115,71){\makebox(0,0){$\alpha_i-\alpha_{i+1}$}}
	\put(-117,119){\makebox(0,0){$\beta_i+\beta_{i+1}$}}
	\put(-225,10){\makebox(0,0){$\alpha_0$}}
	\caption{The fundamental unit $\mathcal T^2$ with cycles that are contractible in the phase of interest labeled. \label{fig:twounit}}
\end{figure}

For simplicity let us assume the $\mathbb Z_n$ replica symmetry to be broken to $\mathbb Z_{n/2}$ by a bulk phase built from $n/2$ fundamental units, each corresponding to $\mathcal T^2$. We consider a phase in which the cycles $\{\alpha_0, \alpha_i - \alpha_{i+1}, \beta_{i}+\beta_{i+1}\}$ are contractible as drawn in figure \ref{fig:twounit}.
We have reason to suspect that there might be a region of moduli space where this phase might dominate over the AdS phases, as this problem is similar to that studied in \cite{MRW} for a genus two Riemann surface.
In our language, they computed a path integral for $\corr{0|\mathcal T^2 |0}$ and found that the above phase dominates for appropriately chosen moduli.  As described in \cite{MRW} and reviewed in appendix \ref{appendix:moduli}, the CFT state $\mathcal T \ket 0$ is then dual to toroidal geon.\footnote{{To gain intuition for this result, note that contractibility of {$\beta_{i} + \beta_{i+1}$} indicated by figure \ref{fig:twounit} implies that we can deform the cycle $\beta_i$ through the bulk until it becomes the cycle $-\beta_{i+1}$.  Thinking of the left half of figure \ref{fig:twounit} as bulk Euclidean time negative infinity and the right half as bulk Euclidean time positive infinity, such a deformation must pass through $t=0$.  But the cycle $\beta_i$ is not required to be contractible on its own (and is shown to be non-contractible by the Schottky analysis in appendix \ref{appendix:moduli}) so the $t=0$ slice must contain a non-contractible cycle.}} Our calculation differs only by the addition of the punctures on either side of figure \ref{fig:twounit}.  In particular, appendix \ref{appendix:moduli} shows that one can indeed find moduli where ${\mathscr T}$ has the ${\mathds Z}_2$ discrete symmetries we require and the bulk dual of the path integral over $\overline {\mathscr T}$ is dominated by the toroidal geon.

Computing the action for this phase explicitly is beyond the scope of the present work, but we can sharpen the above argument to show that there is indeed a region of moduli space where it must dominate over the AdS phases considered in \S \ref{sec:RSP}.  Let us start with the bulk region ${\mathcal M}$ defined by figure \ref{fig:twounit}. Recall that this ${\mathcal M}$ is constructed by cutting a replica-symmetric bulk solution $n{\mathcal M}$ along surfaces that separate the various replica copies.  Since the full $n{\mathcal M}$ has a reflection symmetry across each such surface, the extrinsic curvature of these surfaces must vanish.  We may thus glue half of a Poincar\'e ball to each surface.  On the boundary, this is the same gluing construction used in section \ref{section:intro} to construct $\overline {\mathscr T}$ from ${\mathscr T}$, though we have now used it twice (i.e. once on each boundary).

The AlAdS boundary is now a compact Riemann surface without boundary and with genus $2$.  In our language, the bulk is then a phase of $\langle 0 | {\mathcal T}^2 | 0 \rangle$.  The only issue is that the boundary metric on hemispherical end caps has constant positive curvature, so that the solution is not presented in our standard conformal frame and in fact the boundary Ricci curvature is discontinuous. Nevertheless, comparison with \cite{MRW} shows that the resulting bulk manifold is precisely their Euclidean toroidal geon, and one can perform a conformal transformation to make $R_\text{bndy} = -2$ everywhere. In appendix \ref{appendix:moduli} we show how to tune the moduli of $\mathcal T$ so that resulting bulk is in the toroidal geon phase.

We may repeat this gluing construction in any AdS phase from section \ref{sec:RSP}.  This results in precisely the set of AdS phases from \cite{MRW}.   Furthermore, the difference in actions between any two phases of $\langle 0 | {\mathcal T}^2 |0 \rangle$ is invariant under conformal transformations and, moreover, in the frame with discontinuous $R_{bndy}$ the contributions of the Poincar\'e hemispheres clearly cancel when comparing any two phases of $\corr{0|\mathcal T^2 |0}$ just described. {Therefore, for moduli of $\mathcal T$ identified in appendix \ref{appendix:moduli} where $\langle 0 | {\mathcal T}^2 |0 \rangle$ is in the toroidal geon phase,  we know that there is a corresponding region where the action of figure \ref{fig:twounit} is smaller than that of two copies of figure \ref{fig:unit} in either AdS phase from \S \ref{sec:RSP}.}  For such moduli, the phase described by figure \ref{fig:twounit} will dominate over the AdS phases from \S \ref{sec:RSP} for all $n$.

Let us now consider the implications of the phase described by figure \ref{fig:twounit} for the TFD-like state ${\mathds T} {\mathcal T}^{n/2}$.  For $n$ that are multiples of four, taking the moment of time symmetry to lie between two $\mathcal T^2$ blocks (i.e., passing through an $\alpha_0$ cycle on either the left or right side of figure \ref{fig:twounit}) we find the above saddles to again give an AdS phase. However if we put the moment of time symmetry so that it cuts through a $\mathcal T^2$ block along the dashed line in figure \ref{fig:twounit}, the bulk $t=0$ surface looks like two disconnected toroidal geons, i.e., referring to the toroidal geon as $\ket 1_\text{BH}$,  we have $\text{TFD} \sim \ket 1_\text{BH}\otimes \ket 1_\text{BH}$. By the full replica symmetry of the partition function, both of these configurations must have the same gravitational action, and the TFD-like state $\mathbb T \mathcal T^{n/2}$ is thus a superposition of the two.  On the other hand,  for $n$ congruent to $2$ mod $4$, by similar arguments we  find an equal superposition of $\ket 1_\text{BH} \otimes \ket{\text{AdS}}$ and $\ket{\text{AdS}}\otimes \ket 1_\text{BH} $. So if the phase described by figure \ref{fig:twounit} dominates we have
\ban{
\mathbb T \mathcal T^{n/2}\sim \left \{ \begin{array}{ccc} \frac 12 \ket{\text{AdS}}\ket{\text{AdS}}+\frac 12 \ket 1 _\text{BH} \ket 1_\text{BH}& \hspace{0.5cm} & n \equiv 0 \text{ mod } 4 \\
\\
\frac 12 \ket{1}_\text{BH} \ket{\text{AdS}}+\frac 12 \ket{\text{AdS}} \ket 1_\text{BH} & \hspace{0.5cm}& n \equiv 2 \text{ mod } 4
\end{array}
 \right. \, \label{eq:dens},
}
where the $\sim$ denotes leading behavior at large $c$ up to normalization.
The gravitational action  in this phase is still linear in $n$, and its dominance at large $n$ would again imply that $\mathcal T$ is gapped.

However, the ${\mathbb Z}_2$ symmetry associated with reflections of figure \ref{fig:twounit} across the vertical dashed line implies that ${\mathcal T} = A^\dagger A$ where $A$ is the operator from one copy of the CFT Hilbert space ${\mathcal H}$ to ${\mathcal H} \otimes {\mathcal H}$ defined by the path integral over the right half of figure \ref{fig:twounit}.  Thus $\mathcal T$ is non-negative, and so is ${\mathcal T}^{n/2}$ for even $n$.  Note that while we have suppressed details of the moduli and the order $c^0$ state of bulk quantum fields in \eqref{eq:dens}, these will have some definite values in the phase described and so cannot resolve the problem.  We conclude that the phase associated with figure \ref{fig:twounit} cannot be the most dominant.  Instead, some new phase must become relevant.

Considering phases that repeat more complicated blocks (e.g. based on ${\mathcal T}^4$) appears to lead to similar problems.  However, one possible resolution within the class of gapped phases is that there are additional bulk saddles which, taken by themselves would give
\ban{
\mathbb T\mathcal T^{n/2}\sim \left \{ \begin{array}{ccc}
\frac 12 \ket{1}_\text{BH} \ket{\text{AdS}}+\frac 12 \ket{\text{AdS}} \ket 1_\text{BH} &
\hspace{0.5cm} & n \equiv 0 \text{ mod } 4 \\
\\
\frac 12 \ket{\text{AdS}}\ket{\text{AdS}}+\frac 12 \ket 1 _\text{BH} \ket 1_\text{BH}&
 \hspace{0.5cm}& n \equiv 2 \text{ mod } 4
\end{array}\right.
 \, \label{eq:dens2},
}
at each $n$, and which turn out to be related to the phases above by an unexpected symmetry so that their actions are precisely equal to those just discussed.  In that case, we should sum the contributions \eqref{eq:dens} and \eqref{eq:dens2} with equal weight to give
\ban{
\mathbb T\mathcal T^{n/2} = \frac 12 \left( \ket{\text{AdS}} +\ket 1_\text{BH}\right)\left( \ket{\text{AdS}} +\ket 1_\text{BH}\right) \label{eq:super}\, ,
}
for all $n$.  Interpreting \eqref{eq:super} as an operator, ${\mathcal T}^{n/2}$ is
the projector onto the pure state $\frac 1{\sqrt 2} \left( \ket{\text{AdS}} +\ket 1_\text{BH}\right)$ and is manifestly positive.   The torus ground state  $\ket 0_K$ would then be an equal superposition of empty AdS and the toroidal geon. An obstacle to this resolution is that there are no natural candidates for the missing phases, as all of the phases constructed by acting with the broken elements of the $\mathbb Z_n$ replica symmetry group have been accounted for.

Other possible resolutions involve gapless phases.  Indeed, it might seem most natural to explore the hypothesis that the state ${\mathcal T}^n|0\rangle$ has topology of order $n$ (at least for $n \ll c$).  This would require a ground state $|0\rangle_K$ with topology of order $c$ and, in the semiclassical bulk limit $c \rightarrow \infty$, the action would not be precisely linear at large $n$ and so cannot be gapped.

It is an interesting question then to determine if acting with $\mathcal T$ on $\ket 0$ can generate states of high topology. Indeed, if these states exist, then one could make similar cutting and pasting arguments that there are symmetry breaking phases of higher topology that dominate the path integral which computes $\Tr(\mathcal T^n)$. For this reason, we postpone our consideration of gapless symmetry-breaking phases until after studying the states ${\mathcal T}^2|0\rangle$ in \S \ref{section:states}.  We will return to this topic during the final discussion in \S \ref{section:discuss}.

\section{Single Boundary States}
\label{section:states}

We now consider the action of $\mathcal T$ on both global AdS and toroidal geon states. In the context of the previous section, we seek to understand if states of high topology can dominate the path integrals considered. Explicitly, we investigate whether $\mathcal T^k \ket 0$ can be dual to a black hole with genus $g=2$ behind the horizon. We fail to find a region of moduli space where this is the case, leading to the conjecture that this always fails above genus $g = 1$. Indeed, there also appear to be regions of moduli space where the bulk remains empty global AdS for all $k$.

\subsection{Definitions and Phases}

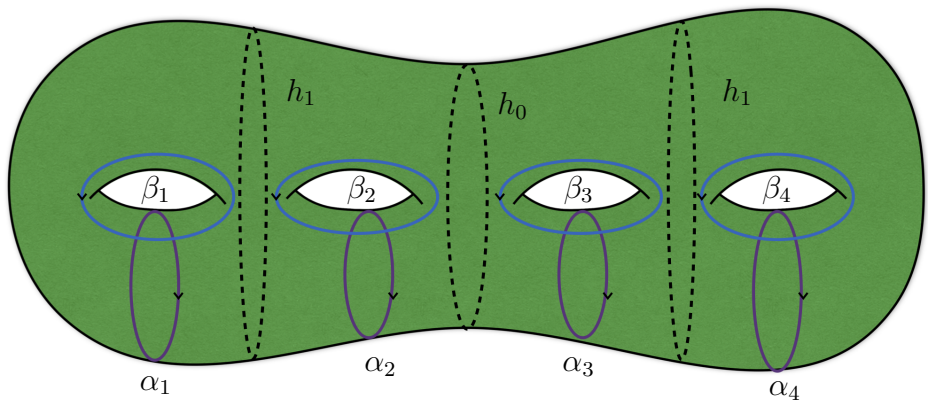
\begin{figure}[ht!]
\centering
\includegraphics[width=0.8\textwidth]{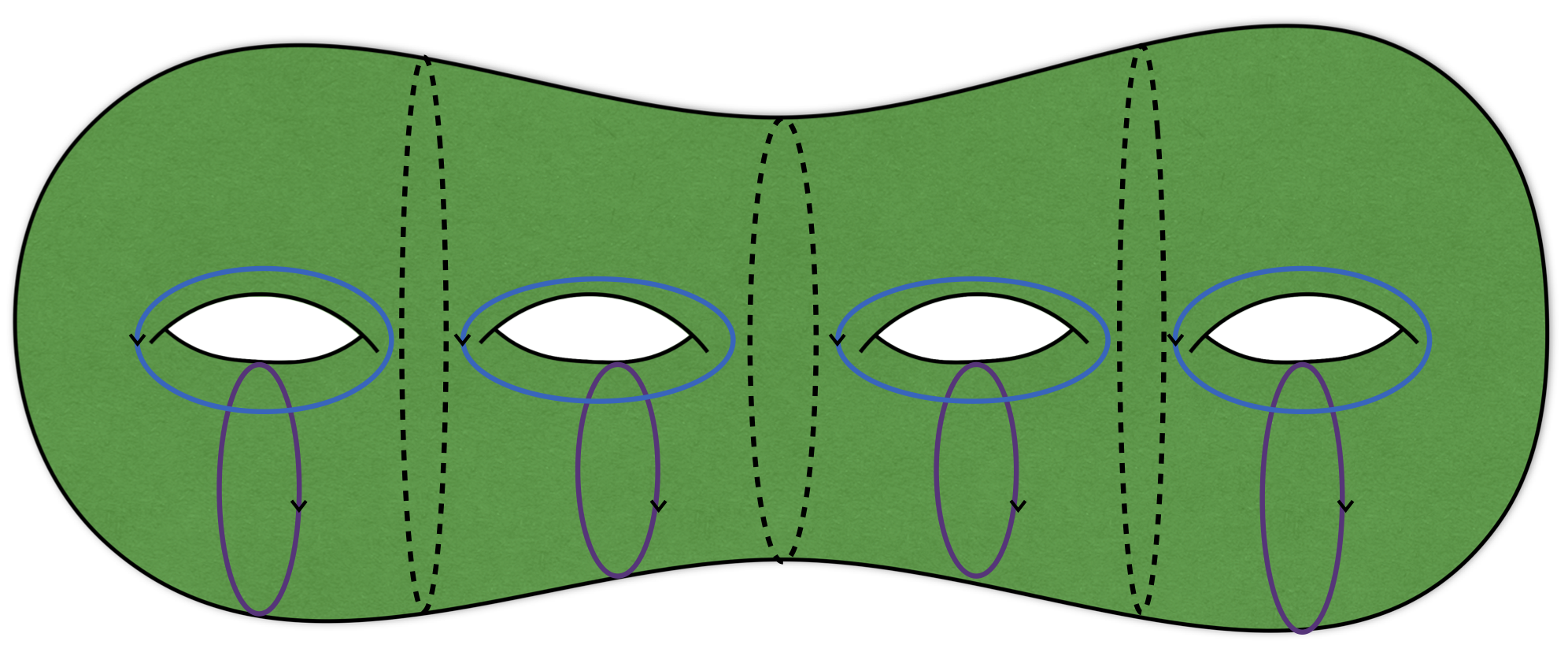}
\put(-295,-1){\makebox(0,0){$\alpha_1$}}
\put(-210,6){\makebox(0,0){$\alpha_2$}}
\put(-135,6){\makebox(0,0){$\alpha_3$}}
\put(-57,-3){\makebox(0,0){$\alpha_4$}}
\put(-295,73){\makebox(0,0){$\beta_1$}}
\put(-217,73){\makebox(0,0){$\beta_2$}}
\put(-135,73){\makebox(0,0){$\beta_3$}}
\put(-60,73){\makebox(0,0){$\beta_4$}}
\put(-75,110){\makebox(0,0){$h_1$}}
\put(-240,110){\makebox(0,0){$h_1$}}
\put(-160,105){\makebox(0,0){$h_0$}}

\caption{
Our genus four Riemann surface with $\alpha$ cycles in purple and $\beta$ cycles in blue. There are three reflection symmetries along the horizontal and vertical axes, as well as in the plane of the page.   \label{fig:nchain}
}
\end{figure}

To study the state $\mathcal T^2 \ket 0$, we consider the partition function defined by the path integral over the genus $g=4$ Riemann surface drawn in figure \ref{fig:nchain} and having  the three ${\mathbb Z}_2$ symmetries described in the caption.
In particular, we study the part of the genus-$4$ moduli space where the vertical ${\mathbb Z}_2$ reflection leaves fixed the geodesics associated with $\beta_i$.  We remark that this is quite different from the part of the $g=4$ moduli space shown in figure \ref{fig:naive}, and that it is figure \ref{fig:naive} rather than figure \ref{fig:nchain} which is relevant in the high temperature limit described in \cite{MBW2}.  Despite our negative results associated with figure \ref{fig:nchain} below, we strongly expect phase where the space behind the horizon has genus $2$ to dominate in such a high-temperature limit.

Returning to figure \ref{fig:nchain},
cutting the path integral along an initial time-slice given by the middle dashed black line $h_0$ defines the state $\mathcal T_1 \mathcal T_2 \ket 0$ given by two torus operators acting on the vacuum. This problem has a $4$-dimensional moduli space, as each torus operator has a two dimensional moduli space associated with it.  Considering this generalization allows us to sidestep the need to study in detail the maps between various conformal frames that would arise in a direct computation of ${\mathcal T}^2|0\rangle$.  While this comes at the cost of both increasing the dimension of moduli space and being unsure of which 2d slice describes ${\mathcal T}^2|0\rangle$, it turns out to be sufficient for our purposes below.

In general, the above initial time-slice for the corresponding bulk state can have genus $0,1$, or $2$, with many possible phases for each genus. Below, we consider the subspace of moduli space preserving our ${\mathbb Z}_2$ symmetries and study whether a bulk state with a genus $2$ initial time slice can dominate in any region of this subspace.   We consider six different phases, chosen as the most numerically tractable among those determined by the algorithm described in \S\ref{section:handlebody}.  At $t=0$, three of the phases have genus $0$, two have genus $1$, and one has genus $2$.  As a shorthand, we refer to the associated three-dimensional bulk solutions as having genus $0$, $1$, and $2$ respectively.

\begin{figure}[ht!]
\centering
\begin{subfigure}{0.32\textwidth}
\includegraphics[width=\textwidth]{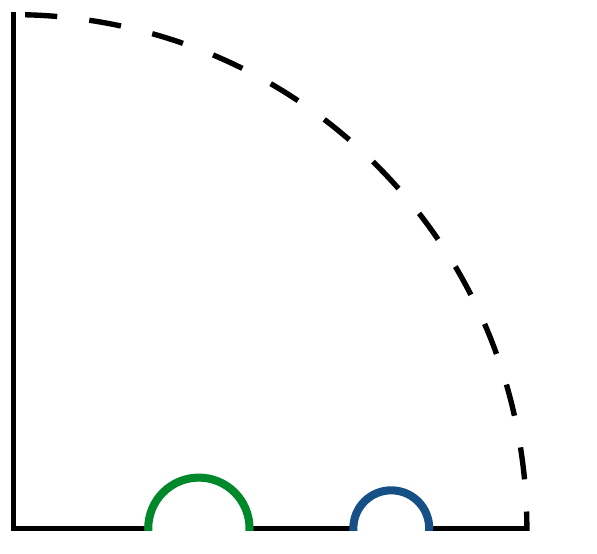}
\put(-150,90){\makebox(0,0){\small$h_0$}}
\put(-50,110){\makebox(0,0){\small$\alpha_{23}$}}
\put(-47,22){\makebox(0,0){\small$\alpha_{34}$}}
\put(-94,26){\makebox(0,0){\small$\alpha_4$}}
\put(-120,13){\makebox(0,0){\small$\beta_{1234}$}}
\put(-68,13){\makebox(0,0){\small$\beta_{4}$}}
\put(-27,13){\makebox(0,0){\small$\beta_{3}$}}
\vspace{0.1cm}
\subcaption{$\{\alpha_1, \alpha_{12} ,\alpha_{34}, \alpha_4\}$ contractible}

\end{subfigure}
\begin{subfigure}{0.32\textwidth}
\includegraphics[width=\textwidth]{ads42}
\put(-150,90){\makebox(0,0){\small$h_0$}}
\put(-40,110){\makebox(0,0){\small$\beta_{1234}$}}
\put(-50,23){\makebox(0,0){\small$\beta_4$}}
\put(-93,25){\makebox(0,0){\small$\beta_3$}}
\put(-120,11.5){\makebox(0,0){\small$\alpha_{23}$}}
\put(-70,11.5){\makebox(0,0){\small$\alpha_{34}$}}
\put(-29,11.5){\makebox(0,0){\small$\alpha_{4}$}}
\subcaption{$\{\beta_1, \beta_2, \beta_3, \beta_4\}$ contractible}

\end{subfigure}
\begin{subfigure}{0.3\textwidth}
\includegraphics[width=\textwidth]{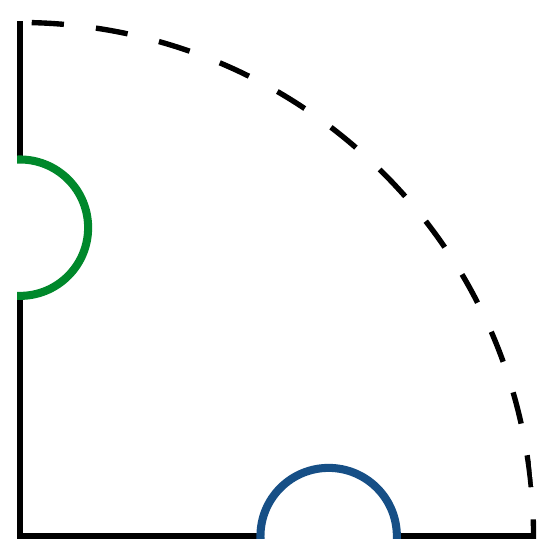}
\put(-116,111){\makebox(0,0){\small$\alpha_{23}$}}
\put(-35,110){\makebox(0,0){\small$h_0$}}
\put(-52,28){\makebox(0,0){\small$\alpha_4$}}
\put(-100,75){\makebox(0,0){\small$\beta_3$}}
\put(-115,33){\makebox(0,0){\small$\alpha_{34}$}}
\put(-90,13){\makebox(0,0){\small$\beta_{4}$}}
\put(-23,13){\makebox(0,0){\small$\beta_{1234}$}}
\subcaption{$\{\alpha_1, \beta_2, \beta_3, \alpha_4\}$ contractible}
\end{subfigure}
\caption{One eighth of the Schottky domain used to construct the three genus $0$ phases ($0_a,0_b,0_c$)  reduced by the reflection symmetries in the x-axis and y-axis and the inversion symmetry through the unit circle (dashed). Various boundary cycles are labeled in each phase.}
\label{fig:AdS}
\end{figure}

The three genus $0$ phases $0_a$, $0_b$, $0_c$ are respectively defined by choosing the sets of cycles $\{\alpha_1, \alpha_{12}, \alpha_{34}, \alpha_4\}$,  $\{\beta_1, \beta_2, \beta_3, \beta_4\}$, or $\{\alpha_1, \beta_2, \beta_3, \alpha_4\}$ to be contractible.\footnote{For simplicity of notation we define $\alpha_{ij}\equiv \alpha_i - \alpha_j$ and $\beta_{ij\cdots k} \equiv \beta_i + \beta_j + \cdots + \beta_k$.} The corresponding Schottky domains are shown in figure \ref{fig:AdS}.
For each phase we can numerically compute the lengths of all the labeled boundary geodesics fixed by symmetry and choose parameters so as to match them to those of the genus 2 phase described below. Note that the bulk surface associated with the $h_0$ cycle (i.e., the surface invariant under the right/left ${\mathbb Z}_2$ reflection symmetry of figure \ref{fig:nchain}) has genus zero by \eqref{eq:bulkgenus}.

\begin{figure}[ht!]
\centering
\begin{subfigure}{0.49\textwidth}
\includegraphics[width=\textwidth]{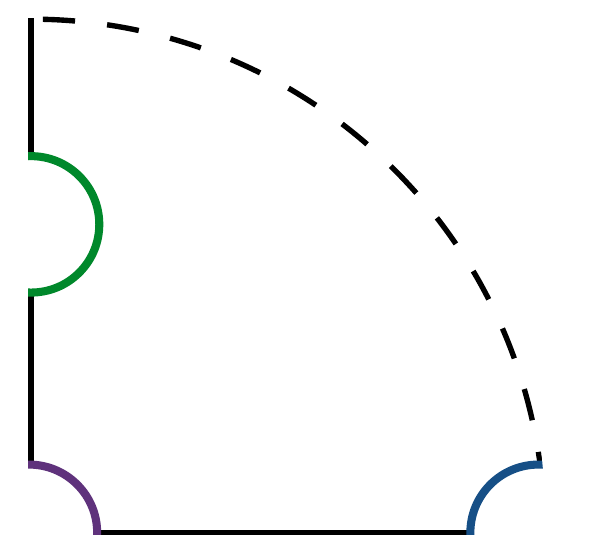}
\put(-60,145){\makebox(0,0){$\alpha_4$}}
\put(-217,175){\makebox(0,0){$\beta_4$}}
\put(-165,120){\makebox(0,0){$\alpha_{34}$}}
\put(-217,65){\makebox(0,0){$\beta_3$}}
\put(-175,30){\makebox(0,0){$\alpha_{23}$}}
\put(-107,17){\makebox(0,0){$h_0$}}
\put(-57,32){\makebox(0,0){$\beta_{1234}$}}
\subcaption{$\{\alpha_{12}, \alpha_{34}, \alpha_{23}, \beta_{1234}\}$ contractible}

\end{subfigure}
\hfill
\begin{subfigure}{0.49\textwidth}
\includegraphics[width=\textwidth]{torus42}
\put(-60,145){\makebox(0,0){$\beta_3$}}
\put(-220,175){\makebox(0,0){$\alpha_{34}$}}
\put(-168,120){\makebox(0,0){$\beta_4$}}
\put(-217,65){\makebox(0,0){$\alpha_4$}}
\put(-168,30){\makebox(0,0){$\beta_{1234}$}}
\put(-107,17){\makebox(0,0){$h_0$}}
\put(-52,32){\makebox(0,0){$\alpha_{23}$}}
\subcaption{$\{\beta_1, \beta_4, \alpha_{23}, \beta_{1234}\}$ contractible}

\end{subfigure}
\caption{One eighth of the Schottky domain used to construct the two genus 1 phases ($1_a,1_b$) reduced by the reflection symmetries in the x-axis and y-axis and the inversion symmetry through the unit circle (dashed). Various boundary cycles are labeled in each phase.}
\label{fig:TPhase}
\end{figure}

The two genus $1$ phases $1_a,1_b$ are respectively defined by choosing the set of cycles $\{\alpha_{12}, \alpha_{34}, \alpha_{23}, \beta_{1234}\}$ or the set $\{\beta_1, \beta_4, \alpha_{23}, \beta_{1234}\}$ to be contractible. The Schottky domains are depicted in figure \ref{fig:TPhase}.
Again, for each phase we can numerically compute the lengths of all the labeled boundary geodesics fixed by symmetry, and choose parameters so as to match them to those of the genus 2 phase. Note that the genus of the time-slice associated with the $h_0$ cycle is now $1$ by \eqref{eq:bulkgenus}.

Finally, we choose the genus 2 phase to have contractible cycles $\{\alpha_{14},\alpha_{23}, \beta_{23}, \beta_{1234}\}$. However, for computational reasons it is more convenient to use the basis $\{\alpha_{23}, \beta_{1234}, \alpha_{14}+\beta_{1234}, \alpha_{23}+\beta_{23}\}$, which gives the same bulk phase.

\begin{figure}[ht!]
\centering
\begin{subfigure}{0.85\textwidth}
\includegraphics[width=\textwidth]{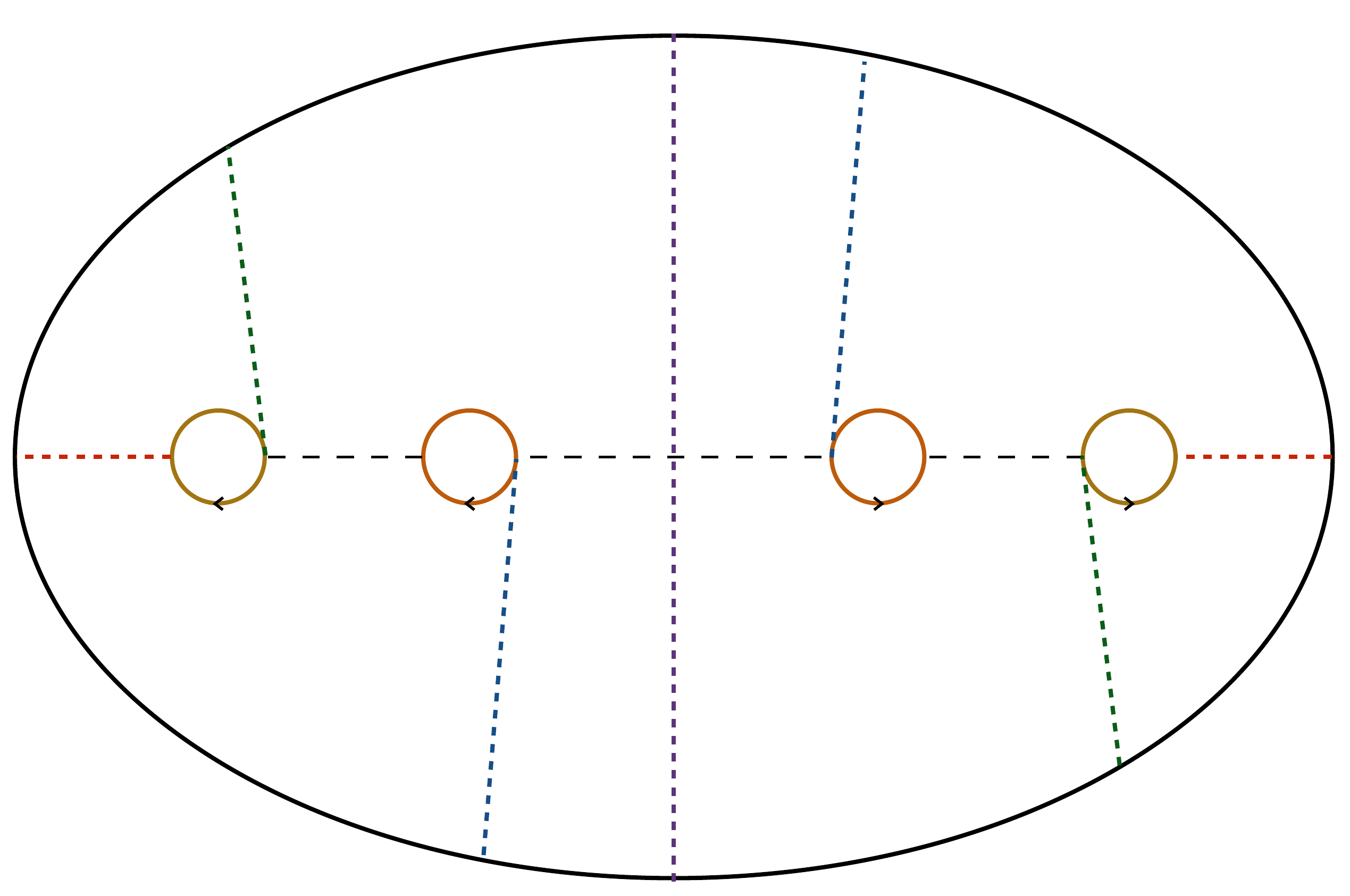}
\put(-345,130){\makebox(0,0){$\alpha_{23}$}}
\put(-30,130){\makebox(0,0){$\alpha_{23}$}}
\put(-280,180){\makebox(0,0){$\alpha_{23}+\beta_{23}$}}
\put(-100,60){\makebox(0,0){$\alpha_{23}+\beta_{23}$}}
\put(-60,145){\makebox(0,0){$\beta_3$}}
\put(-315,97){\makebox(0,0){$\beta_3$}}
\put(-270,65){\makebox(0,0){$\alpha_{14}+\beta_{1234}$}}
\put(-105,175){\makebox(0,0){$\alpha_{14}+\beta_{1234}$}}
\put(-205,180){\makebox(0,0){$\beta_{1234}$}}
\put(-125,23){\makebox(0,0){$h_0$}}
\put(-132,97){\makebox(0,0){$\beta_4$}}
\put(-245,145){\makebox(0,0){$\beta_4$}}
\subcaption{One half of the flattened Riemann surface. The surface has been cut along the cycles $\beta_3$ and $\beta_4$ along the fixed point set of reflection across the vertical purple dashed line. \label{fig:cutA}}

\end{subfigure}
\hfill\\
\vspace{0.25cm}
\begin{subfigure}{0.85\textwidth}
\includegraphics[width=\textwidth]{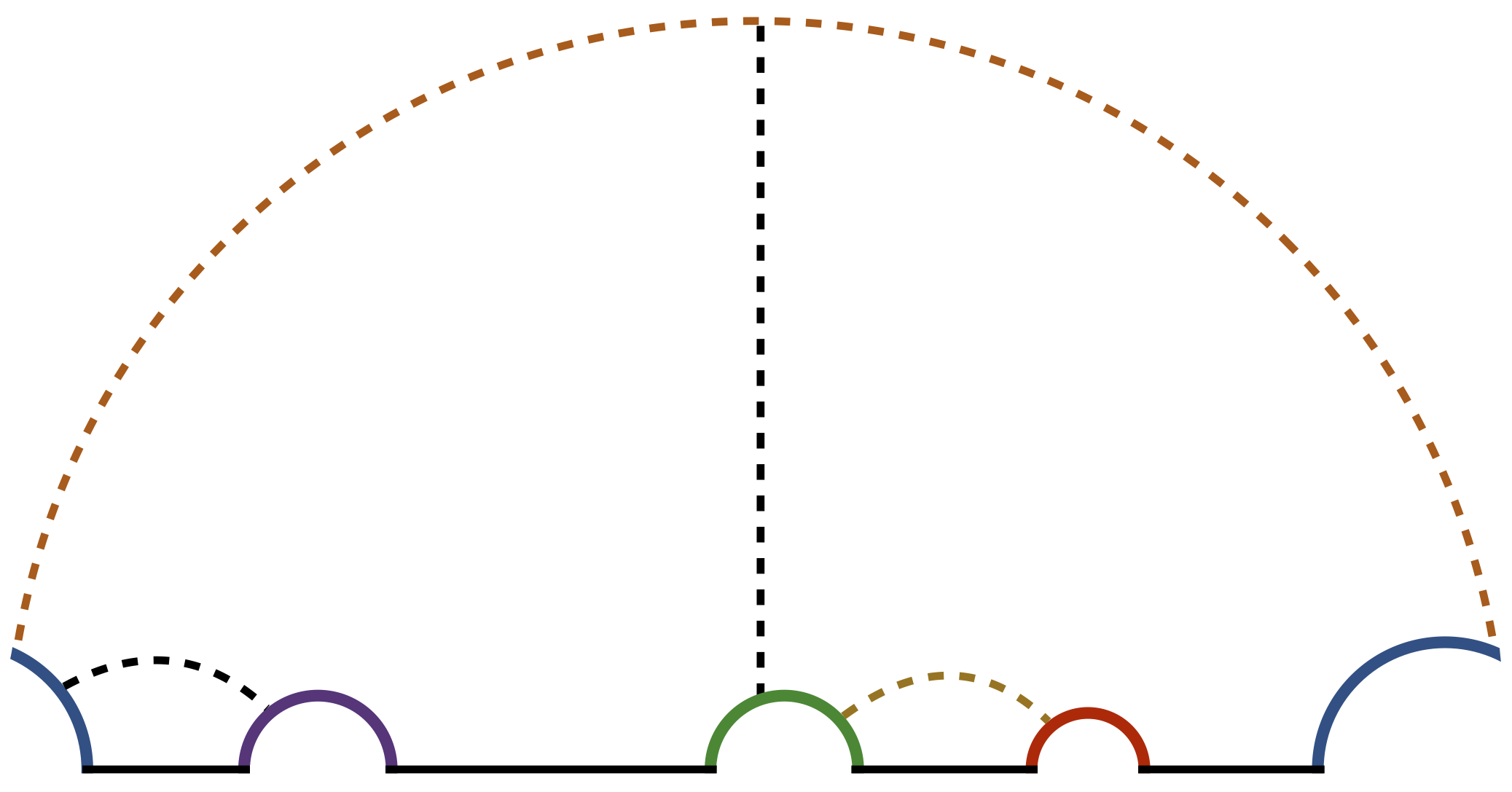}
\put(-90,145){\makebox(0,0){$\beta_4$}}
\put(-10,17){\makebox(0,0){$\alpha_{14}+\beta_{1234}$}}
\put(-390,17){\makebox(0,0){$\alpha_{14}+\beta_{1234}$}}
\put(-220,27){\makebox(0,0){$\alpha_{23}+\beta_{23}$}}
\put(-90,25){\makebox(0,0){$\alpha_{23}$}}
\put(-140,39){\makebox(0,0){$\beta_{3}$}}
\put(-285,33){\makebox(0,0){$\beta_{1234}$}}
\put(-335,-3){\makebox(0,0){$b_1$}}
\put(-245,-3){\makebox(0,0){$b_2$}}
\put(-140,-3){\makebox(0,0){$b_3$}}
\put(-70,-3){\makebox(0,0){$b_4$}}
\vspace{0.15cm}

\subcaption{One fourth of the genus 2 Schottky domain, reduced by two symmetries. The two symmetries of the plane are reflection about the $x-$axis (labeled $h_0$) and the product of the reflection about the $y-$axis (dashed black) and inversion through the unit circle (labeled $\beta_4$). The boundary cycle $h_0$ is broken into four segments $b_i$. \label{fig:Schottky2}}

\end{subfigure}
\caption{}
\label{fig:g2}
\end{figure}

Constructing the Schottky domain is difficult as the domain turns out not to preserve the full set of symmetries of this phase.    We thus use the following procedure to keep track of all the boundary geodesics and symmetries. First, we cut the boundary Riemann surface along $h_0$, keeping only the right hand side, and then cut along $\beta_3$ and $\beta_4$. The result is the surface drawn in figure \ref{fig:cutA}.
Next, following the procedure described in \S\ref{section:handlebody} we cut along the contractible cycles $\{\alpha_{23}, \beta_{1234}, \alpha_{14}+\beta_{1234}, \alpha_{23}+\beta_{23}\}$. Note that while this choice of cycles respects the three $\mathbb Z_2$ symmetries of the boundary Riemann surface, when we choose representative cycles for $\alpha_{14}+\beta_{1234}$ and $\alpha_{23}+\beta_{23}$ we break two $\mathbb Z_2$ symmetries\footnote{One could hope that there exists a basis where we can preserve all of the symmetries, but we find this not to be the case.}  given by reflection in the $x-$axis and $y-$axis in figure \ref{fig:cutA}. However, the product of these symmetries is preserved, and the cycles $\alpha_{14}+\beta_{1234}$ and $\alpha_{23}+\beta_{23}$ are fixed point sets of this product. When we glue everything back together along the cycles $\beta_3$ and $\beta_4$ to construct the Schottky domain, we thus maintain the requirement that the contractible cycles are fixed point sets of a symmetry. In the Schottky domain of figure \ref{fig:Schottky2} this symmetry is the product of reflection about the $y-$axis and inversion through the unit circle.

Finally, we can use this Schottky domain to solve for the metric $\phi$ as described in \S\ref{section:handlebody}. In order to compute the lengths of geodesics on the boundary, we use the trick described at the end of \S \ref{section:handlebody} involving mapping to a subset of the Poincar\'e disk, where the lengths of the boundary segments are determined by the numerical solution for $\phi$.

For the three genus $0$ phases and the two genus $1$ phases, the Schottky domain is defined by four free real parameters, consistent with a four dimensional moduli space for two torus operators with the required symmetries. For the genus $2$ phase our Schottky representation breaks two of the symmetries so a priori we have an $8$ dimensional parameter space.
Carefully tracking the symmetries gives the condition $b_2-b_1=b_4-b_3$ and that the two blue arcs in figure \ref{fig:Schottky2} have equal length.
In the Poincar\'e disk we can also compute the lengths of the images of the cycles $\alpha_{23}+\beta_{23}$ and $\alpha_{14}+\beta_{1234}$ under reflection across the $x-$axis of figure \ref{fig:cutA}. The requirement that the lengths of these cycles be equal to the lengths of their inverse images gives two more conditions. Imposing these symmetry conditions numerically then reduces the parameter space to four dimensions, recovering the same moduli space considered for genus 0 and genus 1.

\subsection{Results}

We would like to find a region of moduli space where the genus $2$ phase dominates. While we do not study all of the possible lower genus phases, the numerical evidence below suggests that the genus $2$ phase never dominates even within the phases we study.  Indeed, to exclude the genus $2$ phase it turns out to suffice to consider only the genus $0_c$ phase (with contractible  cycles $\{\alpha_1, \beta_2, \beta_3, \alpha_4\}$) and the two genus 1 phases.

Our first step was to perform a coarse gradient search in the full four dimensional moduli space to try to minimize the quantity $I_2 - I_\text{dom.}$, i.e. the difference in action between the genus 2 phase and the dominant phase within the set of phases described above. This search identified a region of moduli space which we now study in more detail. For numerical convenience, we parameterize the moduli space using the radii $\rho_i$ of the circles $C_i$ in the Schottky domain for the genus 2 phase (figure \ref{fig:Schottky2}), with $\rho_i$ computed using the flat space metric. These parameters $\{\rho_1, \rho_2, \rho_3, \rho_4\}$ are respectively related to the lengths of the cycles $\{\alpha_{23}, \beta_{1234}, \alpha_{14}+\beta_{1234}, \alpha_{23}+\beta_{23}\}$, with the precise relation determined by the conformal frame $\phi$.

\begin{figure}[ht!]
\centering
\begin{subfigure}{0.49\textwidth}
\includegraphics[width=\textwidth]{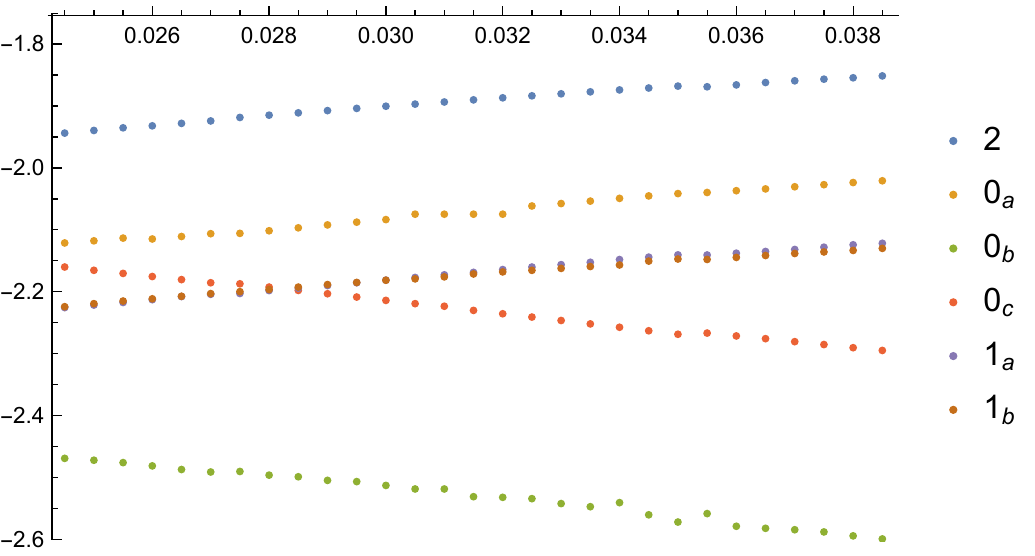}
\put(-17,110){\makebox(0,0){\footnotesize$\rho_1$}}
\put(-207,-7){\makebox(0,0){\footnotesize$I$}}
\subcaption{\label{fig:min1}}
\end{subfigure}
\hfill
\begin{subfigure}{0.49\textwidth}
\includegraphics[width=\textwidth]{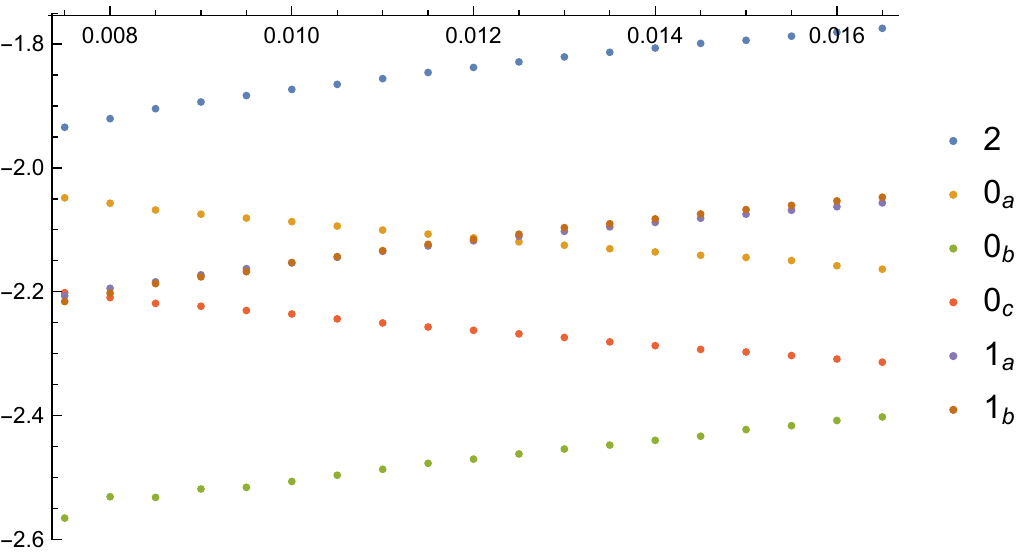}
\put(-17,110){\makebox(0,0){\footnotesize$\rho_2$}}
\put(-207,-7){\makebox(0,0){\footnotesize$I$}}
\subcaption{\label{fig:min2}}
\end{subfigure}\\
\vspace{0.25cm}
\begin{subfigure}{0.49\textwidth}
\includegraphics[width=\textwidth]{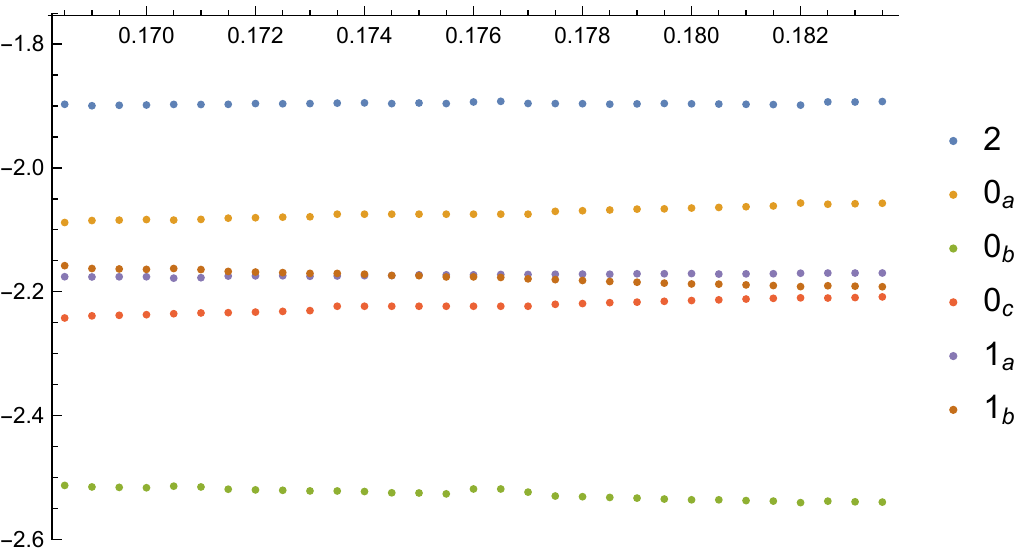}
\put(-17,110){\makebox(0,0){\footnotesize$\rho_3$}}
\put(-207,-7){\makebox(0,0){\footnotesize$I$}}
\subcaption{\label{fig:min3}}
\end{subfigure}
\hfill
\begin{subfigure}{0.49\textwidth}
\includegraphics[width=\textwidth]{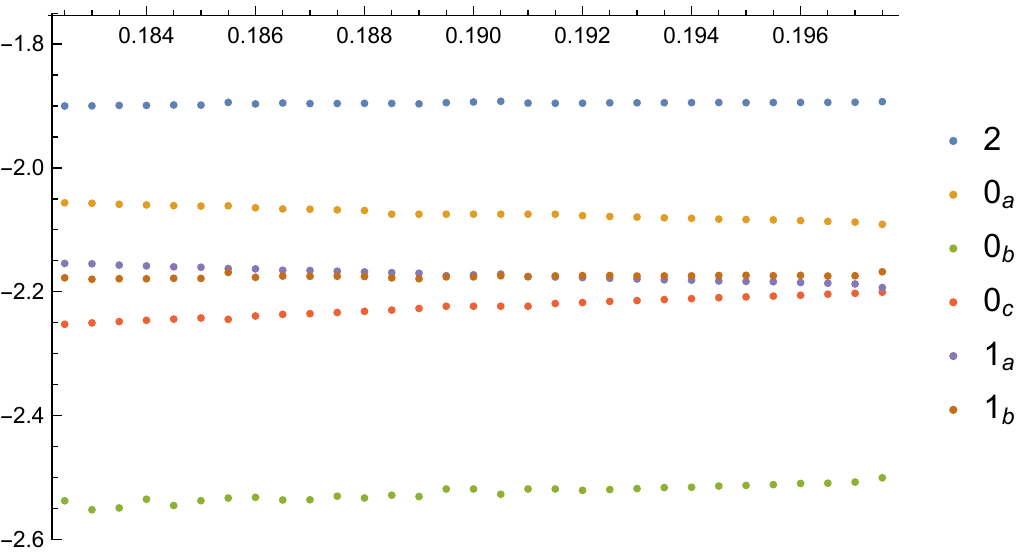}
\put(-17,110){\makebox(0,0){\footnotesize$\rho_4$}}
\put(-207,-7){\makebox(0,0){\footnotesize$I$}}
\subcaption{\label{fig:min4}}
\end{subfigure}
\caption{The action for all six phases along four paths through the region of moduli space picked out by the coarse gradient search. Each phase is labeled by genus with a subscript corresponding to the labels in figures \ref{fig:AdS} and \ref{fig:TPhase}. The numerical errors are less than 1\% in the sense of Appendix A.  This level of error is consistent with what may appear to be stray points along the curves.}
\label{fig:mins}
\end{figure}
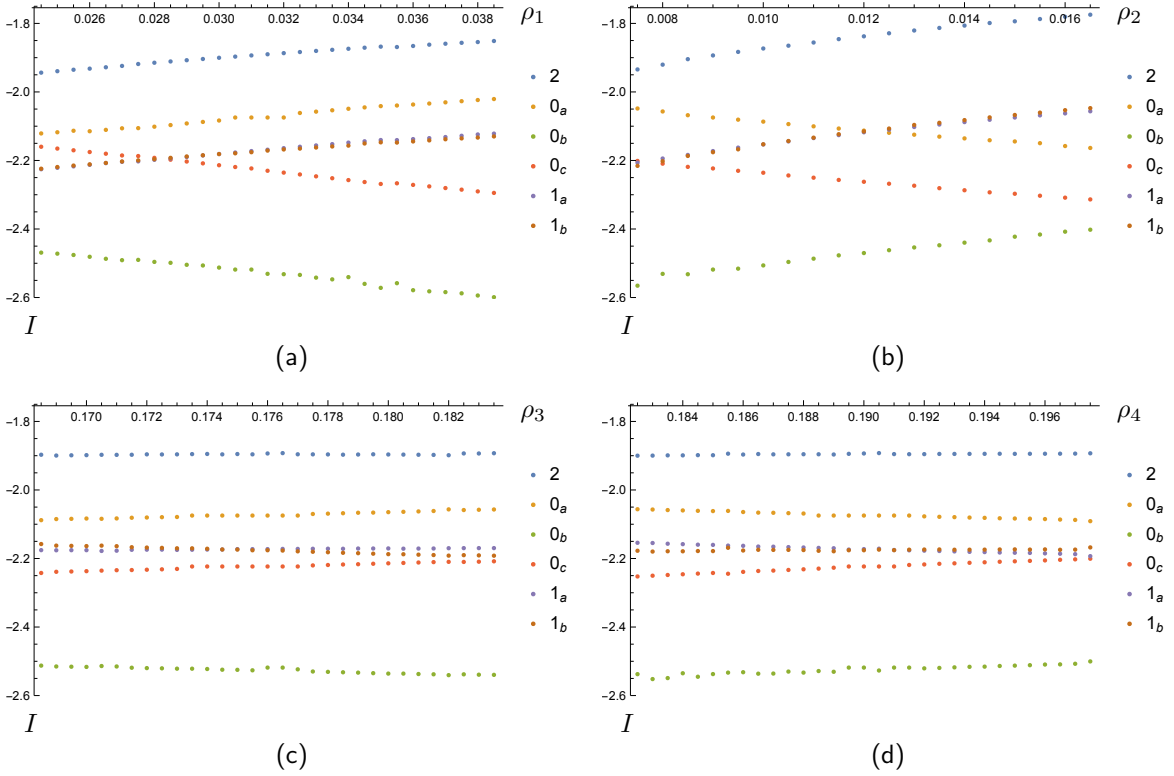

Let us next explore four 1-dimensional trajectories through moduli space each given by varying one of the $\rho_i$ while holding the others fixed. We choose parameters so that these curves lie in the region of moduli space identified by the gradient search above.  The results in figure \ref{fig:mins}
show that the phase $0_{b}$ dominates along all 4 trajectories, though at the right end of figure \ref{fig:min2} we appear to begin to approach a transition to where the $1_b$ phase should dominate.

However, studying the $0_b$ phase turns out to be computationally intensive in this region.  It is thus useful to observe that figure \ref{fig:mins} also shows the computational-more-convenient phases $0_c$, $1_a$, and $1_b$ to dominate over the genus $2$ phase.  Indeed, these phases appear to capture much of the structure of our full set of phases as the actions of the other genus $0$ phases differ from those of $0_c, 1_a, 1_b$  by amounts that are roughly constant along all 4 curves.
We shall return to this constant-offset phenomenon below.

Note that a close study of figure \ref{fig:mins} also suggests that the action difference $(I_2 - I_{0_c})$ increases as either $\rho_3$ or $\rho_4$ decrease, and that
$(I_2 - I_{1_b})$ and $(I_2 - I_{1_a})$ respectively increase as $\rho_3$ and $\rho_4$ increase. It thus appears that -- at least
in the $\rho_3,\rho_4$ directions -- we are near a local minimum of the quantity
\ban{
\Delta I_\text{sub.} = I_2 - \text{min}\left[I_{0_c}, I_{1_a}, I_{1_b} \right]\, .
}

Similarly, as $\rho_1$ and $\rho_2$ increase we find that $(I_2 - I_{0_c})$ increases. However as $\rho_1$ and $\rho_2$ decrease, all 3 actions $I_2$, $I_{1_a}$, and $I_{1_b}$ decrease with $\Delta I_\text{sub.}$ appearing to approach a constant. This latter behavior is consistent with what one expects as the lengths of cycles parameterized by $\rho_1$ and $\rho_2$ tend to zero.   As shown in \cite{MRW}, in phases where a cycle of small boundary length $\ell$ is contractible in the bulk, the action diverges as
\ban{
I \sim - \frac c 6 \frac{\pi^2}{\ell}+ O(\ell^0)\, . \label{eq:pinch}
}
So as $\rho_1$ and $\rho_2$ become small, we expect the phases in which $\alpha_{23}$ and $\beta_{1234}$ are contractible to have actions that diverge to negative infinity. However, the order one behavior depends on the details of the phase so that the difference in action between two such phases naturally approaches a constant. Figures \ref{fig:min1} and \ref{fig:min2} are qualitatively consistent with this behavior, in that the phases whose actions decrease as $\rho_1$ and $\rho_2$ become smaller are precisely those in which a contractible cycle pinches off.

\begin{figure}[ht!]
\centering
\begin{subfigure}{0.47\textwidth}
\includegraphics[width=\textwidth]{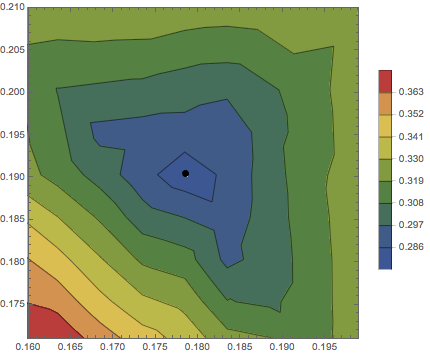}
\put(-120,-10){\makebox(0,0){\footnotesize$\rho_3$}}
\put(-215,90){\makebox(0,0){\footnotesize$\rho_4$}}
\vspace{0.1cm}
\subcaption{$\rho_1,\rho_2$ fixed \label{fig:localminA}}
\end{subfigure}
\hfill
\begin{subfigure}{0.47\textwidth}
\includegraphics[width=\textwidth]{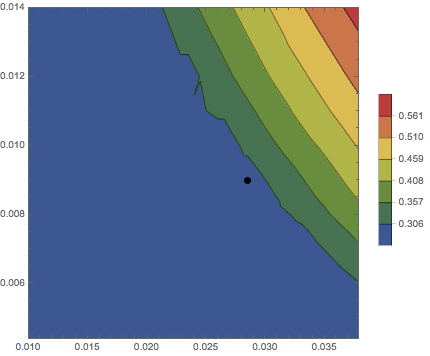}
\put(-120,-10){\makebox(0,0){\footnotesize$\rho_1$}}
\put(-215,90){\makebox(0,0){\footnotesize$\rho_2$}}
\vspace{0.1cm}
\subcaption{$\rho_3,\rho_4$ fixed. \label{fig:localminB}}
\end{subfigure}
\caption{Two 2-dimensional slices of the four dimensional moduli space through $\hat\rho$ (black dot in each figure).   $\rho_1,\rho_2$ are fixed at left, with $\rho_3,\rho)_4$ fixed at right.  Numerical errors are less than 1\% in the sense of appendix A.  This level of error is consistent with some of the jagged features of the contours, though the existence of a local minima is robust to such errors.}
\label{fig:localmin}
\end{figure}

In summary, figure \ref{fig:mins} suggests we are close to a local minimum of $\Delta I_\text{sub.}$ up to possible flat directions associated with approaching the boundary of moduli space.  To confirm this behavior, we now choose a reference point $\hat\rho=(0.0285,0.009,0.1785,0.1905)$ such that $I_{0_c}=I_{1_a}=I_{1_b}$ and explore in more detail how $\Delta I_\text{sub.}$ varies near $\hat \rho$.  Figure \ref{fig:localmin} displays the value of $\Delta I_\text{sub.}$ in two 2-dimensional slices of moduli space through $\hat\rho$. Figure \ref{fig:localminA} varies $(\rho_3,\rho_4)$ with $(\rho_1,\rho_2)$ fixed to match $\hat\rho$, while
\ref{fig:localminB} varies $(\rho_1,\rho_2)$ at fixed $(\rho_3,\rho_4)$.
The data is consistent with the above expectations, indicating a robust local minimum in the $(\rho_3,\rho_4)$-plane while in the $(\rho_1, \rho_2)$-plane $\Delta I_{\text{sub.}}$ either increases or remains roughly constant as one moves away from $\hat \rho$.

\begin{figure}[ht!]
\centering
\begin{subfigure}{0.47\textwidth}
\includegraphics[width=\textwidth]{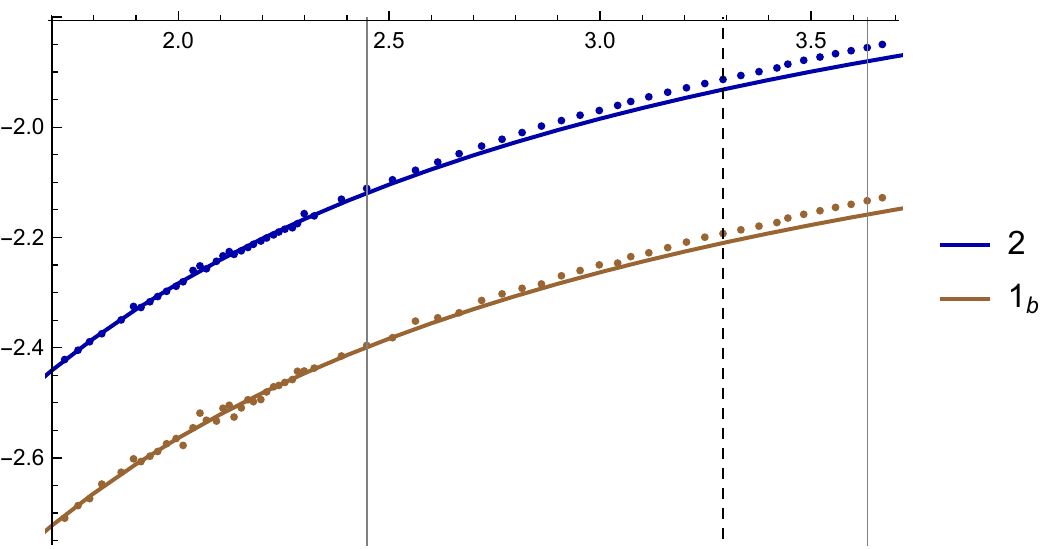}
\put(-10,105){\makebox(0,0){\footnotesize$|\alpha_{23}|$}}
\put(-220,50){\makebox(0,0){\footnotesize$I/c$}}
\vspace{0.1cm}
\subcaption{$\rho_1\to 0$ plateau \label{fig:pinch1}}
\end{subfigure}
\hfill
\begin{subfigure}{0.47\textwidth}
\includegraphics[width=\textwidth]{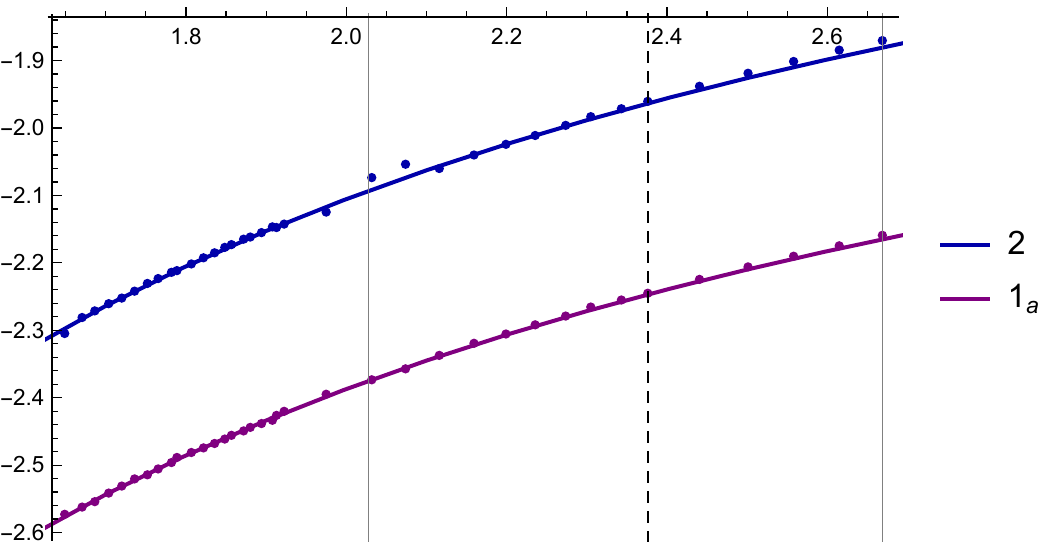}
\put(-10,105){\makebox(0,0){\footnotesize$|\beta_{1234}|$}}
\put(-220,50){\makebox(0,0){\footnotesize$I/c$}}
\vspace{0.1cm}
\subcaption{$\rho_2\to 0$ plateau \label{fig:pinch2}}
\end{subfigure}
\caption{A study of the pinching limits along the plateau in figure \ref{fig:localmin}(b) comparing $I/c$ for the indicated phases (dots) to the form
$-A \frac{\pi^2}6 \frac c \ell + k$ inspired by \eqref{eq:pinch}.  The solid curve shows this function for best-fit values of $A,k$.  Vertical lines indicate $\hat \rho$ and the boundaries of figure \ref{fig:localmin}. (a) $\rho_2,\rho_3, \rho_4$ are fixed to match $\hat \rho$ while $\rho_1$ varies; $A_{\text{best \ fit}}=1.090,1.096$ for $I_2, I_{1_1}$ over the range $1.73< |\alpha_{23}|<2.39$. (b) $\rho_1,\rho_3, \rho_4$ are fixed to match $\hat \rho$ while $\rho_2$ varies; $A_{\text{best \ fit}}= 1.091,1.077$ for $I_2, I_{1_a}$ over the range $1.65<|\beta_{1234}|< 1.97$.
}
\label{fig:pinch}
\end{figure}

Finally, to investigate the flat directions further we compute the action of relevant phases as a function of the length of a pinching cycle along two one-dimensional curves through $\hat\rho$ by varying either $\rho_1$ or $\rho_2$ while holding the other three $\rho_i$ fixed at $\hat \rho_i$.  The results are displayed in figure \ref{fig:pinch} and compared with a fit
to the function $-A \frac{\pi^2}6 \frac c \ell + k$ inspired by \eqref{eq:pinch}. In \ref{fig:pinch1} we use the range $1.73< |\alpha_{23}|<2.39$ and find the best fit to have $A= 1.090$ for $I_2$ and $A=1.096$ for $I_{1_a}$. The difference between these two best-fit values for $A$ is consistent with our estimated $1\%$ numerical error.  However, both differ by $10\%$ from the expected value $A=1$. Since we study the range $\ell \sim 2$ where $\ell$ is not particularly small, we expect that this discrepancy is due to not probing sufficiently far into the asymptotic regime. In practice, it is numerically difficult to compute the action for smaller $\alpha_{23}$.  Similar comments apply to figure \ref{fig:pinch2} where we use the range $1.65<|\beta_{1234}|< 1.97$ to find best-fit parameters $A= 1.091$ for $I_2$ and $A= 1.077$ for $I_{1_a}$.  We take the agreement shown in figure \ref{fig:pinch} to support the above interpretation of figure \ref{fig:localmin} as implying that $\Delta I_{\text{sub.}}$ will remain essentially constant up to the edges of moduli space at $\rho_1=0$, $\rho_2=0$, and thus that the genus 2 phase does not dominate anywhere in moduli space.

\begin{figure}[ht!]
\centering
\begin{subfigure}{0.47\textwidth}
\includegraphics[width=\textwidth]{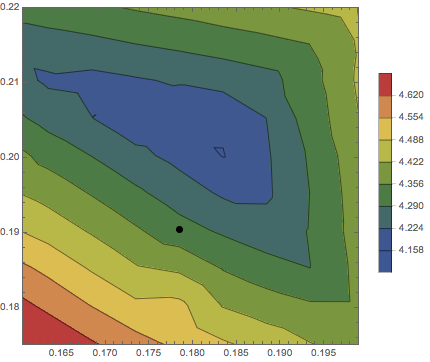}
\put(-120,-10){\makebox(0,0){\footnotesize$\rho_3$}}
\put(-215,90){\makebox(0,0){\footnotesize$\rho_4$}}
\vspace{0.1cm}
\subcaption{$\rho_1,\rho_2$ fixed}
\end{subfigure}
\hfill
\begin{subfigure}{0.47\textwidth}
\includegraphics[width=\textwidth]{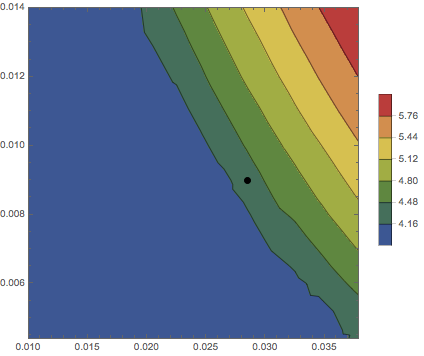}
\put(-120,-10){\makebox(0,0){\footnotesize$\rho_1$}}
\put(-215,90){\makebox(0,0){\footnotesize$\rho_2$}}
\vspace{0.1cm}
\subcaption{$\rho_3,\rho_4$ fixed.}
\end{subfigure}
\caption{The quantity $\Delta \mathscr L$ in the neighborhood of $\hat \rho$ (black dot).}
\label{fig:localH}
\end{figure}

It is interesting to compare the results presented in figure \ref{fig:localmin} with our heuristic based on the total length of contractible cycles. Denoting this total length in a particular phase by $\mathscr L$, figure \ref{fig:localH} plots the quantity,
\ban{
\Delta \mathscr L = \mathscr L_2 - \min[\mathscr L _{0_c},\mathscr L _{1_a},\mathscr L _{1_b}] \,
}
in the neighborhood of $\hat \rho$.
The heuristic suggests that $\Delta I_\text{sub.}$ is smaller (i.e., that the genus 2 phase is less sub-dominant) at small $\Delta {\mathscr L}$, so figures \ref{fig:localmin} and \ref{fig:localH} should be similar.   This is certainly the case at the qualitative level, though -- as seen in other cases -- there are quantitative differences.  In particular, while $\Delta {\mathscr L}$ has a local minimum close to that of
$\Delta I_\text{sub.}$, it clearly differs from $\hat \rho$.

In evaluating our claim that the dominant phase always has genus $0$ or $1$, one should of course ask whether there might be another genus $2$ phase that is more dominant than the one considered here. We have no proof that such phases do not exist, though we have found no natural candidates.   For example, we excluded the phase $\{\alpha_1 - \alpha_3, \alpha_2 - \alpha_4, \beta_1+\beta_3, \beta_2+\beta_4\}$ from our analysis because it breaks a symmetry of the boundary Riemann surface by introducing a twist through the bulk.  On the other hand, in light of the replica symmetry breaking phases for $\Tr(\mathcal T^n)$, it remains possible that this symmetry breaking phase does in fact dominate.  It would be interesting to investigate this further in the future.

Bearing in mind this caveat, the above numerical evidence nevertheless suggests that -- depending on the choice of moduli -- $\mathcal T^2 \ket 0$ can take be either of the form $\ket{\text{AdS}}$ or $\ket 1_\text{BH}$, but that there are no moduli for which it is dual to bulk solutions with higher topology.  If so, the same must hold for ${\mathcal T}^k\ket 0$.

\section{Discussion}
\label{section:discuss}

The results presented in \S\ref{section:replica} and \S\ref{section:states} set up a puzzle with no clear resolution. In \S\ref{section:replica} we presented arguments that if a replica symmetry preserving phase dominates the path integral computing $\Tr(\mathcal T^n)$ for all moduli, then for at least some large $n$ we would have
\ban{
\label{eq:AdS}
\mathcal T^{n/2} \sim \ket{\text{AdS}}\bra{\text{AdS}}	
}
and therefore $\ket 0_K = \ket{\text{AdS}}$. But \eqref{eq:AdS} is inconsistent with the calculations in  $\S\ref{section:states}$ and \cite{MRW} showing that, in certain regions of moduli space, applying $\mathcal T$ to AdS states could produce states described by toroidal geons. It thus seems that replica symmetry breaking phases must dominate the path integral for such moduli.

On the other hand, if the $\mathbb Z_n$ replica symmetry is broken, then acting with appropriate elements of $\mathbb Z_n$ would produce distinct saddles with equal action.  In parallel with the discussion of \S \ref{sec:GRSBP}, if a simple gapped such phase dominates we may then expect to find
\ban{
\label{eq:swap}
\mathcal T^{n/2} \sim \ket \phi \bra \psi + \ket \psi \bra \phi \,
}
for $|\phi \rangle$, $|\psi \rangle$ orthogonal states of differing bulk topology. But $\mathcal T^{n/2}$ would have a negative eigenvalue corresponding to the eigenstate $\frac{1}{\sqrt{2}} \left( |\phi \rangle  - |\psi\rangle\right)$.  We expect that similar issues also arise for more complicated replica-symmetry breaking gapped phases.

However, \S \ref{sec:GRSBP} noted that this issue would be resolved if there are additional phases for ${\rm Tr}({\mathcal T}^n)$ which, taken by themselves would give
\begin{equation}
\mathcal T^{n/2}\sim |\phi\rangle \langle \phi| + |\psi\rangle \langle \psi|  \, \label{eq:dens3},
\end{equation}
for the above $n$, and which somehow turn out to be related to those found thus far above by an unexpected symmetry so that their actions are precisely equal to those just discussed.  In that case, we would sum the contributions \eqref{eq:swap} and \eqref{eq:dens3} with equal weight to give
\ban{
\mathcal T^{n/2} = \frac 12 \left( \ket{\phi} +\ket \psi\right)\left( \bra{\phi} +\bra \psi \right) \label{eq:super2}\, ,
}
the manifestly-positive projector onto $\frac 1{\sqrt 2} \left( \ket{\phi} +\ket \psi\right)$.  If \eqref{eq:super2} holds for all large $n$, the torus ground state  $\ket 0_K$ is then an equal superposition of two different bulk topologies.  Unfortunately, this would also require a similar unexpected new phase for the state ${\mathcal T}^2|0\rangle$ studied in \S \ref{section:states}, or else a fine-tuning that makes the state ${\mathcal T} |0\rangle$ have just the right moduli so that ${\mathcal T}^2 |0\rangle$ lies precisely on the phase boundary at which an AdS and toroidal geon phase exchange dominance.  It would be interesting to investigate both possibilities further in the future.

Other possible resolutions involve gapless phases.  As mentioned at the end of \S \ref{sec:GRSBP}, one option might be for the state ${\mathcal T}^n|0\rangle$ to have genus of order $n$ (at least for $n \ll c$).
But, with the caveat that we leave for the future any investigation of genus-$2$ phases that spontaneously break symmetries of the boundary,
our direct studies of ${\mathcal T}^2|0\rangle$ in \S \ref{section:states} found no evidence that this is the case and instead suggest that ${\mathcal T}^n|0\rangle$ has at most genus $1$.

\begin{figure}[ht!]
	\centering
	\begin{subfigure}{0.47\textwidth}
		\includegraphics[width=\textwidth]{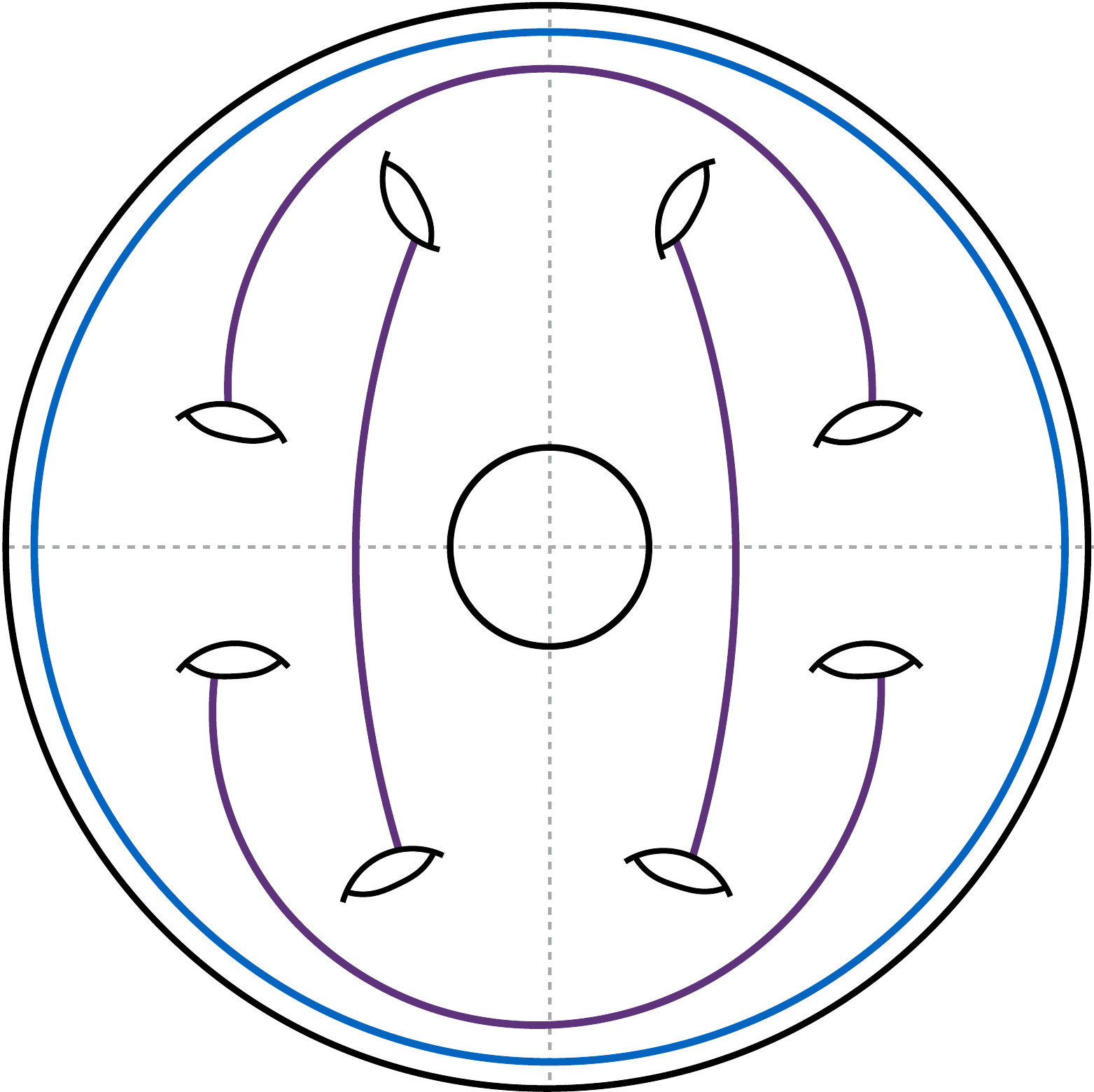}
		\vspace{0.1cm}
		\subcaption{n=8}
		
	\end{subfigure}
	\hfill
	\begin{subfigure}{0.47\textwidth}
		\includegraphics[width=\textwidth]{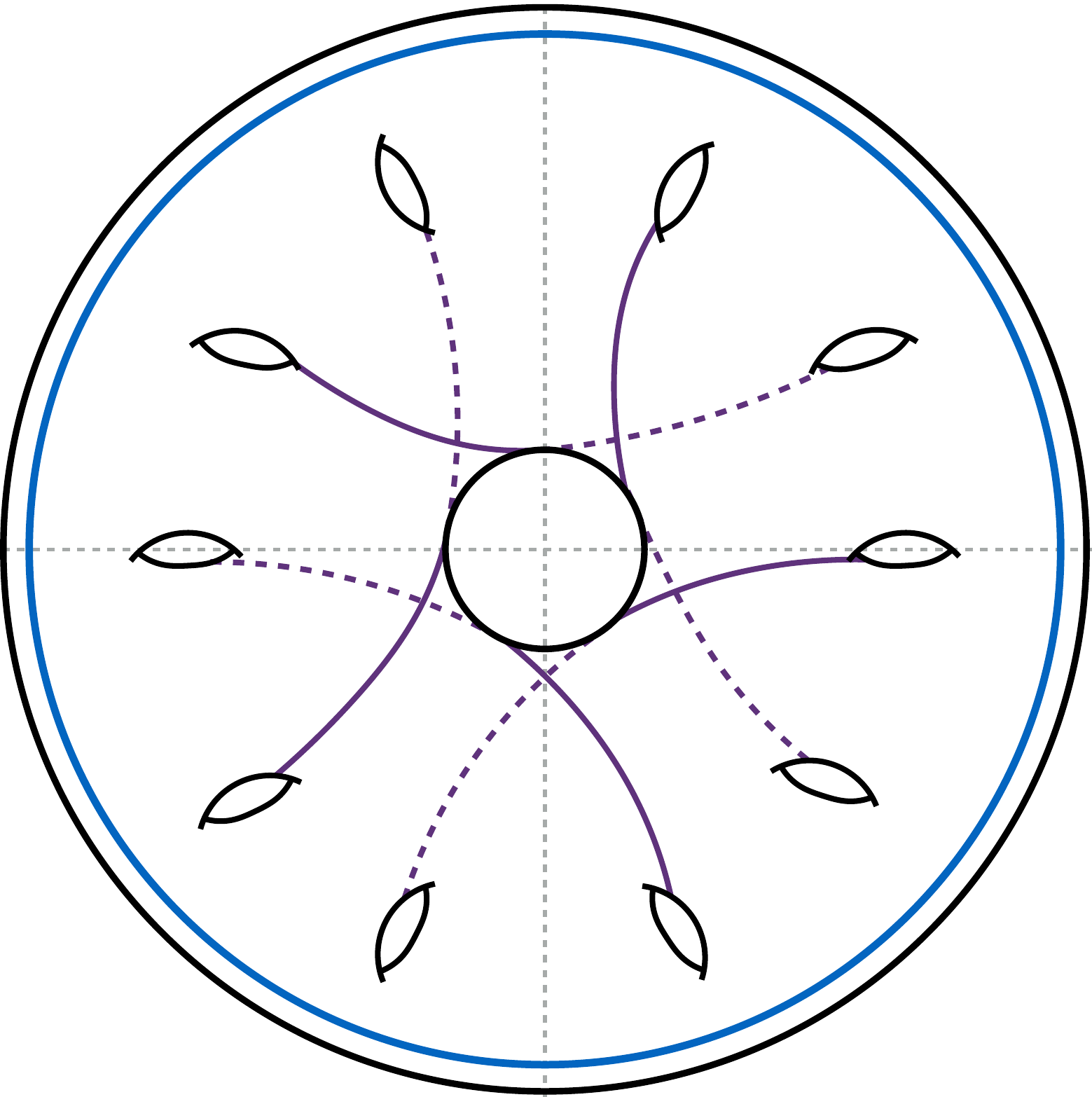}
		\vspace{0.1cm}
		\subcaption{n=10}
		
	\end{subfigure}
	\caption{A gapless phase for $n=8$ and $n=10$. The purple lines joining two holes $i,j$ indicate the pair of cycles $\alpha_i -\alpha_j$ and $\beta_i+\beta_j$ are contractible. Relevant reflection symmetries of the boundary are drawn in dashed grey. Both dashed lines on the left and the vertical dashed line on the right are possible $t=0$ surfaces that can contribute to ${\mathcal T}^{n/2}$ thought of as a state on ${\mathcal H} \otimes {\mathcal H}$.  In contrast, the horizontal dashed line on the right gives a $4$-boundary $t=0$ surface.}
	\label{fig:6block}
\end{figure}

We thus turn to gapless phases where the topology on a $t=0$ surface remains of order $1$ at large $n$. Consider for example the phase defined for $n\geq 6$ with contractible cycles given by $\alpha_i - \alpha_{i+3}$ and $\beta_i + \beta_{i+3}$ for odd $i$, the cycles $\alpha_i -\alpha_{i-3}$ and $\beta_i + \beta_{i-3}$ for even $i$, and the cycle $\beta_0$. We show a depiction of this phase for $n=8$ and $n=10$ in figure \ref{fig:6block}. In particular, the contractibility of $\beta_0$ means that any $t=0$ surface is a connected wormhole with two boundaries.
 As a result, just as for the the BTZ phases of \S \ref{sec:RSP},  the bulk solutions for large $n$ cannot be obtained by cutting and pasting those for small $n$.  The action should thus be only asymptotically linear in $n$ and the phase should be gapless.

\begin{figure}[ht!]
 	\centering
 	\includegraphics[width=0.85\textwidth]{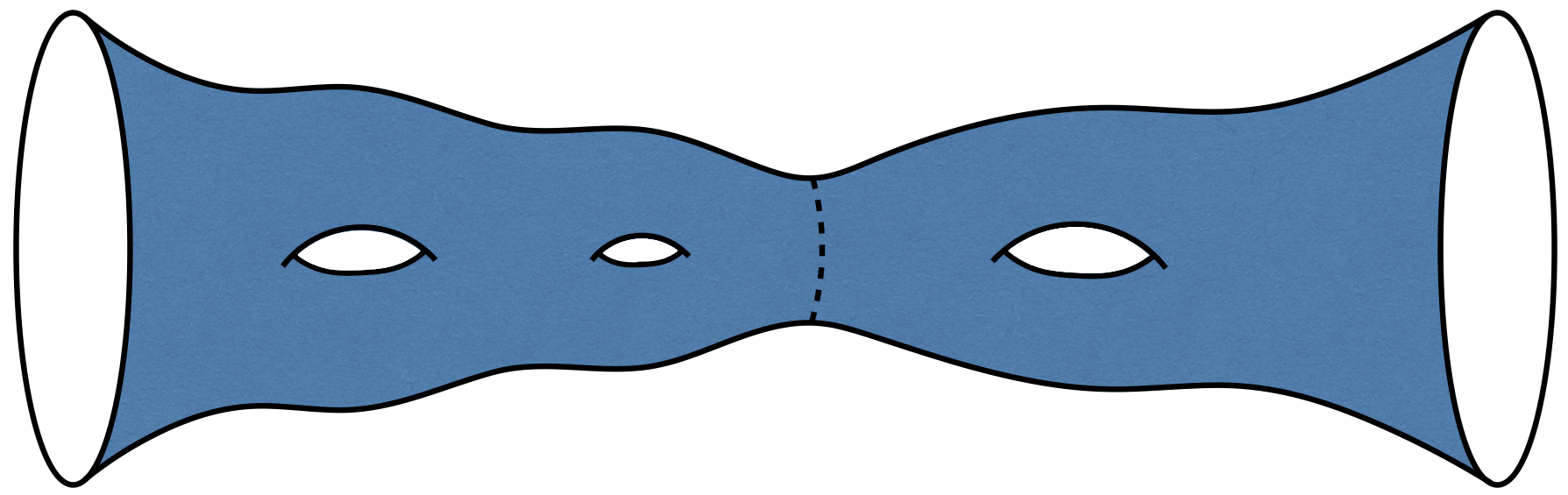}
 	\caption{A cartoon of the $3$-handled wormhole, with a candidate minimal extremal surface drawn in dashed black. \label{fig:wormhole3}}
 \end{figure}

A study of figure \ref{fig:6block} shows that such $t=0$ surfaces have genus $2$, $3$, or $4$. For example, take a $t=0$ surface given by the vertical dashed grey line. For $n=8$, the resulting TFD-like state is a genus 2 wormhole that is symmetric under reflection across the horizontal dashed line, with the symmetry acting to exchange the two boundaries and the two bulk handles. Rotating these dashed lines by $\pi/4$ gives a TFD-like state described by a genus 4 wormhole with a similar symmetry. For $n=10$, the resulting TFD-like state is a genus 3 wormhole that breaks the symmetry of the Riemann surface associated with reflections through the horizontal dashed line. A cartoon of the $t=0$ geometry is depicted in figure \ref{fig:wormhole3}; it is {\it not} symmetric under right-to-left reflections.
Rotating these dashed lines by $\pi/5$ gives the same $t=0$ geometry, but with the two boundaries exchanged. We thus refer to this phase below as the $2,3,4$-handled wormhole.  Note that we use the term wormhole to indicate that two distinct boundaries are connected through the bulk, while we have consistently used the term geon to describe solutions with a single boundary but with non-trivial topology behind a horizon.

As argued previously for the BTZ phases of \S \ref{sec:RSP}, we expect the (log) conformal factors associated with a Schottky description to satisfy the approximate scaling $\phi_{n+2} (w) = \phi_n(w^{\frac{n}{n+2}})$.  And as before, this scaling should result in the length of some bulk cycle on the $t=0$ surface shrinking like $1/n$. So if this phase dominates the state ${\mathcal T}^{n/2}$ on ${\mathcal H}\otimes {\mathcal H}$ will be entangled, with the entropy on each copy of the CFT also scaling like $1/n$.  In particular, as discussed previously, in the limit $n \rightarrow \infty$ it becomes effectively pure on each side, with the would-be entangling surface turning into the infinite throat of an $M=0$ BTZ black hole.    In particular, for $n \equiv 2 \text{ mod } 4$ any $t=0$ surface fails to be symmetric under exchange of the two boundaries.  We thus expect the shrinking cycle to separate a genus-$2$ geon from a toroidal geon; i.e., we expect the shrinking cycle to be the one drawn as a dashed line in fig. \ref{fig:wormhole3}.  At large $n$ (and again interpreting ${\mathcal T}^{n/2}$ as an operator) we thus find
\ban{
\mathcal T^{n/2} \sim \left\{ \begin{array}{ccc}
\frac12 \left(\ket 1_\text{BH} \bra 1_\text{BH} + \ket 2_\text{BH} \bra 2_\text{BH} \right)&\hspace{1cm} & n \equiv 0 \text{ mod } 4\\
 \\
\frac 12  \left(\ket 1_\text{BH} \bra 2_\text{BH} + \frac 12  \ket 2_\text{BH} \bra 1_\text{BH}\right) & \hspace{1cm}& n \equiv 2 \text{ mod } 4
  \end{array}\right. ,
}
so that $\mathcal T$ again has a negative eigenvalue just as in our discussion of gapped phases.

Indeed, it seems unlikely that the above phase will dominate at large $n$ over the gapped replica symmetry breaking phase of figure \ref{fig:twounit}.  In order for it to do so and to respect our heuristic, we must have
\ban{
|\alpha_i - \alpha_{i+3}| + |\beta_i + \beta_{i+3}| < |\alpha_i - \alpha_{i+1}| + |\beta_i + \beta_{i+1}|\, . \label{eq:gapless}
}
While we have not shown this to be impossible, it would be surprising if cycles stretching between replicas $i$ and $i+3$ were shorter than cycles that connect nearest neighbors.  So in the end the $2,3,4$-wormhole phase seems unlikely to be relevant at large $n$.

It thus appears that additional phases not yet studied must become important at large $n$.  It remains interesting to determine whether these are new gapped (or gapless) phases that lead to a ground state of indefinite topology as suggested above, phases that spontaneously break symmetries as mentioned at the end of \S \ref{section:states}, or merely other -- perhaps more complicated -- phases that we were not considered here.  In the latter category one might also consider phases that do not lie in the universal sector, and which are thus not described by pure gravity in the bulk.  In particular, one can ask whether considering condensates of long strings \cite{MaldMaoz} might somehow resolve our puzzle, or other non-gravitational phases as for example in \cite{belin}. However, our analysis suggests that such phases would resolve the puzzle only if they create a replica-symmetric phase of lower action which dominates in the troublesome regime of moduli space. {In a similar vein, one can note that our notation has suppressed information about boundary gravitons and ask if this let us to neglect some important effect. But since boundary gravitons simply dress each of the states appearing in e.g. \eqref{eq:dens}, they cannot affect our discussion of $\Tr(\mathcal T^n)$ .}

In addition to the study of new phases, it would be useful to study in detail the map on the genus-$4$ moduli space associated with passing between the $R_{bndy}=-2$ conformal frame and the conformal frame in which ${\mathcal T}^2 |0\rangle$ is described by an $R_{bndy}=-2$ twice-punctured genus $2$ surface (representing the ${\mathcal T}^2$ with geodesic boundaries) attached to a positive-curvature hemisphere.  This conformal frame was used briefly in \S \ref{sec:GRSBP} to argue that there is a region of moduli space in which the replica-symmetry breaking phase from figure \ref{fig:twounit} dominates over those studied in \S \ref{sec:RSP}.  A study of this map would allow a more precise understanding of ${\mathcal T}^2 |0\rangle$, and also of ${\mathcal T}^k |0 \rangle$ for higher $k$.   In a different direction, it would be interesting to extract a log term of the form shown in \eqref{eq:asymptlin} from the bulk action of a gapless phase at large replica number $n$, or to better understand the relationship between the `total cycle length heuristic' and the actual bulk action.  While we took some steps toward the latter in figures \ref{fig:replicaresults} and \ref{fig:localH}, we leave further investigation for future work.

{Finally, we our treatment here was confined to the holographic case, it would also be interesting to study the analogous torus operators in non-holographic CFTs.  For example, it would be interesting to extract information about the spectrum from known results (see e.g. ) for higher genus free field partition functions.  One might also imagine that one could obtain useful results for rational CFTs.}


\section*{Acknowledgments}
It is a pleasure to thank Henry Maxfield for useful conversations. This work was supported in part by a U.S. National Science Foundation under grant number PHY15-04541 and also by the University of California.

\appendix

\section{Phase Space of $\mathcal T \ket 0$}
\label{appendix:moduli}

In this appendix we compute the gravitational dual of $\mathcal T \ket 0$ in the $\mathbb Z_2$ symmetric subspace of the moduli space of $\mathcal T$ considered in this paper. We show that restricting to this two dimensional subspace does not significantly restrict the phase space of states of $\mathcal T\ket 0$. That is, by varying the moduli of $\mathcal T$, one is able to construct states that are dual to pure AdS or states that are dual to toroidal geons of a wide range of moduli.

Our strategy will be to start with a representation of $\mathcal T^2$ as the path integral over two copies of $\mathscr T$ glued together. We will use this representation to compute the moduli of $\mathcal T$ as lengths of cycles on $\mathscr T$ in the appropriate conformal frame. To compute the inner product $\corr{0|\mathcal T^2 |0}$, we glue hemispheres on to the two cuts in the path integral over $\mathscr T^2$. The resulting path integral is now one over two copies of a once punctured torus $\overline {\mathscr T}$ glued together along the puncture. This path integral is now of the type considered in \cite{MRW}, and we can compute the semi-classical gravitational approximation of $\mathcal T\ket0$ from the saddle points in the usual way.

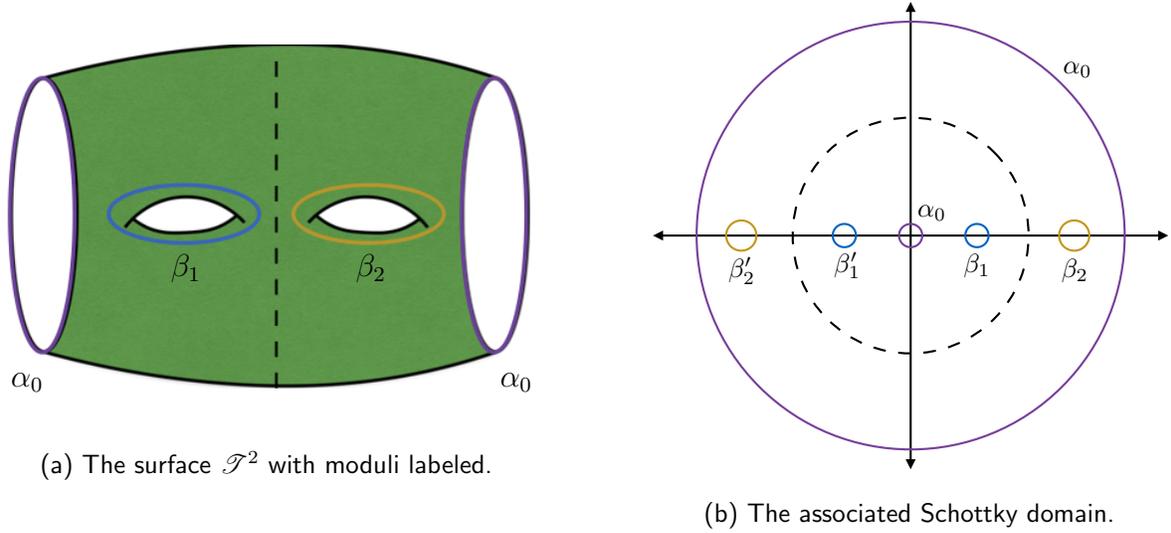
\begin{figure}[ht!]
	\centering
	\begin{subfigure}{0.45\textwidth}
		\includegraphics[width=\textwidth]{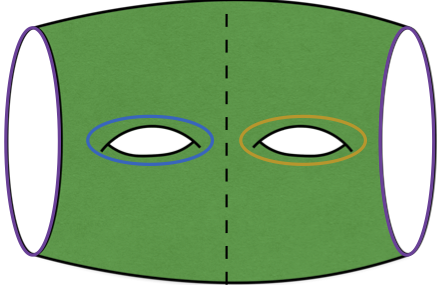}
		\put(-190,2){\makebox(0,0){$\alpha_0$}}
		\put(-130,45){\makebox(0,0){$\beta_1$}}
		\put(-60,45){\makebox(0,0){$\beta_2$}}
		\put(-5,2){\makebox(0,0){$\alpha_0$}}				
		\vspace{0.6cm}
		\subcaption{The surface $\mathscr T^2$ with moduli labeled.}
		
	\end{subfigure}
	\hfill
	\begin{subfigure}{0.45\textwidth}
		\includegraphics[width=\textwidth]{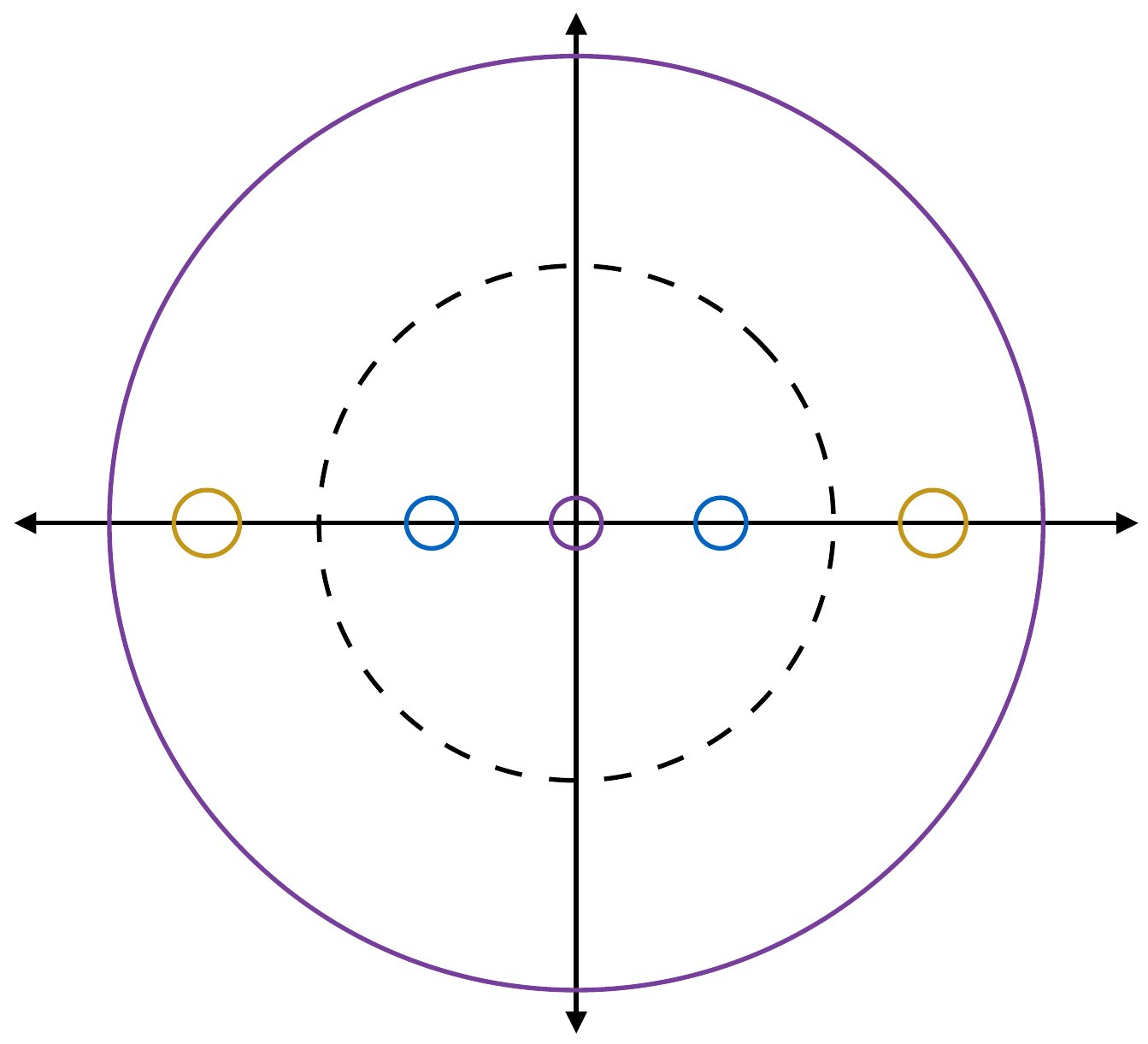}
		\put(-91,99){\makebox(0,0){\footnotesize$\alpha_0$}}
		\put(-36,153){\makebox(0,0){\footnotesize$\alpha_0$}}
		\put(-37,77){\makebox(0,0){\footnotesize$\beta_2$}}
		\put(-74,79){\makebox(0,0){\footnotesize$\beta_1$}}
		\put(-123,79){\makebox(0,0){\footnotesize$\beta_1'$}}
		\put(-163,77){\makebox(0,0){\footnotesize$\beta_2'$}}
		\vspace{0.1cm}
		\subcaption{The associated Schottky domain.}
		
	\end{subfigure}
	\caption{The $\mathbb Z_2$ symmetric surface and Schottky domain used to compute the moduli of $\mathcal T^2$. Note that $|\beta_1| = |\beta_2|$ by the reflection symmetry across the dashed cycle.}
	\label{fig:TSchottky}
\end{figure}

Consider a Schottky uniformization of $\mathscr T^2$ in the conformal frame $R_\text{bndy}=-2$ as shown in figure \ref{fig:TSchottky}. In this conformal frame, we can compute the moduli of $\mathcal T$ by evaluating the length of various boundary cycles as labeled in the figure. We choose to parameterize the $\mathbb Z_2$ symmetric moduli space by $|\alpha_0|$ and $|\beta_1|$.

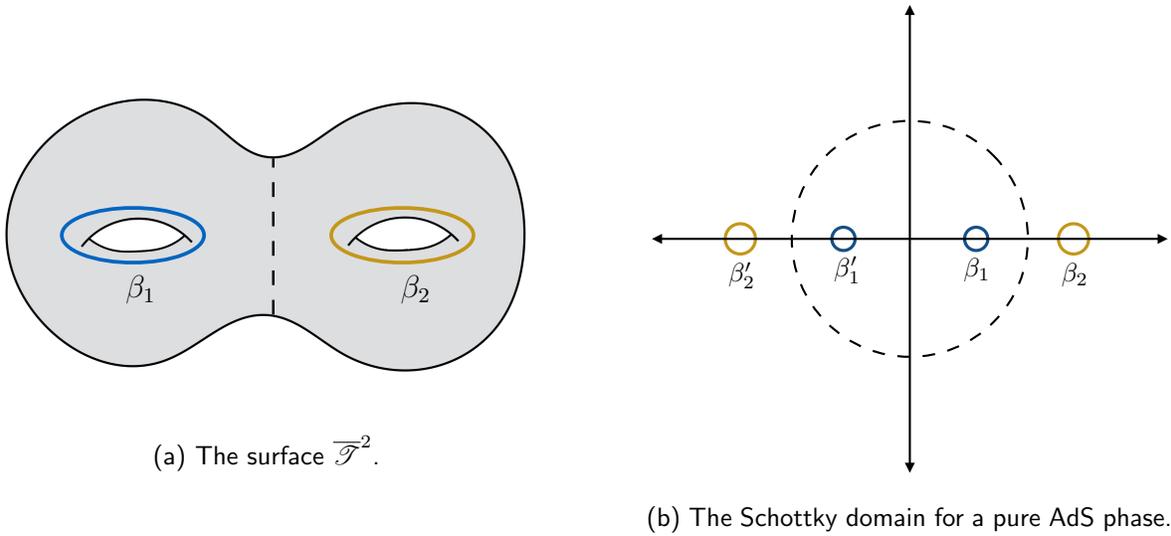
\begin{figure}[ht!]
	\centering
	\begin{subfigure}{0.45\textwidth}
		\includegraphics[width=\textwidth]{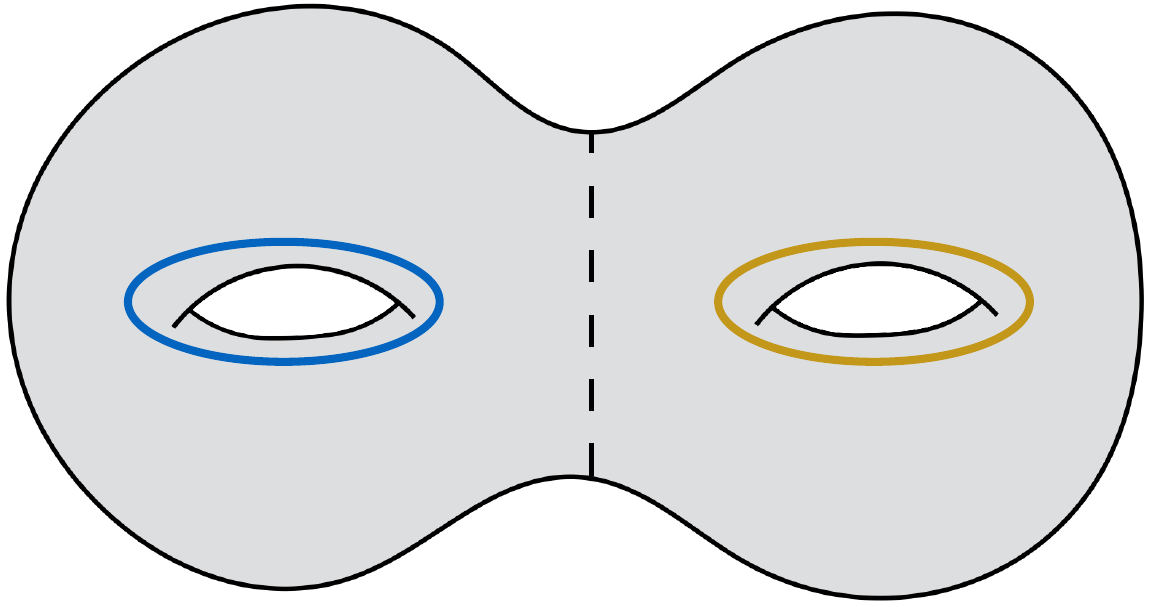}
		\put(-147,32){\makebox(0,0){$\beta_1$}}
		\put(-43,32){\makebox(0,0){$\beta_2$}}
		\vspace{0.6cm}
		\subcaption{The surface $\overline{\mathscr T}^2$.\label{fig:barTSchottkyA}}
		
	\end{subfigure}
	\hfill
	\begin{subfigure}{0.45\textwidth}
		\includegraphics[width=\textwidth]{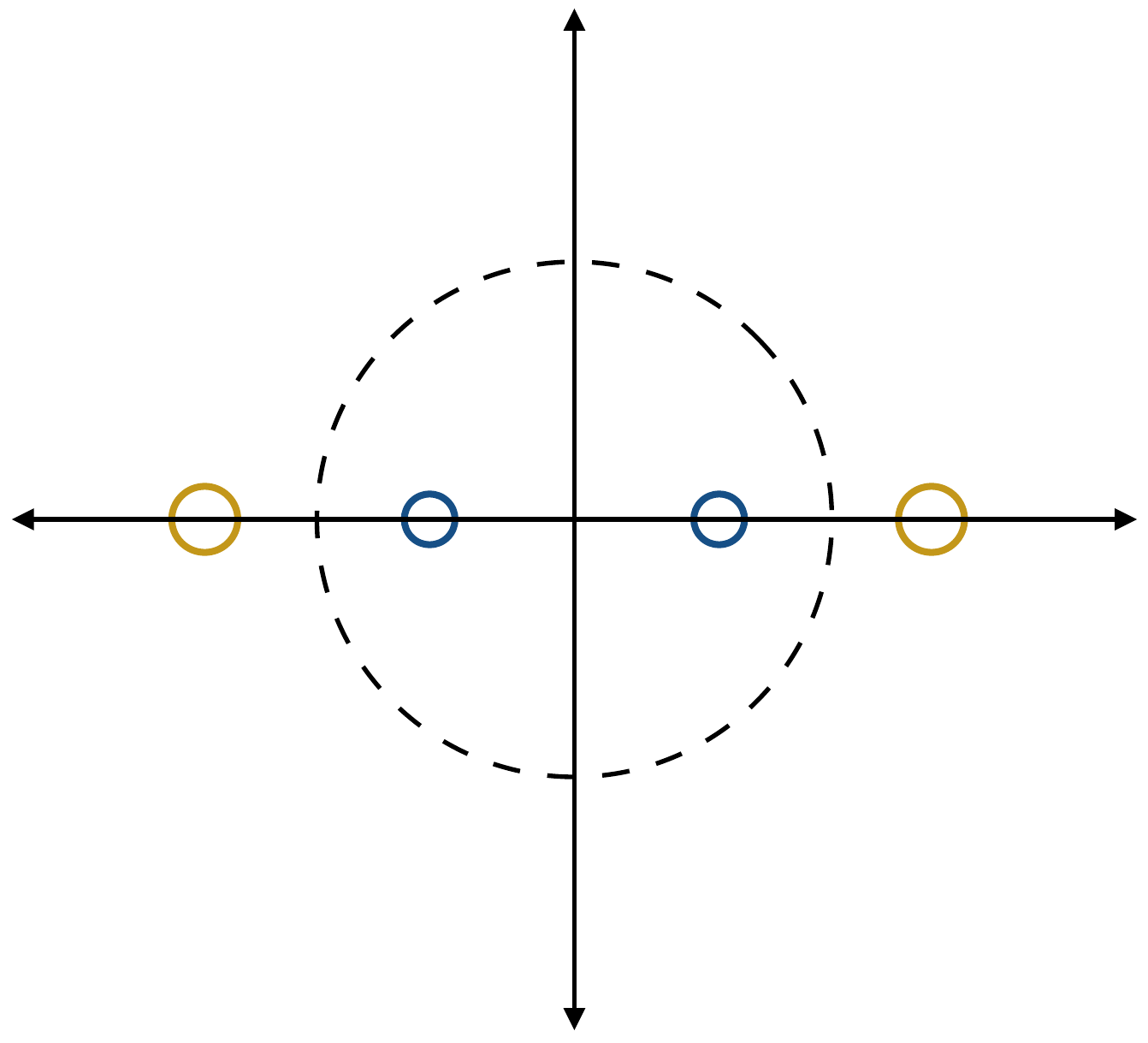}
		\put(-37,77){\makebox(0,0){\footnotesize$\beta_2$}}
		\put(-74,79){\makebox(0,0){\footnotesize$\beta_1$}}
		\put(-123,79){\makebox(0,0){\footnotesize$\beta_1'$}}
		\put(-163,77){\makebox(0,0){\footnotesize$\beta_2'$}}
		\vspace{0.1cm}
		\subcaption{The Schottky domain for a pure AdS phase. \label{fig:barTSchottkyB}}
		
	\end{subfigure}
	\caption{The $\mathbb Z_2$ symmetric surface and Schottky domain used to compute $\corr{0|\mathcal T^2|0}$. The handlebody represented by the Schottky domain is a pure AdS phase.}
	\label{fig:barTSchottky}
\end{figure}

Gluing hemispheres on to the boundaries of ${\mathscr T}^2$ can be achieved by filling in the purple circles $\alpha_0$ in the Schottky domain of figure \ref{fig:TSchottky}. The resulting domain is a representation of the surface $ \overline {\mathscr T}^2$ used to compute the norm of the state $\mathcal T \ket 0$, with the moduli of $\mathcal T$ computed by the original Schottky domain. We draw the surface $\overline {\mathscr T}^2$ and its Schottky uniformization in figure \ref{fig:barTSchottky}. The semi-classical gravitational dual of this state is given by the geometry on the surface fixed by the $\mathbb Z_2$ symmetry exchanging the two halves of $\overline {\mathscr T}$. In this way, we can map out the phase space of states $\mathcal T \ket 0$ in terms of the moduli of $\mathcal T$.

The domain constructed in this manner is a representation of a pure AdS phase of $\mathcal T \ket 0$. The moment of time symmetry on which we define the state is given by the dashed cycle in figure \ref{fig:barTSchottky}. In the associated Schottky domain, this cycle is contractible in the bulk, as it can be lifted off the boundary and shrunk to a point without intersecting the geodesic hemispheres that are identified by the Schottky group. Alternatively, we can use the formula \eqref{eq:bulkgenus}, with $n=0$ and $b=1$ giving $g_{t=0}=0$.

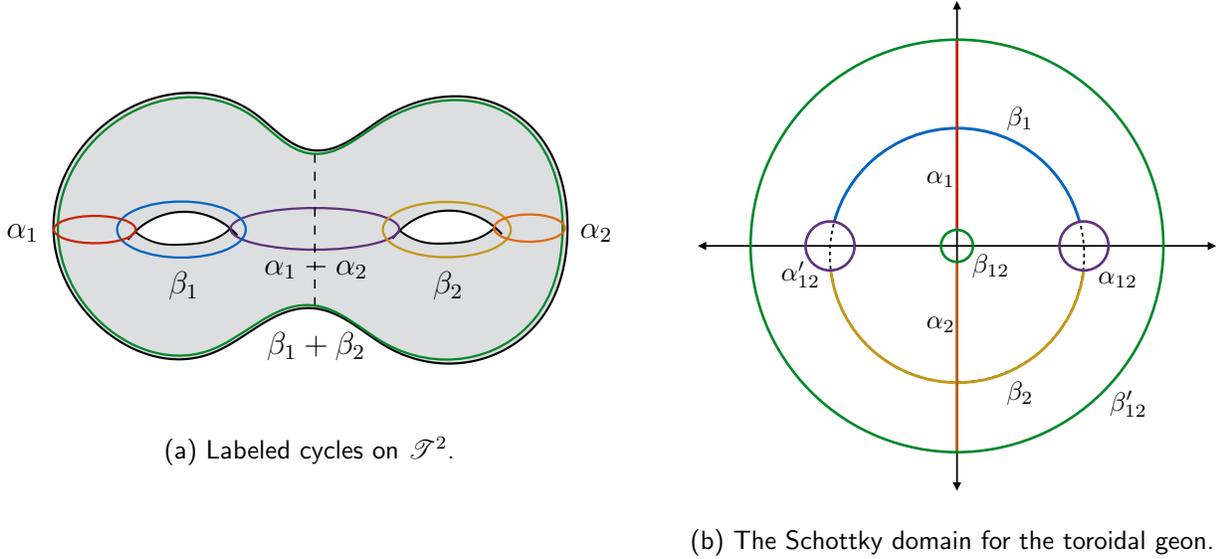
\begin{figure}[ht!]
	\centering
	\begin{subfigure}{0.45\textwidth}
		\includegraphics[width=\textwidth]{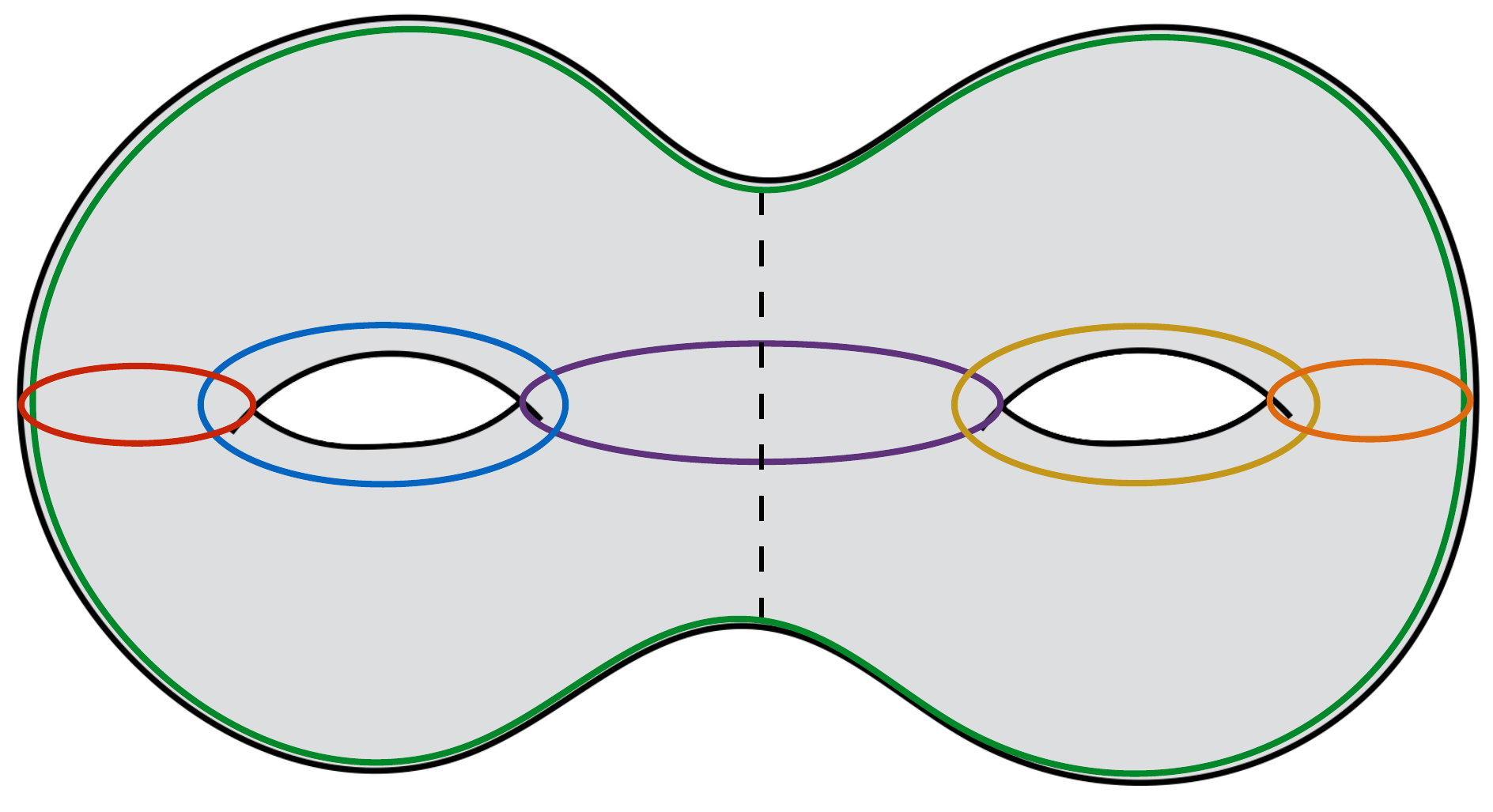}
		\put(-208,53){\makebox(0,0){$\alpha_1$}}
		\put(9,53){\makebox(0,0){$\alpha_2$}}
		\put(-147,33){\makebox(0,0){$\beta_1$}}
		\put(-47,33){\makebox(0,0){$\beta_2$}}
		\put(-97,40){\makebox(0,0){$\alpha_1 - \alpha_2$}}
		\put(-97,10){\makebox(0,0){$\beta_1+\beta_2$}}	
		\vspace{0.6cm}
		\subcaption{Labeled cycles on $\mathscr T^2$.}
		
	\end{subfigure}
	\hfill
	\begin{subfigure}{0.45\textwidth}
		\includegraphics[width=\textwidth]{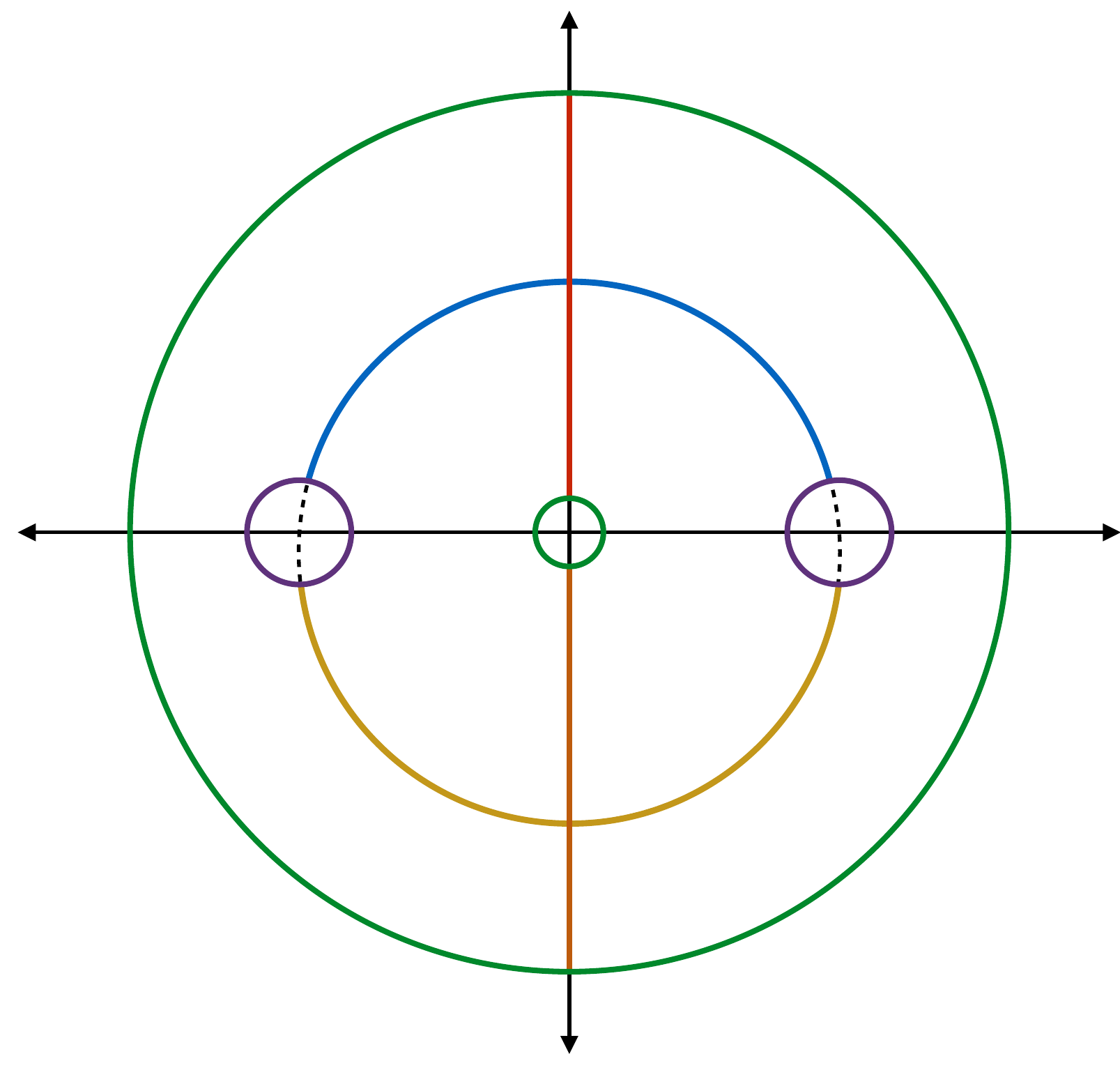}
		\put(-85,89){\makebox(0,0){\footnotesize$\beta_{12}$}}
		\put(-33,37){\makebox(0,0){\footnotesize$\beta_{12}'$}}
		\put(-37,84){\makebox(0,0){\footnotesize$\alpha_{12}$}}
		\put(-157,86){\makebox(0,0){\footnotesize$\alpha'_{12}$}}
		\put(-74,145){\makebox(0,0){\footnotesize$\beta_1$}}
		\put(-74,41){\makebox(0,0){\footnotesize$\beta_2$}}
		\put(-104,122){\makebox(0,0){\footnotesize$\alpha_1$}}
		\put(-104,67){\makebox(0,0){\footnotesize$\alpha_2$}}
		\vspace{0.1cm}
		\subcaption{The Schottky domain for the toroidal geon.}
		
	\end{subfigure}
	\caption{The uniformization of the surface $\mathscr T^2$ used to represent the toroidal geon phase. The contractible cycles are given by $\alpha_{12}=\alpha_1-\alpha_2$ and $\beta_{12}=\beta_1+\beta_2$. To construct the Schottky domain, we cut along these cycles and identify each side of the cut as $C_i$ and $C_i'$.}
	\label{fig:GSchottky}
\end{figure}

To compute the semi-classical gravitational dual of this state, we match the moduli of $\overline{\mathscr T}^2$ to the Schottky uniformizations corresponding to the remaining pure AdS phase and the toroidal geon phase. The remaining pure AdS phase can be described by the Schottky domain in figure \ref{fig:barTSchottkyB}, with the interpretation of the $\alpha$ and $\beta$ cycles flipped. To describe the toroidal geon phase, we can use the Schottky domain in figure \ref{fig:GSchottky}.
In this case, the relevant $t=0$ slice is given by the horizontal axis of the Schottky domain. This slice is broken into $b=1$ boundary segments and intersects $n=2$ pairs of circles given $g_{t=0}=1$. Therefore, this domain is a representation of the toroidal geon phase. {To gain intuition for this result, note that contractibility of {$\beta_1+ \beta_2$} implies that we can deform the cycle $\beta_1$ through the bulk until it becomes the cycle $-\beta_2$.  Thinking of the left half of figure \ref{fig:barTSchottkyA} as bulk Euclidean time negative infinity and the right half as bulk Euclidean time positive infinity, such a deformation must pass through $t=0$.  But the cycle $\beta_1$ is non-contractible, so the $t=0$ slice must contain a non-contractible cycle.}

\begin{figure}[th!]
	\centering
	\includegraphics[width=0.55\textwidth]{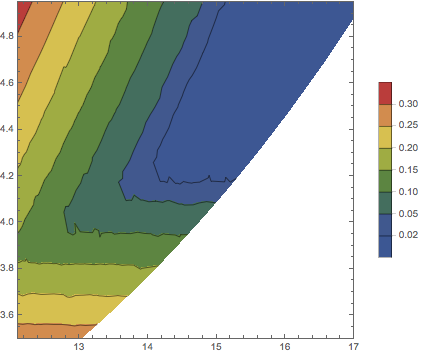}
	\put(-255,100){\makebox(0,0){$|\beta_1|$}}
	\put(-140,-7){\makebox(0,0){$|\alpha_0|$}}
	\put(-140,217){\makebox(0,0){$(I_\text{TG}-I_\text{dom.})/c$}}
	\caption{Phase diagram for $\mathcal T\ket0$. In the blue region $I_\text{TG}-I_\text{dom.}$ vanishes and the toroidal geon phase is dominant. Outside of this region an AdS phase is dominant. In the white region we have no data, as it is numerically difficult to probe.
		 \label{fig:tmoduli}}
\end{figure}

Comparing the actions of these phases, we can determine the dominant saddle and the semi-classical description of $\mathcal T\ket 0$. We display the difference in action between the toroidal geon phase and the dominant phase $\frac 1 c(I_\text{TG}-I_\text{dom.})$ in figure \ref{fig:tmoduli}. When the toroidal geon phase is dominant this quantity vanishes.
In the region where the toroidal geon phase dominates, we compute the length the horizon and the lengths of the cycles of the bulk torus. We find a minimal horizon size matching \cite{MRW} and a range of internal cycles, leading us to conclude the moduli space of states is not significantly restricted by imposing the $\mathbb Z_2$ symmetries of $\mathcal T$.




\section{Estimation of Numerical Error}
\label{appendix:error}

In this appendix we discuss some sources of numerical error. First there is the discretization error from using finite element methods to solve for $\phi$. We can estimate this error as in \cite{MRW} by computing the area of the Riemann surface and comparing it to the Gauss Bonnet theorem. That is, we have
\ban{
A(g) = 4 \pi (g-1) 	
}
in AdS units. We define $\epsilon_A = |1-A/A(g)|$ as an estimate for this error. For domains where the geodesic lengths are computed in the Schottky representation (and not the Poincar\'e disk), we report this value as the overall error.

Further, we perform numeric integration over the boundary circles by adding up the function values on the mesh nodes coinciding with a particular circle. In order to do so, we must set a tolerance for considering a point on the boundary circle, which introduces some numerical error. We can estimate this error by computing the length of a boundary segment using a flat metric and compare it to the analytic formula for the length of the arc of the corresponding circle. The tolerance is chosen to minimize the percent error of each boundary circle. We denote the maximum of all of these errors $\epsilon_C$.

Additionally, we estimate the propagation of these uncertainties in computing the geodesic lengths in the  Poincar\'e disk. To estimate this error, we construct the corresponding domain in the Poincar\'e disk for boundary segment lengths $\ell_0 (1 + \text{max}(\epsilon_A,\epsilon_A))$, where $\ell_0$ is the segment length as computed by the numerical solution $\phi$. The maximum change in geodesic lengths by using different lengths of boundary segments estimates this error $\epsilon_G$. When $\epsilon_G> 0.05$ we find the moduli matching algorithms tend not to converge, and so we take this as a cutoff of numerical error. The overall error is taken to be $\epsilon = \text{max}(\epsilon_A, \epsilon_C, \epsilon_G)$ which is $\epsilon_G$. We find $\epsilon_G$ tends to be 1-2 orders of magnitude larger than $\epsilon_A$ or $\epsilon_C$, and for future work this suggests a way to reduce errors even further. 	

\begin{figure}[th!]
	\centering
	\includegraphics[width=0.55\textwidth]{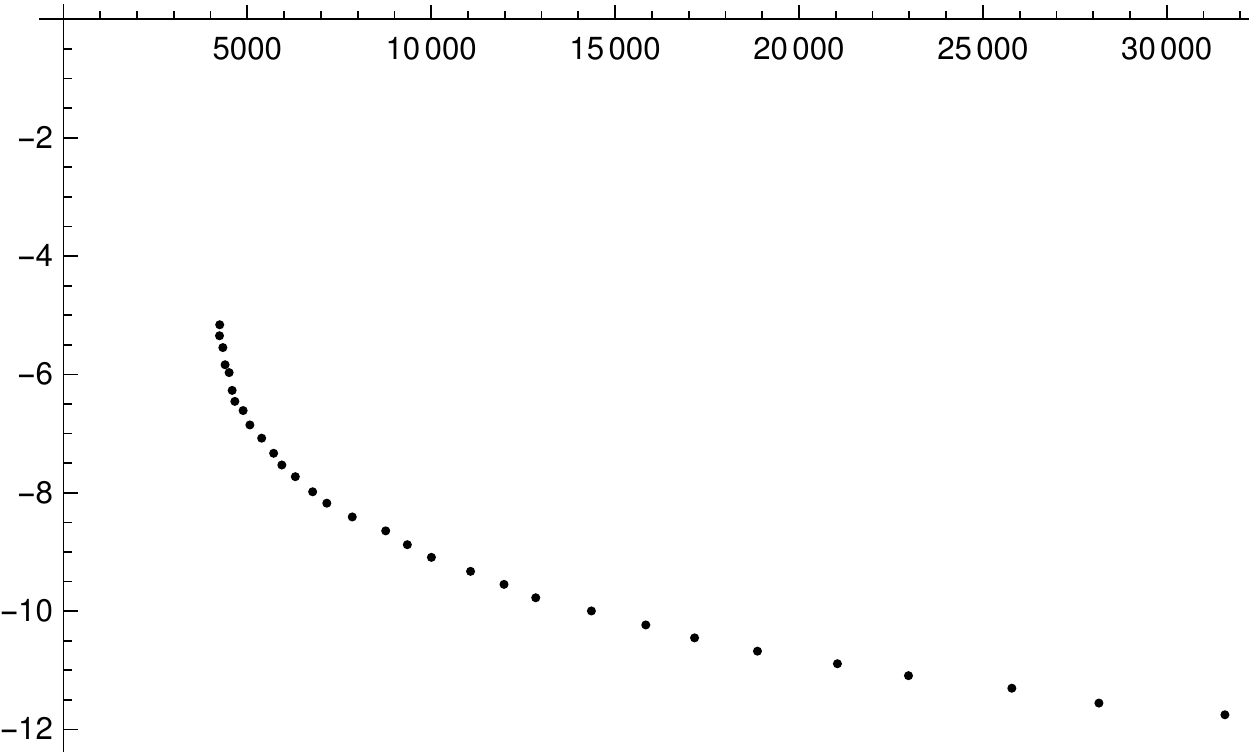}
	\put(-275,70){\makebox(0,0){$\log \epsilon_A$}}
	\put(-100,155){\makebox(0,0){$n_\text{points}$}}
	\caption{Estimation of $\epsilon_A$ as a function of number of lattice points $n_\text{points}$.
		 \label{fig:numerics}}
\end{figure}

Finally, in matching moduli between phases, we only require them to match up to a certain threshold. That is we require the percentage difference between two matching moduli to be less than $\text{max}(\epsilon_G, 3 \times 10^{-3})$.

The error $\epsilon_A$ can be reduced simply by using more lattice points to discretize the domain. To reduce $\epsilon_C$ we must include more boundary points as well as choose the tolerance accordingly. In figure \ref{fig:numerics} we display $\epsilon_A$ for a representative genus 2 phase from \S\ref{section:states} as a function of the number of points.

\bibliographystyle{jhep}
\phantomsection
\renewcommand*{\bibname}{References}

\bibliography{references}

\providecommand{\href}[2]{#2}\begingroup\raggedright\begin{thebibliography}{10}

\bibitem{Hawking:1982dh}
S.~W. Hawking and D.~N. Page, {\it {Thermodynamics of Black Holes in anti-De
  Sitter Space}},  {\em Commun. Math. Phys.} {\bf 87} (1983) 577.

\bibitem{Hartman:2014oaa}
T.~Hartman, C.~A. Keller, and B.~Stoica, {\it {Universal Spectrum of 2d
  Conformal Field Theory in the Large c Limit}},  {\em JHEP} {\bf 09} (2014)
  118, [\href{http://arxiv.org/abs/1405.5137}{{\tt arXiv:1405.5137}}].

\bibitem{MRW}
H.~Maxfield, S.~Ross, and B.~Way, {\it {Holographic partition functions and
  phases for higher genus Riemann surfaces}},  {\em Class. Quant. Grav.} {\bf
  33} (2016), no.~12 125018, [\href{http://arxiv.org/abs/1601.00980}{{\tt
  arXiv:1601.00980}}].

\bibitem{Krasnov1}
K.~Krasnov, {\it {Holography and Riemann surfaces}},  {\em Adv. Theor. Math.
  Phys.} {\bf 4} (2000) 929--979,
  [\href{http://arxiv.org/abs/hep-th/0005106}{{\tt arXiv:hep-th/0005106}}].

\bibitem{Krasnov2}
K.~Krasnov, {\it {Black hole thermodynamics and Riemann surfaces}},  {\em
  Class. Quant. Grav.} {\bf 20} (2003) 2235--2250,
  [\href{http://arxiv.org/abs/gr-qc/0302073}{{\tt arXiv:gr-qc/0302073}}].

\bibitem{maldTFD}
J.~M. Maldacena, {\it {Eternal black holes in anti-de Sitter}},  {\em JHEP}
  {\bf 04} (2003) 021, [\href{http://arxiv.org/abs/hep-th/0106112}{{\tt
  arXiv:hep-th/0106112}}].

\bibitem{HRT}
V.~E. Hubeny, M.~Rangamani, and T.~Takayanagi, {\it {A Covariant holographic
  entanglement entropy proposal}},  {\em JHEP} {\bf 07} (2007) 062,
  [\href{http://arxiv.org/abs/0705.0016}{{\tt arXiv:0705.0016}}].

\bibitem{Galloway:1999bp}
G.~J. Galloway, K.~Schleich, D.~M. Witt, and E.~Woolgar, {\it {Topological
  censorship and higher genus black holes}},  {\em Phys. Rev.} {\bf D60} (1999)
  104039, [\href{http://arxiv.org/abs/gr-qc/9902061}{{\tt
  arXiv:gr-qc/9902061}}].

\bibitem{GKP}
S.~S. Gubser, I.~R. Klebanov, and A.~M. Polyakov, {\it {Gauge theory
  correlators from noncritical string theory}},  {\em Phys. Lett.} {\bf B428}
  (1998) 105--114, [\href{http://arxiv.org/abs/hep-th/9802109}{{\tt
  arXiv:hep-th/9802109}}].

\bibitem{wittenholo}
E.~Witten, {\it {Anti-de Sitter space and holography}},  {\em Adv. Theor. Math.
  Phys.} {\bf 2} (1998) 253--291,
  [\href{http://arxiv.org/abs/hep-th/9802150}{{\tt arXiv:hep-th/9802150}}].

\bibitem{Seiberg:1999xz}
N.~Seiberg and E.~Witten, {\it {The D1 / D5 system and singular CFT}},  {\em
  JHEP} {\bf 04} (1999) 017, [\href{http://arxiv.org/abs/hep-th/9903224}{{\tt
  arXiv:hep-th/9903224}}].

\bibitem{MaldMaoz}
J.~M. Maldacena and L.~Maoz, {\it {Wormholes in AdS}},  {\em JHEP} {\bf 02}
  (2004) 053, [\href{http://arxiv.org/abs/hep-th/0401024}{{\tt
  arXiv:hep-th/0401024}}].

\bibitem{Ford}
L.~R. Ford, {\em Automorphic functions}.
\newblock Chelsea Pub. Co., 1972.

\bibitem{TZ}
L.~Takhtajan and P.~Zograf, {\it {On uniformization of Riemann surfaces and
  Weyl-Peterson metric on Teichmuller and Schottky spaces}},  {\em Math. USSR
  Sb.} {\bf 60} (1988), no.~2 297--131.

\bibitem{Yin}
X.~Yin, {\it {On Non-handlebody Instantons in 3D Gravity}},  {\em JHEP} {\bf
  09} (2008) 120, [\href{http://arxiv.org/abs/0711.2803}{{\tt
  arXiv:0711.2803}}].

\bibitem{Brill1}
D.~R. Brill, {\it {Multi - black hole geometries in (2+1)-dimensional
  gravity}},  {\em Phys. Rev.} {\bf D53} (1996) 4133--4176,
  [\href{http://arxiv.org/abs/gr-qc/9511022}{{\tt arXiv:gr-qc/9511022}}].

\bibitem{Brill2}
S.~Aminneborg, I.~Bengtsson, D.~Brill, S.~Holst, and P.~Peldan, {\it {Black
  holes and wormholes in (2+1)-dimensions}},  {\em Class. Quant. Grav.} {\bf
  15} (1998) 627--644, [\href{http://arxiv.org/abs/gr-qc/9707036}{{\tt
  arXiv:gr-qc/9707036}}].

\bibitem{Skenderis}
K.~Skenderis and B.~C. van Rees, {\it {Holography and wormholes in 2+1
  dimensions}},  {\em Commun. Math. Phys.} {\bf 301} (2011) 583--626,
  [\href{http://arxiv.org/abs/0912.2090}{{\tt arXiv:0912.2090}}].

\bibitem{koebe}
P.~Koebe, {\it {\"U}ber die uniformisierung der algebraischen kurven ii},  {\em
  Math. Ann} {\bf 69} (1910) 1--81.

\bibitem{Maxfield3D}
H.~Maxfield, {\it {Entanglement entropy in three dimensional gravity}},  {\em
  JHEP} {\bf 04} (2015) 031, [\href{http://arxiv.org/abs/1412.0687}{{\tt
  arXiv:1412.0687}}].

\bibitem{cones}
D.~Marolf, M.~Rota, and J.~Wien, {\it {Handlebody phases and the polyhedrality
  of the holographic entropy cone}},
  \href{http://arxiv.org/abs/1705.10736}{{\tt arXiv:1705.10736}}.

\bibitem{MBW1}
V.~Balasubramanian, P.~Hayden, A.~Maloney, D.~Marolf, and S.~F. Ross, {\it
  {Multiboundary Wormholes and Holographic Entanglement}},  {\em Class. Quant.
  Grav.} {\bf 31} (2014) 185015, [\href{http://arxiv.org/abs/1406.2663}{{\tt
  arXiv:1406.2663}}].

\bibitem{FEMgentle}
F.-J. Sayas, {\it A gentle introduction to the finite element method}, .

\bibitem{FEMlecture}
E.~S{\"u}li, {\it Lecture notes on finite element methods for partial
  differential equations}, .

\bibitem{MBW2}
D.~Marolf, H.~Maxfield, A.~Peach, and S.~F. Ross, {\it {Hot multiboundary
  wormholes from bipartite entanglement}},  {\em Class. Quant. Grav.} {\bf 32}
  (2015), no.~21 215006, [\href{http://arxiv.org/abs/1506.04128}{{\tt
  arXiv:1506.04128}}].

\bibitem{belin}
A.~Belin, C.~A. Keller, and I.~G. Zadeh, {\it {Genus two partition functions
  and Rényi entropies of large c conformal field theories}},  {\em J. Phys.}
  {\bf A50} (2017), no.~43 435401, [\href{http://arxiv.org/abs/1704.08250}{{\tt
  arXiv:1704.08250}}].

\end{thebibliography}\endgroup

\end{document}